\documentclass{LMCS}

\def\dOi{9(3:23)2013}
\lmcsheading%
{\dOi}
{1--21}
{}
{}
{Feb.~16, 2012}
{Sep.~18, 2013}
{}

\ACMCCS{[{\bf Theory of computation}]: Logic; Semantics and
  reasoning---Program semantics---Denotational semantics}

\usepackage{amsmath}
\usepackage{amssymb}
\usepackage{amstext}
\usepackage{xspace}
\usepackage{url}
\usepackage{xy}
\xyoption{all}
\usepackage{enumerate,hyperref}

\newdimen\proofrulebreadth \proofrulebreadth=.05em
\newdimen\proofdotseparation \proofdotseparation=1.25ex
\newdimen\proofrulebaseline \proofrulebaseline=2ex
\newcount\proofdotnumber \proofdotnumber=3
\let\then\relax
\def\hfi{\hskip0pt plus.0001fil}
\mathchardef\squigto="3A3B
\newif\ifinsideprooftree\insideprooftreefalse
\newif\ifonleftofproofrule\onleftofproofrulefalse
\newif\ifproofdots\proofdotsfalse
\newif\ifdoubleproof\doubleprooffalse
\let\wereinproofbit\relax
\newdimen\shortenproofleft
\newdimen\shortenproofright
\newdimen\proofbelowshift
\newbox\proofabove
\newbox\proofbelow
\newbox\proofrulename
\def\shiftproofbelow{\let\next\relax\afterassignment\setshiftproofbelow\dimen0 }
\def\shiftproofbelowneg{\def\next{\multiply\dimen0 by-1 }%
\afterassignment\setshiftproofbelow\dimen0 }
\def\setshiftproofbelow{\next\proofbelowshift=\dimen0 }
\def\setproofrulebreadth{\proofrulebreadth}

\def\prooftree{
\ifnum  \lastpenalty=1
\then   \unpenalty
\else   \onleftofproofrulefalse
\fi
\ifonleftofproofrule
\else   \ifinsideprooftree
        \then   \hskip.5em plus1fil
        \fi
\fi
\bgroup
\setbox\proofbelow=\hbox{}\setbox\proofrulename=\hbox{}%
\let\justifies\proofover\let\leadsto\proofoverdots\let\Justifies\proofoverdbl
\let\using\proofusing\let\[\prooftree
\ifinsideprooftree\let\]\endprooftree\fi
\proofdotsfalse\doubleprooffalse
\let\thickness\setproofrulebreadth
\let\shiftright\shiftproofbelow \let\shift\shiftproofbelow
\let\shiftleft\shiftproofbelowneg
\let\ifwasinsideprooftree\ifinsideprooftree
\insideprooftreetrue
\setbox\proofabove=\hbox\bgroup$\displaystyle 
\let\wereinproofbit\prooftree
\shortenproofleft=0pt \shortenproofright=0pt \proofbelowshift=0pt
\onleftofproofruletrue\penalty1
}

\def\eproofbit{
\ifx    \wereinproofbit\prooftree
\then   \ifcase \lastpenalty
        \then   \shortenproofright=0pt  
        \or     \unpenalty\hfil         
        \or     \unpenalty\unskip       
        \else   \shortenproofright=0pt  
        \fi
\fi
\global\dimen0=\shortenproofleft
\global\dimen1=\shortenproofright
\global\dimen2=\proofrulebreadth
\global\dimen3=\proofbelowshift
\global\dimen4=\proofdotseparation
\global\count255=\proofdotnumber
$\egroup  
\shortenproofleft=\dimen0
\shortenproofright=\dimen1
\proofrulebreadth=\dimen2
\proofbelowshift=\dimen3
\proofdotseparation=\dimen4
\proofdotnumber=\count255
}

\def\proofover{
\eproofbit 
\setbox\proofbelow=\hbox\bgroup 
\let\wereinproofbit\proofover
$\displaystyle
}%
\def\proofoverdbl{
\eproofbit 
\doubleprooftrue
\setbox\proofbelow=\hbox\bgroup 
\let\wereinproofbit\proofoverdbl
$\displaystyle
}%
\def\proofoverdots{
\eproofbit 
\proofdotstrue
\setbox\proofbelow=\hbox\bgroup 
\let\wereinproofbit\proofoverdots
$\displaystyle
}%
\def\proofusing{
\eproofbit 
\setbox\proofrulename=\hbox\bgroup 
\let\wereinproofbit\proofusing
\kern0.3em$
}

\def\endprooftree{
\eproofbit 
  \dimen5 =0pt
\dimen0=\wd\proofabove \advance\dimen0-\shortenproofleft
\advance\dimen0-\shortenproofright
\dimen1=.5\dimen0 \advance\dimen1-.5\wd\proofbelow
\dimen4=\dimen1
\advance\dimen1\proofbelowshift \advance\dimen4-\proofbelowshift
\ifdim  \dimen1<0pt
\then   \advance\shortenproofleft\dimen1
        \advance\dimen0-\dimen1
        \dimen1=0pt
        \ifdim  \shortenproofleft<0pt
        \then   \setbox\proofabove=\hbox{%
                        \kern-\shortenproofleft\unhbox\proofabove}%
                \shortenproofleft=0pt
        \fi
\fi
\ifdim  \dimen4<0pt
\then   \advance\shortenproofright\dimen4
        \advance\dimen0-\dimen4
        \dimen4=0pt
\fi
\ifdim  \shortenproofright<\wd\proofrulename
\then   \shortenproofright=\wd\proofrulename
\fi
\dimen2=\shortenproofleft \advance\dimen2 by\dimen1
\dimen3=\shortenproofright\advance\dimen3 by\dimen4
\ifproofdots
\then
        \dimen6=\shortenproofleft \advance\dimen6 .5\dimen0
        \setbox1=\vbox to\proofdotseparation{\vss\hbox{$\cdot$}\vss}%
        \setbox0=\hbox{%
                \advance\dimen6-.5\wd1
                \kern\dimen6
                $\vcenter to\proofdotnumber\proofdotseparation
                        {\leaders\box1\vfill}$%
                \unhbox\proofrulename}%
\else   \dimen6=\fontdimen22\the\textfont2 
        \dimen7=\dimen6
        \advance\dimen6by.5\proofrulebreadth
        \advance\dimen7by-.5\proofrulebreadth
        \setbox0=\hbox{%
                \kern\shortenproofleft
                \ifdoubleproof
                \then   \hbox to\dimen0{%
                        $\mathsurround0pt\mathord=\mkern-6mu%
                        \cleaders\hbox{$\mkern-2mu=\mkern-2mu$}\hfill
                        \mkern-6mu\mathord=$}%
                \else   \vrule height\dimen6 depth-\dimen7 width\dimen0
                \fi
                \unhbox\proofrulename}%
        \ht0=\dimen6 \dp0=-\dimen7
\fi
\let\doll\relax
\ifwasinsideprooftree
\then   \let\VBOX\vbox
\else   \ifmmode\else$\let\doll=$\fi
        \let\VBOX\vcenter
\fi
\VBOX   {\baselineskip\proofrulebaseline \lineskip.2ex
        \expandafter\lineskiplimit\ifproofdots0ex\else-0.6ex\fi
        \hbox   spread\dimen5   {\hfi\unhbox\proofabove\hfi}%
        \hbox{\box0}%
        \hbox   {\kern\dimen2 \box\proofbelow}}\doll%
\global\dimen2=\dimen2
\global\dimen3=\dimen3
\egroup 
\ifonleftofproofrule
\then   \shortenproofleft=\dimen2
\fi
\shortenproofright=\dimen3
\onleftofproofrulefalse
\ifinsideprooftree
\then   \hskip.5em plus 1fil \penalty2
\fi
}

\newcommand{\pullback}[1][dr]{\save*!/#1-1.2pc/#1:(-1,1)@^{|-}\restore}
\makeatother
 \newdir{ >}{{}*!/-7.5pt/@{>}}
 \newdir{|>}{!/4.5pt/@{|}*:(1,-.2)@^{>}*:(1,+.2)@_{>}}
 \newdir{ |>}{{}*!/-3pt/@{|}*!/-7.5pt/:(1,-.2)@^{>}*!/-7.5pt/:(1,+.2)@_{>}}
\newcommand{\xyline}[2][]{\ensuremath{\smash{\xymatrix@1#1{#2}}}}
\newcommand{\xyinline}[2][]{\ensuremath{\smash{\xymatrix@1#1{#2}}}^{\rule[8.5pt]{0pt}{0pt}}}
\makeatletter

\newif\ifignore 
\ignorefalse
\newcommand{\auxproof}[1]{
\ifignore\mbox{}\newline
\textbf{PROOF:} \dotfill\newline
{\it #1}\mbox{}\newline
\textbf{ENDPROOF}\dotfill
\fi}

\newenvironment{myproof}[1][Proof]%
   { \begin{trivlist}%
     \item[\hskip \labelsep {\bfseries #1}]%
   }%
   { \end{trivlist}%
   }

\newcommand{\after}{\mathrel{\circ}}
\newcommand{\cat}[1]{\ensuremath{\mathbf{#1}}}
\newcommand{\Cat}[1]{\ensuremath{\mathbf{#1}}}

\newcommand{\idmap}[1][]{\ensuremath{\mathrm{id}_{#1}}}

\newcommand{\support}{\ensuremath{\mathrm{supp}}}
\newcommand{\st}{\ensuremath{\mathsf{st}}}
\newcommand{\dst}{\ensuremath{\mathsf{dst}}}
\newcommand{\dis}{\ensuremath{\mathsf{dis}}}
\newcommand{\Alg}{\textsl{Alg}\xspace}
\newcommand{\CoAlg}{\textsl{CoAlg}\xspace}
\newcommand{\StMnd}{\textsl{StMnd}\xspace}

\newcommand{\Idl}{\textsl{Idl}\xspace}
\newcommand{\Dwn}{\textsl{Dwn}\xspace}
\newcommand{\Mlt}{\ensuremath{\mathcal{M}}}
\newcommand{\Dstr}{\ensuremath{\mathcal{D}}}
\newcommand{\E}{\ensuremath{\mathcal{E}}}

\newcommand{\scalar}{\mathrel{\bullet}}

\newcommand{\powerset}{\mathcal{P}}
\newcommand{\Pow}{\powerset}

\newcommand{\Sets}{\Cat{Sets}\xspace}
\newcommand{\PoSets}{\Cat{PoSets}\xspace}
\newcommand{\MSL}{\Cat{MSL}\xspace}
\newcommand{\Dcpo}{\Cat{Dcpo}\xspace}
\newcommand{\Frm}{\Cat{Frm}\xspace}
\newcommand{\BifMRel}{\Cat{BifMRel}\xspace}
\newcommand{\sotimes}{\mathrel{\raisebox{.05pc}{$\scriptstyle \otimes$}}}
\newcommand{\tuple}[1]{\ensuremath{\langle #1 \rangle}}
\newcommand{\set}[2]{\{#1\;|\;#2\}}
\newcommand{\setin}[3]{\{#1\in#2\;|\;#3\}}
\newcommand{\conjun}{\mathrel{\wedge}}

\newcommand{\ex}[2]{\exists{#1}.\,#2}
\newcommand{\exin}[3]{\exists{#1\in#2}.\,#3}

\newcommand{\downset}{\mathop{\downarrow}\!}
\def\ddownset{\mathop{\rlap{$\downarrow$}\raisebox{.4ex}{$\downarrow$}}}

\newcommand{\congrightarrow}{\mathrel{\stackrel{
           \raisebox{.5ex}{$\scriptstyle\cong\,$}}{
           \raisebox{0ex}[0ex][0ex]{$\rightarrow$}}}}
\newcommand{\conglongrightarrow}{\mathrel{\stackrel{
           \raisebox{.5ex}{$\scriptstyle\cong\,$}}{
           \raisebox{0ex}[0ex][0ex]{$\longrightarrow$}}}}

\renewcommand{\arraycolsep}{3pt}

\begin{document}

\title{Bases as Coalgebras}
\author[B. Jacobs]{Bart Jacobs}

\address{Institute for Computing and Information Sciences (iCIS), 
Radboud University Nijmegen, The Netherlands.} 
\urladdr{www.cs.ru.nl/B.Jacobs}


\keywords{category, monad, algebra, coalgebra, basis, 
  Kock-Z\"oberlein monad, comonoid, no-cloning}
\subjclass{F.4.1}

\begin{abstract}
The free algebra adjunction, between the category of algebras of a
monad and the underlying category, induces a comonad on the category
of algebras. The coalgebras of this comonad are the topic of study in
this paper (following earlier work).  It is illustrated how such
coalgebras-on-algebras can be understood as bases, decomposing each
element $x$ into primitives elements from which $x$ can be
reconstructed via the operations of the algebra. This holds in
particular for the free vector space monad, but also for other monads,
like powerset or distribution. For instance, continuous dcpos or
stably continuous frames, where each element is the join of the
elements way below it, can be described as such coalgebras. Further,
it is shown how these coalgebras-on-algebras give rise to a comonoid
structure for copy and delete, and thus to diagonalisation of endomaps
like in linear algebra.
\end{abstract}

\maketitle

\section{Introduction}\label{IntroSec}

The concept of basis in mathematics is best known for a vector space.
It involves a way of writing an arbitrary vector $v$ as finite linear
combination $v = \sum_{i} v_{i}e_{i}$, using special base vectors
$e_i$, which are mutually independent. In the current paper this
phenomenon will be studied at a more general level, using algebras and
coalgebras.  A vector space is an algebra --- to be precise, an
Eilenberg-Moore algebra of a particular monad --- and the
decompositions can be described as a coalgebra $v \mapsto \sum_{i}
v_{i}e_{i}$. The fact that bases are coalgebras is the main, novel
observation. It applies to other structures than vector spaces, like
directed complete partial orders and convex sets.

In general, algebras are used for composition and coalgebras for
decomposition.  An algebra $a\colon T(X)\rightarrow X$, for a functor
or a monad $T$, can be used to produce elements in $X$ from
ingredients structured by $T$. Conversely, a coalgebra $c\colon
X\rightarrow T(X)$ allows one to decompose an element in $X$ into its
ingredients with structure according to $T$. This is the fundamental
difference between algebraic and coalgebraic data structures. 

Assume an arbitrary category $\cat{A}$, carrying a monad $T$, as
in the lower left corner of the next diagram.
\begin{equation}
\label{MonadIterDiag}
\vcenter{\xymatrix@C-1pc{
& \Alg(T)
   \ar@(ul,ur)^-{\mbox{comonad }\overline{T}}
   \ar[dl]_{\dashv}\ar@/^3.5ex/[dr]
& &
\Alg(-)\ar@{--}[dl]\ar@(ul,ur)^-{\mbox{comonad}}
& & \cdots \\
\quad\cat{A}\quad\ar@(dr,dl)[]^-{\mbox{monad }T}\ar@/^3.5ex/[ur]
& &
\CoAlg(\overline{T})\ar[ul]_{\dashv}
   \ar@(dr,dl)[]^-{\mbox{monad }\overline{\overline{T}}}
& &
\CoAlg(-)\ar[ul]\ar@(dr,dl)[]^-{\mbox{monad}}\ar@{--}[ur] &
}}
\end{equation}

\noindent The category $\Alg(T)$ of algebras of the monad $T$ comes
with a standard adjunction $\Alg(T) \rightleftarrows \cat{A}$. This
adjunction induces a comonad $\overline{T}$ on $\Alg(T)$, see
Section~\ref{ComonadSec} for details. Then we can form the category
$\CoAlg(\overline{T})$ of coalgebras of the comonad $\overline{T}$,
with standard adjunction $\CoAlg(\overline{T}) \leftrightarrows
\Alg(T)$, inducing a monad $\smash{\overline{\overline{T}}}$ on
$\CoAlg(\overline{T})$. This process can be continued, and gives rise
to an alternating sequence of monads and comonads.

This situation~\eqref{MonadIterDiag} has been studied by various
authors, see
\textit{e.g.}~\cite{Barr69,RoseburghW91,Jacobs94b,Kock95}. One obvious
question is: does the sequence~\eqref{MonadIterDiag} stabilise?
Stabilisation after 2 steps is proven in~\cite{Barr69} for monads on
$\Sets$. Here we prove stabilisation in 3 steps for special monads (of
so-called Kock-Z{\"o}berlein type) on $\PoSets$, see below.

But more importantly, here it is proposed that a
$\overline{T}$-coalgebra on a $T$-algebra can be seen as a
\emph{basis} for this algebra, see Section~\ref{ComonadSec}. In
particular, in Section~\ref{SetSec} it will be shown that the concept
of basis in linear algebra gives rise to such a coalgebra
$X\rightarrow\overline{\Mlt}(X)$ for the multiset monad $\Mlt$; this
coalgebra decomposes an element $x$ of a vector space $X$ into a
formal sum $\sum_{i}x_{i}e_{i}\in\Mlt(X)$ given by its coefficients
$x_i$ for a Hamel basis $(e_{i})$, see Theorem~\ref{OperatorCoalgThm}
for more details. In the same vein, the operation that sends an
element of a convex set to a formal convex sum of extreme elements is
an instance of such a coalgebra.

Other examples arise in an order-theoretic setting, see
Section~\ref{OrderSec}.  Here one uses the notion of monad of
Kock-Z{\"o}berlein type --- where $T(\eta_{X}) \leq \eta_{TX}$,
see~\cite{Kock95,Escardo98}. We describe how such monads fit in the
present setting (with continuous dcpos as coalgebras), and add a new
result (Theorem~\ref{KZCoAlgBasisThm}) about
algebras-on-coalgebras-on-algebras, see Section~\ref{OrderSec}.  This
builds on rather old (little noticed) work of the
author~\cite{Jacobs94b}.

The first two steps of the sequence~\eqref{MonadIterDiag} are also
relevant in the semantics of effectful programming based on monads.
In~\cite{Levy06} it is shown that the exception monad transformer is a
monad itself---so it can play the role of $T$
in~(\ref{MonadIterDiag})---and that its algebras provide a syntax for
raising exceptions, whereas the associated coalgebra/basis takes care
of exception handling. These results are intriguing and need to be
tested and investigated further, but that is beyond the scope of this
paper.  We mention them only briefly in Section~\ref{ExceptionSec}.

In recent work~\cite{CoeckePV12} in the categorical foundations of
quantum mechanics it is shown that orthonormal bases in
finite-dimensional Hilbert spaces are equivalent to comonoids
structures (in fact, Frobenius algebras). These como\-noids are used
for copying and deleting elements. In Section~\ref{ComonoidSec} it is
shown how bases as coalgebras (capturing bases-as-decomposition) also
give rise to such comonoids (capturing
bases-as-copier-and-deleter). These comonoids can be used to formulate
in general terms what it means for an endomap to be diagonalised. This
is illustrated for a.o.\ the Pauli functions.

\section{Comonads on categories of algebras}\label{ComonadSec}

In this preliminary section we investigate the situation of a monad
and the induced comonad on its category of algebras. We shall see that
coalgebras of this comonad capture the notion of basis, in a very
general sense. This will be illustrated later in several situations
see in particular Subsection~\ref{VectorSpaceSubsec}.

For an arbitrary monad $T\colon \cat{A} \rightarrow \cat{A}$, with
unit $\eta\colon \idmap \Rightarrow T$ and multiplication $\mu\colon
T^{2} \Rightarrow T$, there is a category $\Alg(T)$ of
(Eilenberg-Moore) algebras, together with a left adjoint $F$ (for free
algebra functor) to the forgetful functor $U\colon
\Alg(T)\rightarrow\cat{A}$. This adjunction $\Alg(T) \leftrightarrows
\cat{A}$ induces a comonad on the category $\Alg(T)$, which we shall
write as $\overline{T} = FU$ in:
\vspace*{-1em}
\begin{equation}
\label{ComonadFromMonadDiag}
\vcenter{\xymatrix@R-.5pc@C-1pc{
& & \\
\Alg(T)\ar[d]_{\dashv}^{U}
   \ar `u[r] `[rd] `[u]^{\overline{T}=FU\;\mbox{\small comonad}} [] & \\
\cat{A}\ar@/^3ex/[u]^{F}\ar `r[rd] `[r]^{T=UF\;\mbox{\small monad}} `[lr] [] 
   & \\
& & 
}}
\end{equation}

\noindent For an algebra $\smash{(TX\stackrel{a}{\rightarrow} X)} \in
\Alg(T)$ there are counit $\varepsilon\colon \overline{T} \Rightarrow
\idmap$ and comultiplication $\delta\colon \overline{T} \Rightarrow
\overline{\overline{T}}$ maps in $\Alg(T)$ given by:
\begin{equation}
\label{ComonadFromMonadStructEqn}
\vcenter{\xymatrix{
\ensuremath{\left(\xy
(0,4)*{TX};
(0,-4)*{X};
{\ar^{a} (0,2); (0,-2)};
\endxy\right)}
& &
\ensuremath{\left(\xy
(0,4)*{T^{2}X};
(0,-4)*{TX};
{\ar^{\mu\rlap{$\scriptscriptstyle{_X}$}} (0,2); (0,-2)};
\endxy\right)}\ar[ll]_-{\varepsilon = a}\ar[rr]^-{\delta = T(\eta_{X})}
& &
\ensuremath{\left(\xy
(0,4)*{T^{3}X};
(0,-4)*{T^{2}X};
{\ar^{\mu_{T\!\rlap{$\scriptscriptstyle X$}}} (0,2); (0,-2)};
\endxy\right)}
}}
\end{equation}


\auxproof{
In general, for an adjunction $F\dashv G$ the induced comonad $FG$ 
comes with counit $\varepsilon \colon FG \rightarrow \idmap$
inherited from the adjunction and comultiplication $\delta \colon
FG \rightarrow FGFG$ given by $\delta = F\eta G$.

In this case, starting from the free algebra adjunction $F\dashv U$
the unit is the unit of the monad, and the counit $FU \rightarrow
\idmap$, at an algebra, is this algebra itself, as a map of algebras:
$$\xymatrix{
FU\ensuremath{\left(\xy
(0,4)*{TX};
(0,-4)*{X};
{\ar^{a} (0,2); (0,-2)};
\endxy\right)}
= \ensuremath{\left(\xy
(0,4)*{T^{2}X};
(0,-4)*{TX};
{\ar^{\mu_X} (0,2); (0,-2)};
\endxy\right)}\ar[rr]^-{a}
& &
\ensuremath{\left(\xy
(0,4)*{TX};
(0,-4)*{X};
{\ar^{a} (0,2); (0,-2)};
\endxy\right)}
}$$
}

\begin{defi}
\label{BasisDef}
Consider a monad $T \colon \cat{A} \rightarrow \cat{A}$ together with
the comonad $\overline{T} \colon \Alg(T) \rightarrow \Alg(T)$ induced by $T$,
as in~(\ref{ComonadFromMonadDiag}). A \emph{basis} for a $T$-algebra
$\smash{(TX\stackrel{a}{\rightarrow} X) \in \Alg(T)}$ is a
$\smash{\overline{T}}$-coalgebra on this algebra, given by a map of
algebras $b$ of the form:
$$\xymatrix@C-.5pc{
\ensuremath{\left(\xy
(0,4)*{TX};
(0,-4)*{X};
{\ar^{a} (0,2); (0,-2)};
\endxy\right)}\ar[rr]^-{b}
& &
\overline{T}\ensuremath{\left(\xy
(0,4)*{TX};
(0,-4)*{X};
{\ar^{a} (0,2); (0,-2)};
\endxy\right)}
=
FU\ensuremath{\left(\xy
(0,4)*{TX};
(0,-4)*{X};
{\ar^{a} (0,2); (0,-2)};
\endxy\right)}
=
\ensuremath{\left(\xy
(0,4)*{T^{2}X};
(0,-4)*{TX};
{\ar^{\mu\rlap{$\scriptscriptstyle{_X}$}} (0,2); (0,-2)};
\endxy\right)}
}$$

\noindent Thus, a basis $b$ is a map $\smash{X\xrightarrow{b} TX}$ in
$\cat{A}$ satisfying $b \after a = \mu_{X} \after T(b)$ and $a \after
b = \idmap$ and $T(\eta_{X}) \after b = T(b) \after b$ in:
$$\xymatrix@R-.5pc{
T(X)\ar[d]_{a}\ar[r]^-{T(b)} & T^{2}(X)\ar[d]^{\mu_X}
& &
X\ar@{=}[dr]\ar[r]^-{b} & T(X)\ar[d]^{\varepsilon = a}
& &
T(X)\ar[r]^-{T(b)} & T^{2}(X) \\
X\ar[r]_-{b} & T(X)
& &
& X
& &
X\ar[u]^-{b}\ar[r]_{b} & T(X)\ar[u]_{\delta=T(\eta_{X})} 
}$$
\end{defi}

\noindent As we shall see a basis as described above may be understood as
providing a decomposition of each element $x$ of an algebra into a
collection $b(x)$ of basic elements that together form $x$.  The
actual basic elements $X_{b} \rightarrowtail X$ involved can be
obtained as the indecomposable ones, via the following equaliser in
the underlying category.
\begin{equation}
\label{BasisEqualiserDiag}
\vcenter{\xymatrix@C+1pc{
X_{b}\ar@{ >->}[r]^-{e} & X \ar@<+.5ex>@/^1ex/[r]^-{b}
   \ar@<-.5ex>@/_1ex/[r]_-{\eta} & T\rlap{$(X)$}
}}
\end{equation}

\noindent One can then ask in which cases the map of algebras
$T(X_{b}) \rightarrow X$, induced by the equaliser $e\colon X_{b}
\rightarrow U(TX\rightarrow X)$, is an isomorphism. This is (almost
always) the case for monads on \Sets, see
Proposition~\ref{SetsBasisProp} below. But first we observe that free
algebras always carry a basis.

\begin{lem}
\label{FreeAlgCoalgLem}
Free algebras have a canonical basis: each $\smash{FX =
  \big(T^{2}X \stackrel{\mu}{\rightarrow} TX\big)} \in \Alg(T)$
carries a $\overline{T}$-coalgebra, namely given by $T(\eta_{X})$. This
gives a situation:
$$
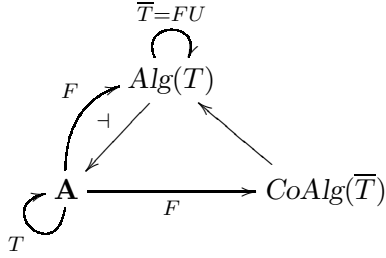
\xymatrix@C-1pc{
& \Alg(T)
   \ar@(ul,ur)^-{\overline{T}=FU}
   \ar[dl]_{\dashv}
& \\
\cat{A}\ar@(d,l)[]^-{T}\ar@/^3.5ex/[ur]^{F}\ar[rr]_-{F}
   & & \CoAlg\rlap{$(\overline{T})$}\ar[ul]
}$$
\end{lem}

\begin{myproof}
It is easy to check that $T(\eta_{X})$ is a morphism in $\Alg(T)$ and
a $\overline{T}$-coalgebra:
$$\vcenter{\xymatrix{
F(X) = \ensuremath{\left(\xy
(0,4)*{T^{2}X};
(0,-4)*{TX};
{\ar^{\mu\rlap{$\scriptscriptstyle{_X}$}} (0,2); (0,-2)};
\endxy\right)}\ar[rr]^-{T(\eta_{X})}
& &
\ensuremath{\left(\xy
(0,4)*{T^{3}X};
(0,-4)*{T^{2}X};
{\ar^{\mu_{T\!\rlap{$\scriptscriptstyle X$}}} (0,2); (0,-2)};
\endxy\right)} = \overline{T}(FX).
}}\eqno{\qEd}$$

\auxproof{
$$\begin{array}{rcl}
\varepsilon \after T(\eta_{X})
& = &
\mu_{X} \after T(\eta_{X}) \\
& = &
\idmap \\
\delta \after T(\eta_{X})
& = &
T(\eta_{TX}) \after T(\eta_{X}) \\
& = &
T(\eta_{TX} \after \eta_{X}) \\
& = &
T(T(\eta_{X}) \after \eta_{X}) \\
& = &
T^{2}(\eta_{X}) \after T(\eta_{X}) \\
& = &
\overline{T}(T(\eta_{X})) \after T(\eta_{X}).
\end{array}$$
}
\end{myproof}

\noindent The object $X_b$ of basic elements, as
in~(\ref{BasisEqualiserDiag}), in the situation of this lemma is the
original set $X$ in case the monad $T$ satisfies the so-called
\emph{equaliser requirement}~\cite{Moggi88}, which says precisely that
$\eta_{X} \colon X \rightarrow TX$ is the equaliser of $T(\eta_{X}),
\eta_{TX} \colon TX \rightrightarrows T^{2}X$. This requirement does
\emph{not} hold, for instance, for the powerset monad.


There is some redundancy in the data described in
Definition~\ref{BasisDef}. This is implicitly used in the description
of syntax for exception in~\cite{SchroderM04} (with only `handle' as
coalgebra and no `raise' algebra, see also~\cite{Levy06} and
Section~\ref{ExceptionSec}).

\begin{lem}
Assume a monad $T\colon\cat{A} \rightarrow \cat{A}$, with induced
comonad $\smash{\overline{T}} \colon \Alg(T) \rightarrow \Alg(T)$.
Having:
\begin{iteMize}{$\bullet$}
\item a $T$-algebra $a\colon T(X) \rightarrow X$ together with a
  $\smash{\overline{T}}$-coalgebra $b\colon a \rightarrow
  \overline{T}(a)$
\end{iteMize}

\noindent is the same as having:
\begin{iteMize}{$\bullet$}
\item a map $c\colon X \rightarrow T(X)$ in $\cat{A}$ which forms an
  equaliser diagram in $\cat{A}$:
$$\xymatrix@C+.5pc{
X\ar@{ >->}[r]^-{c} & T(X) \ar@<+.5ex>@/^1ex/[r]^-{T(c)}
   \ar@<-.5ex>@/_1ex/[r]_-{T(\eta)}
    & T(X)
}$$
\end{iteMize}
\end{lem}

\begin{myproof}
Assuming an algebra and coalgebra $(a,b)$ as above, it is easy to check
that $b\colon X\rightarrow T(X)$ is the equaliser of $T(b)$ and
$T(\eta)$.

\auxproof{
Recall that $\delta = T(\eta)$, so $T(\eta) \after b = \delta \after b
= T(b) \after b$. If also $f\colon Y\rightarrow T(X)$ satisfies
$T(b) \after f = T(\eta) \after f$, then $a \after f\colon Y \rightarrow
X$ satisfies:
$$\begin{array}{rcccccl}
b \after a \after f
& = &
\mu \after T(b) \after f
& = &
\mu \after T(\eta) \after f
& = &
f.
\end{array}$$

\noindent Moreover, $a \after f$ is the only such map: if $g\colon Y
\rightarrow X$ also satisfies $b \after g = f$, then $g = a \after b
\after g = a \after f$.
}

The other direction is a bit more work: assume we have $c\colon X
\rightarrow T(X)$ forming an equaliser of $T(c)$ and $T(\eta)$. 
Consider the map $a$ defined in the diagram:
$$\xymatrix@C+.5pc@R-.5pc{
X\ar@{ >->}[r]^-{c} & T(X) \ar@<+.5ex>[r]^-{T(c)}\ar@<-.5ex>[r]_-{T(\eta)}
    & T(X) \\
T(X)\ar@{..>}[u]^{a}\ar[r]_-{T(c)} & T^{2}(X)\ar[u]_{\mu} 
}$$

\noindent Using the equaliser property one checks that $a$ is a
$T$-algebra, and that $c\colon X\rightarrow T(X)$ satisfies the
$\overline{T}$-coalgebra requirements from
Definition~\ref{BasisDef}. \qed

\auxproof{
First we need to check that the map $\mu \after T(c)$ actually
equalises:
$$\begin{array}{rcl}
T(c) \after \mu \after T(c)
& = &
\mu \after T^{2}(c) \after T(c) \\
& = &
\mu \after T^{2}(\eta) \after T(c) \\
& = &
T(\eta) \after \mu \after T(c).
\end{array}$$

\noindent Further, $a\colon T(X) \rightarrow X$ is an algebra:
$$\begin{array}{rcl}
c \after a \after \eta
& = &
\mu \after T(c) \after \eta \\
& = &
\mu \after \eta \after c \\
& = &
c \\
& = &
c \after \idmap \\
c \after a \after T(a) 
& = &
\mu \after T(c) \after T(a) \\
& = &
\mu \after T(\mu \after T(c)) \\
& = &
\mu \after \mu \after T^{2}(c) \\
& = &
\mu \after T(c) \after \mu \\
& = &
c \after a \after \mu.
\end{array}$$

\noindent Obviously, $\delta \after c = T(c) \after c$, since $\delta
= T(\eta)$. By construction of $a$, $c$ is a map of algebras $a
\rightarrow \mu$. Hence we only need to verify:
$$\begin{array}{rcl}
c \after a \after c
& = &
\mu \after T(c) \after c \\
& = &
\mu \after T(\eta) \after c \\
& = &
c \\
& = &
c \after \idmap.
\end{array}$$
}
\end{myproof}

The comonad $\overline{T}\colon\Alg(T)\rightarrow\Alg(T)$
from~(\ref{ComonadFromMonadDiag}) gives rise to a category of
coalgebras $\CoAlg(\overline{T})\rightarrow\Alg(T)$, where this
forgetful functor has a right adjoint, which maps an algebra
$TY\rightarrow Y$ to the diagonal coalgebra $\delta\colon\mu_{Y}
\rightarrow\mu_{TY}$ as in~(\ref{ComonadFromMonadStructEqn}).  Thus we
obtain a monad on the category $\CoAlg(\overline{T})$, written as
$\overline{\overline{T}}$, like in the
sequence~\eqref{MonadIterDiag}. On a basis $c\colon a\rightarrow
\overline{T}(a)$, for an algebra $a\colon TX\rightarrow X$, there is a
unit $\eta_{c} = c\colon c\rightarrow\delta$ and multiplication
$\mu_{c} = T(c)\colon \delta\rightarrow\delta$ in
$\CoAlg(\overline{T})$.

\auxproof{ 
Explicitly, for algebras $a\colon TX\rightarrow X$ and $b\colon
TY\rightarrow Y$ and a basis $c\colon a\rightarrow \overline{T}$
there is a bijective correspondence:
$$\begin{prooftree}
{\xymatrix{
\Big(a\stackrel{c}{\rightarrow}\overline{T}(a)\Big)\ar[r]^-{f} & 
   \Big(\mu_{Y}\stackrel{\delta}{\rightarrow}\overline{T}(\mu_{Y})\Big)}}
\Justifies
{\xymatrix{
\Big(TX\stackrel{a}{\rightarrow}X\Big)\ar[r]_-{g} &
   \Big(TY\stackrel{b}{\rightarrow}Y\Big)}}
\end{prooftree}$$

\noindent It is given by:
\begin{iteMize}{$\bullet$}
\item for $f\colon X\rightarrow TY$ take $\overline{f} = b \after f$.
It is a map of algebras $a\rightarrow b$ since:
$$\begin{array}{rcll}
b \after T(\overline{f})
& = &
b \after T(b) \after T(f) \\
& = &
b \after \mu \after T(f) \\
& = &
b \after f \after a
  & \mbox{since $f$ is a map of algebras $a\rightarrow\mu_Y$} \\
& = &
\overline{f} \after a.
\end{array}$$

\item for $g\colon X\rightarrow Y$ take $\overline{g} = T(g) \after c$.
This is both a map of algebras $a\rightarrow\mu_{Y}$ and of coalgebras
$c\rightarrow\delta$:
$$\begin{array}{rcll}
\mu \after T(\overline{g})
& = &
\mu \after T^{2}(g) \after T(c) \\
& = &
T(g) \after \mu \after T(c) \\
& = &
T(g) \after c \after a 
  & \mbox{since $c$ is a map of algebra $a\rightarrow\mu_{Y}$} \\
& = &
\overline{g} \after a \\
\delta \after \overline{g}
& = &
T(\eta) \after T(g) \after c \\
& = &
T^{2}(g) \after T(\eta) \after c \\
& = &
T^{2}(g) \after T(c) \after c 
   & \mbox{since $\delta \after c = T(c) \after c$} \\
& = &
T(\overline{g}) \after c
\end{array}$$
\end{iteMize}

\noindent Finally,
$$\begin{array}{rcl}
\overline{\overline{f}}
& = &
T(\overline{f}) \after c \\
& = &
T(b \after f) \after c \\
& = &
T(b) \after \delta \after f \\
& = &
T(b \after \eta) \after f \\
& = &
f \\
\overline{\overline{g}}
& = &
b \after \overline{g} \\
& = &
b \after T(g) \after c \\
& = &
g \after a \after c \\
& = &
g.
\end{array}$$

In particular, for a $\overline{T}$-coalgebra $c\colon X\rightarrow
T(X)$ on an algebra $a\colon T(X)\rightarrow X$ we have unit $\eta_{c} =
c\colon c\rightarrow\delta$ in $\CoAlg(\overline{T})$ and counit
$\varepsilon_{b} = b\colon \mu\rightarrow b$ in $\Alg(T)$ of this
adjunction $\CoAlg(\overline{T}) \rightleftarrows \Alg(T)$. The induced
monad $\overline{\overline{T}} \colon \CoAlg(\overline{T}) \rightarrow
\CoAlg(\overline{T})$ thus has multiplication $\mu_{c} = T(c)\colon
\delta\rightarrow\delta$ in a situation:
$$\xymatrix{
\Big(\mu_{TX}\stackrel{\delta}{\longrightarrow}\mu_{T^{2}X}\Big)
   \ar[rr]^-{\mu = T(a)} & &
\Big(\mu_{X}\stackrel{\delta}{\longrightarrow}\mu_{TX}\Big)
}$$

\noindent using that $T(a) \after \mu_{X} = \mu_{TX} \after T^{2}(a)$
and $\delta \after T(a) = T(\eta \after a) = T(T(a) \after \eta) =
T^{2}(a) \after \delta$.
}

By iterating this construction as in~\eqref{MonadIterDiag} one obtains
alternating monads and comonads. Such iterations are studied for
instance in~\cite{Barr69,RoseburghW91,Jacobs94b,Kock95}. In special
cases it is known that the iterations stop after a number of
cycles. This happens after 2 iterations for monads on sets, as we
shall see next, and after 3 iterations for Kock-Z{\"o}berlein monads
in Section~\ref{OrderSec}. This stabilisation means that in presence
of sufficiently many iterated (co)algebraic operations, the algebraic
structure that we start from becomes free --- typically on some atoms
or basic elements.

\section{Set-theoretic examples}\label{SetSec}

It turns out that for monads on the category \Sets only free algebras
have bases. This result goes back to~\cite{Barr69}. We repeat it in
the present context, with a sketch of proof.  Subsequently we describe
the situation for the powerset monad (from~\cite{Jacobs94b}), the
free vector space monad, and the distribution monad.

\begin{prop}
\label{SetsBasisProp}
For a monad $T$ on \Sets, if an algebra
$\smash{TX\stackrel{a}{\rightarrow} X}$ has a basis
$\smash{X\stackrel{b}{\rightarrow} \overline{T}X}$ with non-empty
equaliser $X_{b} \rightarrowtail X \rightrightarrows TX$ as
in~(\ref{BasisEqualiserDiag}), then the induced map $T(X_{b})
\rightarrow X$ is an isomorphism of algebras and coalgebras. In
particular, in the set-theoretic case any algebra with a non-empty
basis is free.
\end{prop}

\begin{myproof}
Let's consider the equaliser $X_{b} \rightarrowtail X$ of
$b,\eta\colon X\rightrightarrows T(X)$ from~(\ref{BasisEqualiserDiag}) in
\Sets. It is a so-called coreflexive equaliser, because there is a map
$TX\rightarrow X$, namely the algebra $a$, satisfying $a \after b =
\idmap = a \after \eta$. It is well-known---see
\textit{e.g.}~\cite[Lemma~6.5]{Mesablishvili06} or the dual result
in~\cite[Volume~I, Example~2.10.3.a]{Borceux94}---that if
$X_{b}\neq\emptyset$ such coreflexive equalisers in \Sets are split,
and thus absolute. The latter means that they are preserved under any
functor application. In particular, by applying $T$ we obtain a new
equaliser in \Sets, of the form:
\begin{equation}
\label{RestrictedBaseDiag}
\vcenter{\xymatrix@R-1pc@C+0.5pc{
T(X_{b})\ar@{ >->}[r]^-{T(e)} & T(X) 
   \ar@<+.5ex>[r]^-{T(b)}\ar@<-.5ex>[r]_-{T(\eta) = \delta} & T^{2}(X) \\
X\ar[ur]_{b}\ar@{-->}[u]^{b'}
}}
\end{equation}

\noindent The resulting map $b'$ is the inverse to the adjoint transpose
$a \after T(e) \colon T(X_{b}) \rightarrow X$, since:
\begin{iteMize}{$\bullet$}
\item $a \after T(e) \after b' = a \after b = \idmap$;

\item the other equation follows because $T(e)$ is equaliser, and thus
  mono:
$$\begin{array}{rcll}
T(e) \after b' \after a \after T(e)
& = &
b \after a \after T(e) \\
& = &
\mu \after T(b) \after T(e) \quad & \mbox{see Definition~\ref{BasisDef}} \\
& = &
\mu \after T(\eta) \after T(e) & \mbox{since $e$ is equaliser} \\
& = &
T(e) \\
& = &
T(e) \after \idmap.
\end{array}$$
\end{iteMize}

\noindent Hence the homomorphism of algebras $a \after T(e)$, from
$F(X_{b}) = \mu_{X_{b}}$ to $a$ is an isomorphism. In particular,
$b'\colon X\rightarrow T(X_{b})$ in~(\ref{RestrictedBaseDiag}) is a
map of algebras, as inverse of an isomorphism of algebras. It is not
hard to see that it is also an isomorphism between the coalgebras
$b\colon X\rightarrow T(X)$ and $T(\eta) \colon T(X_{b}) \rightarrow
T^{2}(X_{b})$, as in Lemma~\ref{FreeAlgCoalgLem}. \qed

\auxproof{ 
We include the split equaliser argument in \Sets,
  following~\cite[Lemma~6.5]{Mesablishvili06}.
Consider a diagram:
$$\vcenter{\xymatrix{
X\ar[r]^-{h} & Y\ar@<+1ex>[r]^-{f}\ar@<-1ex>[r]_-{g} & Z
}}\eqno{(*)}$$

\noindent It is called a \textit{split equaliser} if there are maps
$k\colon Z\rightarrow Y$ and $\ell\colon Y\rightarrow X$ with
$$\ell \after h = \idmap
\qquad
k \after f = \idmap
\qquad
k\after g = h \after \ell.$$

\noindent Such a split equaliser is indeed an equaliser:
\begin{iteMize}{$\bullet$}
\item if $h'\colon X'\rightarrow Y$ satisfies $f \after h' = g \after
  h'$, then $e = \ell \after h'\colon X'\rightarrow X$ satisfies: $h
  \after e = h \after \ell \after h' = k \after g \after h' = k \after
  f \after h' = \idmap \after h' = h'$.

\item if also $d\colon X'\rightarrow X$ satisfies $h \after d = h'$,
then $e = \ell \after h' = \ell \after h \after d = \idmap \after d = d.$
\end{iteMize}

\noindent Clearly, split equalisers are preserved under functor
application, and are thus absolute.

Now assume in~$(*)$ that $X\neq \emptyset$, $h$ is mono and satisfies
$f \after h = g \after h$, and that $f,g$ form a coreflexive pair, via
$s\colon Z\rightarrow Y$ satisfying $s \after f = \idmap = s \after
g$. Thus we can form the cokernel pair $q_{1},q_{2}\colon
Y\rightrightarrows D$ of $h$ as pushout:
$$\xymatrix@R-1pc{
X\ar[r]^-{h}\ar[d]_{h} & Y\ar[d]^{q_2}\ar@/^2ex/[ddr]^{g} \\
Y\ar[r]_-{q_1}\ar@/_2ex/[drr]_{f} & D\pullback[lu]\ar@{-->}[dr]^-{r} \\
& & Z
}$$

\noindent The first claim is that this induced map $r$ is injective.
We can construct $D$ explicitly as $Y+Y/\!\equiv$, where $\equiv$ is
the least equivalence relation on $Y+Y$ containing the pairs
$\tuple{\kappa_{1}h(x), \kappa_{2}h(x)}$, for $x\in X$. Thus,
$\equiv$ contains the following elements:
\begin{iteMize}{$\bullet$}
\item $\tuple{\kappa_{1}y, \kappa_{1}y}$, for $y\in Y$;
\item $\tuple{\kappa_{2}y, \kappa_{2}y}$, for $y\in Y$;
\item $\tuple{\kappa_{1}h(x), \kappa_{2}h(x)}$, for $x\in X$;
\item $\tuple{\kappa_{2}h(x), \kappa_{1}h(x)}$, for $x\in X$.
\end{iteMize}

\noindent Reflexivity and symmetry are obvious, by construction.  For
transitivity, assume we have pairs $\tuple{\kappa_{1}h(x),
  \kappa_{2}h(x)}$ and $\tuple{\kappa_{2}h(x'), \kappa_{1}h(x')}$,
where $h(x) = h(x')$. The composed pair $\tuple{\kappa_{1}h(x),
  \kappa_{1}h(x')}$ is then also covered by the abover description,
namely by the first item.

An element $[w]$ of $Y+Y/\!\equiv$ is thus of the form:
\begin{iteMize}{$\bullet$}
\item $[w] = \{\kappa_{1}h(x), \kappa_{2}h(x)\}$, in which case
$r([w]) = f(h(x)) = g(h(x))$, and thus $(s \after r)([w]) = h(x)$;
\item $[w] = \{\kappa_{1}y\}$, where $y\neq h(x)$, and then $r([w]) =
  f(y)$, and thus $(s \after r)([w]) = y$;
\item $[w] = \{\kappa_{2}y\}$, where $y\neq h(x)$, and then $r([w]) =
  g(y)$, and thus $(s \after r)([w]) = y$.
\end{iteMize}

\noindent If $r([w]) = r([w'])$, then by applying $s\colon
Z\rightarrow Y$, we see:
\begin{iteMize}
\item if $[w] = \{\kappa_{1}h(x), \kappa_{2}h(x)\}$, then the second
and third cases cannot occur for $w'$, so $[w'] = \{\kappa_{1}h(x'),
\kappa_{2}h(x')\}$ for some $x'$. But then $h(x) = h(x')$, and
so $[w]=[w']$.

\item if $[w] = \{\kappa_{1}y\}$ with $y\neq h(x)$, the first case for
  $[w']$ cannot occur; the second case immediatly yields $[w] = [w']$,
  so we concentrate on the third: if $[w'] = \{\kappa_{2}y'\}$, then
  $f(y) = g(y')$, and thus $y = y'$. Now we use that $h$ is the
  equaliser of $f,g$, so that there is an $x$ with $h(x) \neq y$.  But
  this is impossible.

\item The case $[w] = \{\kappa_{2}y\}$ is handled similarly.
\end{iteMize}

Since $X\neq\emptyset$ and $h\colon X\rightarrow Y$ is injective (as
equaliser) there is a section $i\colon Y\rightarrow X$ with $i \after
h = \idmap$, making $h$ a split mono. Also, $D\neq \emptyset$, so that
there is an $r'\colon Z\rightarrow D$ with $r' \after r = \idmap$,
using that $r$ is injective. We know form a map $j$ in:
$$\xymatrix@R-1pc{
X\ar[r]^-{h}\ar[d]_{h} & Y\ar[d]^{q_2}\ar@/^2ex/[ddr]^{h \after i} \\
Y\ar[r]_-{q_1}\ar@/_2ex/[drr]_{\idmap} & D\pullback[lu]\ar@{-->}[dr]^-{j} \\
& & Y
}$$

\noindent By construction, this pair $i\colon Y\rightarrow X, j\colon
D\rightarrow Y$ makes $h\colon X\rightarrowtail Y \rightrightarrows D$
a split equaliser.

In a next step $i\colon Y\rightarrow X, j\after r'\colon Z\rightarrow
Y$ makes $(*)$ a split equaliser, since $j \after r' \after f
= j \after r' \after r \after q_{1} = j \after q_{1} = \idmap$ and
$j \after r' \after g = j \after r' \after r \after q_{2} = j \after
q_{2} = h \after i$.

\medskip

Finally we also check that $\varphi = a \after T(e) \colon T(B)
\congrightarrow X$ is a map of coalgebras, in:
$$\xymatrix@R-1pc{
T^{2}(B)\ar[rr]^-{T(\varphi)}_-{\cong} & & T(X) \\
T(B)\ar[u]^{T(\eta)}\ar[rr]^-{\varphi}_-{\cong} & & X\ar[u]_{b}
}$$

\noindent This is simple:
$$\begin{array}{rcl}
b \after \varphi
& = &
b \after a \after T(e) \\
& = &
\mu \after T(b) \after T(e) \\
& = &
\mu \after T(\eta) \after T(e) \\
& = &
T(e) \\
& = &
T(a) \after T(\eta) \after T(e) \\
& = &
T(a) \after T^{2}(e) \after T(\eta) \\
& = &
T(\varphi) \after T(\eta).
\end{array}$$
}
\end{myproof}

\subsection{Complete lattices}\label{CompleteLatticeSubsec}

Consider the powerset monad $\Pow$ on \Sets, with the category
$\Cat{CL} = \Alg(\Pow)$ of complete lattices and join-preserving maps
as its category of algebras.  The induced comonad
$\smash{\overline{\Pow}} \colon \Cat{CL} \rightarrow \Cat{CL}$ as
in~(\ref{ComonadFromMonadDiag}) sends a complete lattice $(L,\leq)$ to
the lattice $(\Pow(L), \subseteq)$ of subsets, ignoring the original
order $\leq$.  The counit $\varepsilon \colon
\smash{\overline{\Pow}(L)} \rightarrow L$ sends a subset $U\in\Pow(L)$
to its join $\varepsilon(U) = \bigvee U$; the comultiplication $\delta
\colon {\overline{\Pow}(L)} \rightarrow
\smash{\overline{\Pow}^{2}(L)}$ sends $U\in\Pow(L)$ to the subset of
singletons $\delta(U) = \set{\{x\}}{x\in U}$.

An (Eilenberg-Moore) coalgebra of the comonad
$\smash{\overline{\Pow}}$ on $\Cat{CL}$ is a map $b\colon L
\rightarrow \smash{\overline{\Pow}(L)}$ in $\Cat{CL}$ satisfying
$\varepsilon \after b = \idmap$ and $\delta \after b =
\overline{\Pow}(b) \after b$. More concretely, this says that $\bigvee
b(x) = x$ and $\set{\{y\}}{y\in b(x)} = \set{b(y)}{y\in b(x)}$. It is
shown in~\cite{Jacobs94b} that a complete lattice $L$ carries
such a coalgebra structure $b$ if and only if $L$ is \emph{atomic},
where:
$$\begin{array}{rcl}
b(x)
& = &
\setin{a}{L}{a \mbox{ is an atom with }a \leq x}.
\end{array}$$

\noindent Thus, such a coalgebra of the comonad
$\smash{\overline{\Pow}}$, if it exists, is uniquely determined and
gives a decomposition of lattice elements into the atoms below it. The
atoms in the lattice thus form a basis.

(Recall: the complete lattice $L$ is atomic when each element is the
join of the atoms below it. And an atom $a\in L$ is a non-zero element
with no non-zero elements below it, satisfying: $a \leq \bigvee U$
implies $a \leq x$ for some $x\in U$.)

The equaliser~(\ref{BasisEqualiserDiag}) for the basic elements in
this situation, for an atomic complete lattice $L$, is the set of
atoms: 
$$X_{b} 
= 
\setin{x}{L}{\{x\} = b(x)} 
= 
\setin{x}{L}{x\mbox{ is an atom}}.$$ 

\noindent If $X_{b}\neq \emptyset$, the induced map $\Pow(X_{b})
\rightarrow L$ is an isomorphism, by Lemma~\ref{SetsBasisProp}.

\subsection{Vector spaces}\label{VectorSpaceSubsec}

For a semiring $S$ one can define the multiset monad $\Mlt_{S}$ on
$\Sets$ by $\Mlt_{S}(X) = \set{\varphi\colon X\rightarrow
  S}{\support(\varphi)\mbox{ is finite}}$. Such an element $\varphi$
can be identified with a formal finite sum $\sum_{i}s_{i}x_{i}$ with
multiplicities $s_{i}\in S$ for elements $x_{i}\in X$. The unit of
this monad $\eta\colon X \rightarrow \Mlt_{S}(X)$ is given by
singleton multisets: $\eta(x) = 1x$. The multiplication $\mu \colon
\Mlt_{S}^{2}(X) \rightarrow \Mlt_{S}(X)$ involves (matrix)
multiplication: $\mu(\sum_{i}s_{i}\varphi_{i})(x) = \sum_{i}
s_{i}\cdot \varphi_{i}(x)$, where $\cdot$ is the multiplication of the
semiring $S$.

The category of algebras $\Alg(\Mlt_{S})$ of the multiset monad
$\Mlt_S$ is the category of $\Cat{Mod}_S$ of modules over $S$:
commutative monoids with $S$-scalar multiplication, see
\textit{e.g.}~\cite{CoumansJ13} for more information. The induced
comonad $\overline{\Mlt_S} \colon \Cat{Mod}_{S} \rightarrow
\Cat{Mod}_{S}$ from~(\ref{ComonadFromMonadDiag}) sends such a module
$X = (X, +, 0, \scalar)$ to the free module $\Mlt_{S}(X)$ of finite
multisets (formal sums) on the underlying set $X$, ignoring the
existing module structure on $X$. The counit and comultiplication are
given by:
\begin{equation}
\label{MltComonadStructDiag}
\vcenter{\xymatrix@R-2pc{
X & & \Mlt_{S}(X)\ar[ll]_-{\varepsilon}\ar[rr]^-{\delta} & & \Mlt_{S}^{2}(X) \\
\big(\sum_{j} s_{j}\scalar x_{j}\big)  & & 
   \big(\sum_{j}s_{j}x_{j}\big)\ar@{|->}[ll]\ar@{|->}[rr] & &
   \big(\sum_{j}s_{j}(1x_{j})\big).
}}
\end{equation}

\noindent The formal sum (multiset) in the middle is mapped by the
counit $\varepsilon$ to an actual sum in $X$, namely to its
interpretation.  The comultiplication $\delta$ maps this formal sum to
a multiset of multisets, with the inner multisets given by singletons
$1x_{j} = \eta(x_{j})$.

The following is a novel observation, motivating the view of
coalgebras on algebras as bases.

\begin{thm}
\label{OperatorCoalgThm}
Let $X$ be a vector space, say over $S=\mathbb{R}$ or $S=\mathbb{C}$.
Coalgebras $X\rightarrow \overline{\Mlt_{S}}(X)$ correspond to (Hamel)
bases on $X$.
\end{thm}

\begin{myproof}
Suppose we have a basis $B\subseteq X$ for the vector space $X$. Then
we can define a coalgebra $\smash{b \colon X\rightarrow
  \overline{\Mlt_{S}}(X)}$ via (finite) formal sums $b(x) =
\sum_{j}s_{j}a_{j}$, where $s_{j}\in S$ is the $j$-th coefficient of
$x$ wrt\ $a_{j}\in B\subseteq X$.  By construction we have
$\varepsilon \after b = \idmap$. The equation $\delta \after b =
\Mlt_{S}(b) \after b$ holds because $b(a) = 1a$, for basic elements
$a\in B$.

Conversely, given a coalgebra $\smash{b \colon X\rightarrow
  \overline{\Mlt_{S}}(X)}$ take $X_{b} = \setin{a}{X}{b(a) = 1a}$ as
in~(\ref{BasisEqualiserDiag}). Any finite subset of elements of
$X_{b}$ is linearly independent: if $\sum_{j}s_{j}\scalar a_{j} = 0$,
for finitely many $a_{j}\in X_{b}$, then in $\Mlt_{S}(X)$,
$$\textstyle 0 
\hspace*{\arraycolsep} = \hspace*{\arraycolsep}
b(0)
\hspace*{\arraycolsep} = \hspace*{\arraycolsep}
b(\sum_{j}s_{j}\scalar a_{j})
\hspace*{\arraycolsep} = \hspace*{\arraycolsep}
\sum_{j}s_{j}b(a_{j})
\hspace*{\arraycolsep} = \hspace*{\arraycolsep}
\sum_{j}s_{j}(1a_{j})
\hspace*{\arraycolsep} = \hspace*{\arraycolsep}
\sum_{j}s_{j}a_{j}.$$

\noindent Hence $s_{j} = 0$, for each $j$. Next, since $\delta \after
b = \Mlt_{S}(b) \after b$, each $a_{j}$ in $b(x) = \sum_{j}s_{j}a_{j}$
satisfies $b(a_{j}) = 1a_{j}$, so that $a_{j} \in X_{b}$. Because
$\varepsilon \after b = \idmap$, each element $x\in X$ can be
expressed as sum of such basic elements. \qed
\end{myproof}

A basis for complete lattices in
Subsection~\ref{CompleteLatticeSubsec}, if it exists, is uniquely
determined. In the context of vector spaces bases are unique up to
isomorphism.

Our next example involves convex sets, where extreme points play the
role of base vectors. Via the language of coalgebras we can make the
similarity with vector spaces explicit.

\subsection{Convex sets}\label{ConvexSubsec}

The (discrete probability) distribution monad $\Dstr$ on $\Sets$ is
given by $\Dstr(X) = \set{\varphi\colon X \rightarrow [0,1]}{
  \support(\varphi) \mbox{ is finite, and } \sum_{x\in X}\varphi(x) =
  1}$.  The unit and multiplication of this monad are as for the
multiset monad $\Mlt_S$, described above.

Algebras of the distribution monad can be identified with ``convex
sets'' (see \textit{e.g.}~\cite{Jacobs10e}), where convex sums exist:
the mapping $\Dstr(X) \rightarrow X$ sends a formal convex combination
to an actual convex sum. A typical example is the unit interval
$[0,1]$. Notice that it does not have arbitrary sums; but convex sums
exist in $[0,1]$.  An algebra homomorphism preserves such convex
sums. Such a map is usually called `affine'. We write $\Cat{Conv}$ for
this category $\Alg(\Dstr)$ of convex sets and affine maps.

A point $x\in X$ in a convex set $X$ is called \textit{extreme} if it
does not occur as non-trivial convex combination: if $x =
\sum_{i}r_{i}x_{i}$, then $r_{j}=1$ and $x_{j}=x$, for some $j$, and
thus $r_{i}=0$ for $i\neq j$. One usually writes $\partial X \subseteq
X$ for the subset of extreme points. In a free convex set $\Dstr(Y)$,
for a set $Y$, the extreme points are the singletons $\eta(y) = 1y$,
for $y\in Y$. Thus $\partial\Dstr(X) \cong X$.

Now assume we have a coalgebra $\smash{b\colon X \rightarrow
  \overline{\Dstr}(X)}$ for the induced comonad $\overline{\Dstr}
\colon \Cat{Conv} \rightarrow \Cat{Conv}$. We form the subset $X_{b} =
\setin{x}{X}{b(x) = 1x}$ as in~(\ref{BasisEqualiserDiag}), and claim
$X_{b} = \partial X$, that is, these basic elements in $X_{b}$ are
precisely the extreme points.

It is easy to see that there is an inclusion $\partial X \subseteq
X_{b}$: if $x$ is extreme, and $b(x)$ is a formal sum
$\sum_{i}r_{i}x_{i}$, then $x$ equals the actual sum
$\sum_{i}r_{i}x_{i}\in X$. But then $r_{j} = 1$ and $x = x_{j}$, for
some $j$ --- and $r_{i}=0$ for $i\neq j$. Hence $b(x) = 1x_{j} = 1x$.

For the reverse inclusion $X_{b} \subseteq \partial X$, assume $x\in
X$ satisfies $b(x) = 1x$, and $x = \sum_{i}r_{i}x_{i}$, where $r_{i}
\neq 0$ and $\sum_{i}r_{i}=1$. Since $b$ is an algebra homomorphism,
it preserves convex sums:
$$\begin{array}{rcccccccl}
1
& = &
(1x)(x)
& = &
b(x)(x)
& = &
b(\sum_{i}r_{i}x_{i})(x)
& = &
\sum_{i}r_{i}b(x_{i})(x).
\end{array}$$

\noindent But then $b(x_{i})(x)=1$, and so $x_{i} = x$ for each $i$.
Hence $x = \sum_{i}r_{i}x_{i}$ is a singleton sum, making $x$ extreme.

We thus see that a $\smash{\overline{\Dstr}}$-coalgebra $b\colon X
\rightarrow \Dstr(X)$ determines a coalgebra $b'\colon X \rightarrow
\Dstr(\partial X)$, as in~(\ref{RestrictedBaseDiag}), that describes
each element as convex sum of extreme points. As we have seen, this
map $b'$ is an isomorphism $\smash{X \conglongrightarrow
  \Dstr(\partial X)}$ describing each convex set with a basis as a
free convex set on its extreme points. This is the essence of the
equivalence of categories $\CoAlg(\overline{\Dstr}) \simeq \Sets$.

The situation is reminiscent of the Krein-Milman theorem, which states
that a convex and compact subset $S$ of a locally convex space is
equal to the closed convex hull of its extreme points: $S =
\overline{\Dstr(\partial S)}$, where $\overline{(-)}$ is the closure
operation. What we have here is a non-topological version of such a
result.

\section{Order-theoretic examples}\label{OrderSec}

Assume $\cat{A}$ is a poset-enriched category. This means that all
homsets $\cat{A}(X,Y)$ are posets, and that pre- and post-composition
are monotone. In this context maps $f\colon X\rightarrow Y$ and
$g\colon Y\rightarrow X$ in opposite direction form an adjunction
$f\dashv g$ (or Galois connection) if there are inequalities
$\idmap[X] \leq g\after f$ and $f\after g \leq \idmap[Y]$,
corresponding to unit and counit of the adjunction. In such a
situation the adjoints $f,g$ determine each other.

A monad $T = (T, \eta, \mu)$ on such a poset-enriched category
$\cat{A}$ is said to be of \textit{Kock-Z{\"o}berlein type} or just a
\textit{Kock-Z{\"o}berlein monad} if $T\colon \cat{A}(X,Y)
\rightarrow \cat{A}(TX,TY)$ is monotone and $T(\eta_{X}) \leq
\eta_{TX}$ holds in the homset $\cat{A}\big(T(X), T^{2}(X)\big)$. This
notion is introduced in~\cite{Kock95} in proper 2-categorical form.
Here we shall use the special `poset' instance---like
in~\cite{Escardo98} where the dual form occurs. The following result
goes back to~\cite{Kock95}; for convenience we include the proof.

\begin{thm}
\label{KZAlgThm}
Let $T$ be a Kock-Z{\"o}berlein monad on a poset-enriched category
$\cat{A}$. For a map $a\colon T(X)\rightarrow X$ in $\cat{A}$ the
following statements are equivalent.
\begin{enumerate}[\em(1)]
\item $a\colon T(X)\rightarrow X$ is an (Eilenberg-Moore) algebra
of the monad $T$;

\item $a\colon T(X)\rightarrow X$ is a left-adjoint-left-inverse of the
  unit $\eta\colon X\rightarrow T(X)$; this means that $a\dashv
  \eta_{X}$ is a reflection.
\end{enumerate}
\end{thm}

\begin{myproof}
First assume $a\colon T(X)\rightarrow X$ is an algebra,
\textit{i.e.}~satisfies $a\after \eta = \idmap$ and $a \after \mu = a
\after T(a)$. It suffices to prove $\idmap \leq \eta \after a$,
corresponding to the unit of the reflection, since the equation
$a\after \eta = \idmap$ is the counit (isomorphism). This is easy, by
naturality: $\eta \after a = T(a) \after \eta \geq T(a) \after T(\eta)
= \idmap$.

In the other direction, assume $a\colon T(X)\rightarrow X$ is
left-adjoint-left-inverse of the unit $\eta\colon X\rightarrow T(X)$,
so that $a\after\eta = \idmap$ and $\idmap \leq \eta \after a$. We
have to prove $a \after \mu = a \after T(a)$. In one direction, we
have:
\begin{equation}
\label{KZMuEqn}
\begin{array}{rcl}
\mu
& \leq &
T(a),
\end{array}
\end{equation}

\noindent since $\mu \leq \mu \after T(\eta \after a) = T(a)$, and
thus $a \after \mu \leq a \after T(a)$. For the reverse inequality we
use:
$$\begin{array}[b]{rcll}
a \after T(a)
\hspace*{\arraycolsep}=\hspace*{\arraycolsep}
a \after T(a) \after T(\idmap)
& = &
a \after T(a) \after T(\mu) \after T(\eta) \\
& \leq &
a \after T(a) \after T(\mu) \after \eta \qquad
   & \mbox{since }T(\eta) \leq \eta \\
& = &
a \after \eta \after a \after \mu 
   & \mbox{by naturality} \\
& = &
a\after \mu.
\end{array}\eqno{\qEd}$$
\end{myproof}

In a next step we consider the induced comonad $\overline{T}$ on the
category $\Alg(T)$ of algebra of a Kock-Z{\"o}berlein monad $T$, as
in~\eqref{MonadIterDiag}. A first, trivial but important, observation
is that the category $\Alg(T)$ is also poset enriched. It is not hard
to see that the comonad $\overline{T}$ is also of Kock-Z{\"o}berlein
type, in the sense that for each algebra
$\smash{(TX\stackrel{a}{\rightarrow} X)}$ we have:
$$\varepsilon_{\overline{T}(a)}
= 
\mu
\leq
T(a)
=
\overline{T}(\varepsilon_{a})$$

\noindent by~(\ref{KZMuEqn}). Thus one may expect a result similar to
Theorem~\ref{KZAlgThm} for coalgebras of this comonad
$\overline{T}$. It is formulated in~\cite[Thm.~4.2]{Kock95} (and
attributed to the present author). We repeat the poset version in the
current context.

\begin{thm}
\label{KZCoAlgThm}
Let $T$ be a Kock-Z{\"o}berlein monad on a poset-enriched category
$\cat{A}$, with induced comonad $\overline{T}$ on the category of
algebras $\Alg(T)$. Assume an algebra $a\colon T(X)\rightarrow X$. For
a map $c\colon X\rightarrow T(X)$, forming a map of algebras in,
\begin{equation}
\label{KZCoAlgEqn}
\vcenter{\xymatrix@C-.5pc{
\ensuremath{\left(\xy
(0,4)*{TX};
(0,-4)*{X};
{\ar^{a} (0,2); (0,-2)};
\endxy\right)}\ar[rr]^-{c}
& &
\overline{T}\ensuremath{\left(\xy
(0,4)*{TX};
(0,-4)*{X};
{\ar^{a} (0,2); (0,-2)};
\endxy\right)}
=
\ensuremath{\left(\xy
(0,4)*{T^{2}X};
(0,-4)*{TX};
{\ar^{\mu\rlap{$\scriptscriptstyle{_X}$}} (0,2); (0,-2)};
\endxy\right)}
}}
\end{equation}

\noindent the following statements are equivalent.
\begin{enumerate}[\em(1)]
\item $c\colon a\rightarrow \overline{T}(a)$ is an
  (Eilenberg-Moore) coalgebra of the comonad $\overline{T}$;

\item $c\colon a\rightarrow \overline{T}(a)$ is a
  left-adjoint-right-inverse of the counit $a\colon
  \overline{T}(a)\rightarrow a$; this means that $c\dashv
  a$ is a coreflection.
\end{enumerate}
\end{thm}

\begin{myproof}
Assume $c$ is a $\overline{T}$-coalgebra, \textit{i.e.}~$c \after a =
\mu \after T(c)$, $a \after c = \idmap$ and $T(\eta) \after c = T(c)
\after c$, like in Definition~\ref{BasisDef}. We have to prove $c
\after a \leq \idmap$, which is obtained in:
$$c \after a 
=
\mu \after T(c) 
\;\stackrel{(\ref{KZMuEqn})}{\leq}\;
T(a) \after T(c) 
=
\idmap.$$

\noindent Conversely, assume a coreflection $c\dashv a$, so that $a
\after c = \idmap$ and $c \after a \leq \idmap$. We have to prove
$T(\eta) \after c = T(c) \after c$. In one direction we have $T(c)
\leq T(\eta \after a) \after T(c) = T(\eta)$, and thus $T(c) \after c
\leq T(\eta) \after c$.  In the other direction, we use:
$$\begin{array}[b]{rcll}
T(c) \after c
\hspace*{\arraycolsep}=\hspace*{\arraycolsep}
T^{2}(\idmap) \after T(c) \after c
& = &
T^{2}(a \after \eta) \after T(c) \after c \\
& \leq &
T^{2}(a) \after T(\eta) \after T(c) \after c \quad
   & \mbox{since }\rlap{$T(\eta) \leq \eta$} \\
& = &
T^{2}(a) \after T^{2}(c) \after T(\eta) \after c 
   & \mbox{by naturality} \\
& = &
T(\eta) \after c.
\end{array}\eqno{\qEd}$$\vspace{3 pt}
\end{myproof}

\noindent One can iterate the $\overline{(-)}$ construction, as
in~\eqref{MonadIterDiag}. Below we show that for Kock-Z{\"o}berlein
monads the iteration stops after 3 steps. First we need another
characterisation. The proof is as before.

\begin{lem}
\label{KZAlgCoAlgLem}
Let $T$ be a Kock-Z{\"o}berlein monad on a poset-enriched category
$\cat{A}$, giving rise to comonad $\smash{\overline{T}}$ on $\Alg(T)$
and monad $\smash{\overline{\overline{T}}}$ on
$\CoAlg(\smash{\overline{T}})$. Assume:
\begin{iteMize}{$\bullet$}
\item an algebra $a\colon T(X)\rightarrow X$ in $\Alg(T)$;
\item a coalgebra $c\colon X\rightarrow T(X)$ on $a$ in
  $\smash{\CoAlg(\overline{T})}$;
\item an algebra $b\colon T(X)\rightarrow X$ on $c$ in
  $\smash{\Alg(\overline{\overline{T}})}$, where:
\begin{iteMize}{$-$}
\item $b \after c = \idmap$ and $b \after T(b) = b \after T(a)$,
since $b$ is a $\smash{\overline{\overline{T}}}$-algebra;

\item $a \after T(b) = b \after \mu$, since $b$ is a map of algebras
$\smash{a \rightarrow \overline{T}(a) = \mu}$;

\item $c \after b = T(b) \after T(\eta)$, since $b$ is a map of
algebras $\smash{\delta = \overline{\overline{c}} \rightarrow c}$.
\end{iteMize}
\end{iteMize}


\noindent The following statements are then equivalent.
\begin{enumerate}[\em(1)]
\item $\smash{b\colon \overline{\overline{T}}(c)\rightarrow c}$ is an
  algebra of the monad $\smash{\overline{\overline{T}}}$;

\item $\smash{b\colon \overline{\overline{T}}(c)\rightarrow c}$ is a
  left-adjoint-left-inverse of the unit $\smash{c\colon c \rightarrow
  \overline{\overline{T}}(c)}$. \qed
\end{enumerate}
\end{lem}

\auxproof{
First assume $b\colon T(X) \rightarrow X$ is a
$\smash{\overline{\overline{T}}}$-algebra. We have to show that there is a
reflection $b\dashv c$, \textit{i.e.}~that $b \after c = \idmap$ and
$\idmap \leq c \after b$. The former holds by assumption, and the
latter holds because:
$$\begin{array}{rcll}
c \after b
& = &
T(b) \after T(\eta) 
   & \mbox{since $b$ is a map of coalgebras} \\
& \geq &
T(b) \after T(c)
   & \mbox{since $T(c) \leq T(\eta)$, see the proof of
    Theorem~\ref{KZCoAlgThm}} \\
& = &
\idmap
\end{array}$$

In the other direction, assume $b$ is a map of algebras
$\overline{T}(c) \rightarrow c$ with a reflection $b \dashv c$. We have
to show that $b$ is a $\smash{\overline{\overline{T}}}$-algebra,
\textit{i.e.}~satisfies $b \after c = \idmap$ and $b \after T(b) = b
\after T(a)$. Again, the former holds by assumption, and for the
latter we have $T(a) \leq T(b)$, since $T(b) \geq T(b) \after T(c
\after a) = T(a)$, and thus $b \after T(a) \leq b\after T(b)$. For the
reverse inequality we use:
$$\begin{array}{rcll}
b \after T(b)
& = &
b \after T(b) \after T^{2}(\idmap) \\
& = &
b \after T(b) \after T^{2}(a \after c) \\
& \leq &
b \after T(b) \after T^{2}(a) \after T(\eta) 
   & \mbox{since $T(c) \leq T(\eta)$} \\
& \leq &
b \after T(b) \after T(\eta) \after T(a) 
   & \mbox{by naturality} \\
& = &
b \after c \after b \after T(a) 
   & \mbox{since $b$ is a map of algebras} \\
& = &
b \after T(a).
\end{array}$$
}

The next result shows how such series of adjunctions can arise.

\begin{lem}
\label{KZAlgCoAlgAlgLem}
Assume an algebra $a\colon T(X)\rightarrow X$ of a Kock-Z{\"o}berlein
monad. The free algebra $T(X)$ then carries multiple (co)reflections
(algebras and coalgebras) in a situation:
\begin{equation}
\label{KZAlgCoAlgAlgEqn}
\vcenter{\def\labelstyle{\scriptstyle}\xymatrix{
T^{2}(X)\ar@/_3ex/[d]_{\dashv\;}|{\mathstrut\mu}\ar@/_12ex/[d]_{T(a)} \\
T(X)\ar[u]^{\dashv}_{\eta}
   \ar@/^7.5ex/[u]^{\dashv\;\;\;}|{\mathstrut T(\eta)} \\
}}
\end{equation}

\noindent This yields a functor $T\colon \Alg(T)
\rightarrow \smash{\Alg(\overline{\overline{T}})}$ between categories of
algebras.
\end{lem}

\begin{myproof}
We check all (co)reflections from right to left.
\begin{iteMize}{$\bullet$}
\item In the first case the counit is the identity since $\mu \after
  \eta = \idmap$; because $T(\eta) \leq \eta$ for a Kock-Z{\"o}berlein
  monad, we get a unit $\eta \after \mu = T(\mu) \after \eta \geq
  T(\mu) \after T(\eta) = \idmap$. (This follows already from
  Theorem~\ref{KZAlgThm}.)

\item In the next case we have a coreflection $T(\eta) \dashv \mu$
  since the unit is the identity $\mu \after T(\eta) = \idmap$, and:
  $T(\eta) \after \mu = \mu \after T^{2}(\eta) \leq \mu \after T(\eta)
  = \idmap$.

\item Finally one gets a reflection $T(a) \dashv T(\eta)$ from the
  reflection $a\dashv\eta$ from Theorem~\ref{KZAlgThm}: $T(a) \after
  T(\eta) = T(a \after \eta) = \idmap$ and $T(\eta) \after T(a) =
  T(\eta \after a) \geq T(\idmap) = \idmap$. \qed
\end{iteMize}

\auxproof{
To be precise, we also have to check:
\begin{iteMize}{$\bullet$}
\item $T(\eta)$ is a map of algebras $\mu_{TX}\rightarrow\mu_{X}$;
this is obvious.

\item $T(a)$ is a map of coalgebras $T(\eta_{X}) \rightarrow
  T(\eta_{TX})$, and a map of algebras $\mu_{TX}\rightarrow\mu_{X}$;
  the latter is trivial, and for the former we use naturality:
  $T(\eta_{X}) \after T(a) = T^{2}(a) \after T(\eta_{TX})$.
\end{iteMize}

Additionally, if $f\colon X\rightarrow X'$ is a map of algebras
$a\rightarrow a'$ in $\Alg(T)$, then $T(f)\colon T(X) \rightarrow
T(X')$ is a morphsm of algebras $T(a)\rightarrow T(a')$ in
$\smash{\Alg(\overline{\overline{T}})}$. This is obvious.
}
\end{myproof}

\noindent This lemma describes the only form that such structures can have.
This is the main (new) result of this section.

\begin{thm}
\label{KZCoAlgBasisThm}
If we have a reflection-coreflection-reflection chain $b \dashv c
\dashv a \dashv \eta_{X}$ on an object $X$, like in
Lemma~\ref{KZAlgCoAlgLem}, then $X$ is a free algebra.

Thus: for a Kock-Z{\"o}berlein monad $T$, the functor $T\colon \Alg(T)
\rightarrow \smash{\Alg(\overline{\overline{T}})}$ is an equivalence of
categories.
\end{thm}

\begin{myproof}
Assume $b \dashv c \dashv a \dashv \eta_{X}$ on $X$, and consider the
equaliser~(\ref{BasisEqualiserDiag}) in:
\begin{equation}
\label{KZCoAlgBasisEqual}
\vcenter{\xymatrix@R-2pc@C-1pc{
X_{c}\ar@{ >->}[rr]^-{e} & & X
   \ar@<+.5ex>[rr]^-{c}\ar@<-.5ex>[rr]_-{\eta} & & T(X) \\
 & T(X)\ar[ur]_-{b} \\
X\ar[ur]_-{\eta}\ar@{-->}[uu]^{k}
}}
\end{equation}

\noindent We use the letter `$k$' because the elements in $X_c$ will
turn out to be compact elements, in the examples later on. The first
thing we note is:
\begin{equation}
\label{KZCoAlgBasisEqn}
\begin{array}{rcl}
k \after e
& = &
\idmap[X_c].
\end{array}
\end{equation}

\noindent This follows since $e$ is a mono, and:
$$\begin{array}{rcll}
e \after k \after e
& = &
b \after \eta \after e \qquad
  & \mbox{by construction of $k$} \\
& = &
b \after c \after e
   & \mbox{since $e$ is equaliser} \\
& = &
e & \mbox{since $b$ is a $\smash{\overline{\overline{T}}}$-algebra and
   $c$ is unit.}
\end{array}$$

Next we observe that the object $X_c$ carries a $T$-algebra structure
$a_c$ inherited from $a\colon T(X)\rightarrow X$, as in:
$$\xymatrix{
a_{c} \stackrel{\textrm{def}}{=} \Big(T(X_{c})\ar[r]^-{T(e)} &
   T(X)\ar[r]^-{a} & X\ar[r]^-{k} & X_{c}\Big)
}$$

\noindent It is an algebra indeed, since:
$$a_{c} \after \eta
=
k \after a \after T(e) \after \eta
=
k \after a \after \eta \after e 
=
k \after e 
\;\smash{\stackrel{(\ref{KZCoAlgBasisEqn})}{=}}\;
\idmap.$$

\noindent The other algebra equation is left to the reader.

\auxproof{
First we make explicit what it means for a map $b\colon
T(X)\rightarrow X$ in to be an algebra (on coalgebra $c$ on
algebra $a$):
\begin{iteMize}{$\bullet$}
\item $b \after c = \idmap$ and $b \after T(b) = b \after T(a)$,
since $b$ is a $\smash{\overline{\overline{T}}}$-algebra;

\item $a \after T(b) = b \after \mu$, since $b$ is a map of algebras
$\smash{a \rightarrow \overline{T}(a) = \mu}$;

\item $c \after b = T(b) \after T(\eta)$, since $b$ is a map of
algebras $\smash{\delta = \overline{\overline{c}} \rightarrow c}$.
\end{iteMize}

\noindent Now we use that the equaliser map $e$ is a mono in:
$$\begin{array}{rcll}
\lefteqn{e \after a_{c} \after T(a_{c})} \\
& = &
e \after k \after a \after T(e) \after T(k \after a \after T(e)) \\
& = &
b \after \eta \after a \after T(b \after \eta) \after T(a) \after T^{2}(e)
   \quad
  & \mbox{by~(\ref{KZCoAlgBasisEqual})} \\
& = &
b \after \eta \after b \after \mu \after T(\eta) \after T(a)\after T^{2}(e)
   & \mbox{see the properties above} \\
& = &
b \after \eta \after b \after T(a) \after T^{2}(e) \\
& = &
b \after T(b) \after \eta \after T(a) \after T^{2}(e)
   & \mbox{by naturality} \\
& = &
b \after T(a) \after \eta \after T(a) \after T^{2}(e)
   & \mbox{see the properties above} \\
& = &
b \after \eta \after a \after T(a) \after T^{2}(e)
   & \mbox{by naturality} \\
& = &
e \after k \after a \after T(a) \after T^{2}(e)
  & \mbox{by~(\ref{KZCoAlgBasisEqual})} \\
& = &
e \after k \after a \after \mu \after T^{2}(e)
  & \mbox{since $a$ is a $T$-algebra} \\
& = &
e \after k \after a \after T(e) \after \mu
   & \mbox{by naturality} \\
& = &
e \after a_{c} \after \mu.
\end{array}$$
}

Next we show that the transpose $a\after T(e) \colon T(X_{c})
\rightarrow X$ of the equaliser $e\colon X_{c} \rightarrowtail X$ is
an isomorphism of algebras $\mu_{X_c} \cong a$. The inverse is $T(k)
\after c \colon X\rightarrow T(X)\rightarrow T(X_{c})$, since:
$$\begin{array}{rcll}
\lefteqn{\big(a\after T(e)\big) \after \big(T(k) \after c\big)} \\
& = &
a \after T(b \after \eta) \after c \qquad
  & \mbox{by~(\ref{KZCoAlgBasisEqual})} \\
& = &
b \after \mu \after T(\eta) \after c 
   & \mbox{see the assumptions in Lemma~\ref{KZAlgCoAlgLem}} \\
& = &
b \after c \\
& = &
\idmap
   & \mbox{see Theorem~\ref{KZCoAlgThm}} \\
\lefteqn{\big(T(k) \after c\big) \after \big(a\after T(e)\big)} \\
& = &
T(k) \after \mu \after T(c) \after T(e) \quad
   & \mbox{since $c$ is a map 
      $\smash{a \rightarrow \overline{T}(a) = \mu}$} \\
& = & 
T(k) \after \mu \after T(\eta) \after T(e)
   & \mbox{since $e$ is equaliser of $c$ and $\eta$} \\
& = & 
T(k) \after T(e) \\
& = &
\idmap & \mbox{by~(\ref{KZCoAlgBasisEqn}).}
\end{array}$$
We continue to check that the assumed chain of adjunctions $b \dashv c
\dashv a \dashv \eta_{X}$ is related to the chain $T(a_{c}) \dashv
T(\eta) \dashv \mu \dashv \eta$ in~(\ref{KZAlgCoAlgAlgEqn}) via these
isomorphisms. In particular we still need to check that the
following two square commute.
$$\xymatrix@R1pc{
T^{2}(X_{c})\ar[rr]^-{T(a \after T(e))}_-{\cong} & & T(X)
& &
T^{2}(X_{c})\ar[d]_{T(a_{c})}\ar[rr]^-{T(a \after T(e))}_-{\cong} & & 
   T(X)\ar[d]^{b} \\
T(X_{c})\ar[u]^{T(\eta)}\ar[rr]_-{a \after T(e)}^-{\cong} & & X\ar[u]_{c}
& &
T(X_{c})\ar[rr]_-{a \after T(e)}^-{\cong} & & X
}$$

\noindent These square commute since:
$$\begin{array}{rcll}
T(a \after T(e)) \after T(\eta)
& = &
T(a) \after T(\eta) \after T(e)
   & \mbox{by naturality} \\
& = &
T(e) \\
& = &
\mu \after T(\eta) \after T(e) \\
& = &
\mu \after T(c) \after T(e) \\
& = &
c \after a \after T(e)
   & \mbox{see Theorem~\ref{KZCoAlgThm}} \\
a \after T(e) \after T(a_{c})
& = &
a \after T(e) \after T(k \after a \after T(e)) \\
& = &
a \after T(b \after \eta) \after T(a \after T(e)) 
  & \mbox{by~(\ref{KZCoAlgBasisEqual})} \\
& = &
b \after \mu \after T(\eta) \after T(a \after T(e)) \quad
   & \mbox{see in Lemma~\ref{KZAlgCoAlgLem}} \\
& = &
b \after T(a \after T(e)).
\end{array}$$
We still have to check that the functor $T\colon \Alg(T) \rightarrow
\smash{\Alg(\overline{\overline{T}})}$ is an equivalence.  In the
reverse direction, given a coalgebra $c\colon
\overline{\overline{T}}(b) \rightarrow b$ on $X$, we take the induced
algebra $T(X_{c}) \rightarrow X_{c}$ on the
equaliser~(\ref{KZCoAlgBasisEqual}). Then $T(X_{c}) \congrightarrow X$
is an isomorphism of $\overline{\overline{T}}$-algebras, as we have
    seen.

For the isomorphism in the other direction, assume we start from an
algebra $a\colon T(X)\rightarrow X$, obtain the
$\overline{\overline{T}}$-algebra $T(a)$ described in the chain $T(a)
\dashv T(\eta) \dashv \mu \dashv \eta$ in~(\ref{KZAlgCoAlgAlgEqn}),
and then form the equaliser~(\ref{KZCoAlgBasisEqual}); it now looks as
follows.
$$\xymatrix@R-2pc@C-1pc{
X\ar@{ >->}[rr]^-{\eta_X} & & T(X)
   \ar@<+.5ex>[rr]^-{T(\eta)}\ar@<-.5ex>[rr]_-{\eta} & & T^{2}(X)
}$$

\noindent This is the equaliser requirement~\cite{Moggi88}, which
holds since $X$ carries an algebra structure. Clearly, $\eta \after
\eta = T(\eta) \after \eta$ by naturality. And if a map $f\colon
Y\rightarrow T(X)$ satisfies $\eta \after f = T(\eta) \after f$, then
$f$ factors through $\eta\colon X\rightarrow T(X)$ via $f' = a \after
f$, since
$$\eta \after f'
= 
\eta \after a \after f
=
T(a) \after \eta \after f
=
T(a) \after T(\eta) \after f
=
f.$$

\noindent This $f'$ is unique with this property, since if $g\colon
Y\rightarrow X$ also satisfies $\eta \after g = f$, then $f' = a
\after f = a \after \eta \after g = g$. \qed
\end{myproof}

In the remainder of this section we review some examples.

\subsection{Dcpos over Posets}\label{DcpoSubsec}

The main example from~\cite{Jacobs94b} involves the ideal monad $\Idl$
on the category \PoSets of partially ordered sets with monotone
functions between them. In the light of Theorems~\ref{KZAlgThm}
and~\ref{KZCoAlgThm} we briefly review the essentials.

For a poset $X = (X,\leq)$ let $\Idl(X)$ be the set of directed
downsets in $X$, ordered by inclusion. This $\Idl$ is in fact a monad
on \PoSets with unit $X\rightarrow\Idl(X)$ given by principal downset
$x\mapsto\downset x$ and multiplication $\Idl^{2}(X) \rightarrow
\Idl(X)$ by union.  This monad is of Kock-Z{\"o}berlein type since for
$U\in\Idl(X)$ we have:
$$\begin{array}{rcl}
\Idl(\downset)(U)
\hspace*{\arraycolsep}=\hspace*{\arraycolsep}
\downset\set{\downset x}{x\in U}
& = &
\setin{V}{\Idl(X)}{\exin{x}{U}{V\subseteq \downset x}} \\
& \subseteq &
\setin{V}{\Idl(X)}{V\subseteq U} \qquad
    \mbox{since $U$ is a downset} \\
& = &
\downset U.
\end{array}$$

\auxproof{
For a poset $(X,\leq)$ a subset $U\subseteq X$ is called directed if
it is non-empty and for each pair $x_{1},x_{2}\in U$ there is a $y\in
U$ with $x_{1},x_{2}\leq y$. The set $\Idl(X)$ contains the downclosed
directed subsets, ordered by inclusion. The mapping $X \mapsto
\Idl(X)$ yields a functor $\PoSets \rightarrow \PoSets$: for $f\colon
X\rightarrow Y$ one defines $\Idl(f)(U) = \downset\set{f(x)}{x\in U}$.
}

\noindent Applying Theorem~\ref{KZAlgThm} to the ideal monad yields the
(folklore) equivalence of the following points.
\begin{enumerate}[(1)]
\item $X$ is a directed complete partial order (dcpo): each directed
  subset $U\subseteq X$ has a join $\bigvee U$ in $X$; 

\item The unit $\downset\; \colon  X \rightarrow  \Idl(X)$
  has a left adjoint---which is the join;

\item $X$ carries a (necessarily unique) algebra structure $\Idl(X)
  \rightarrow X$, which is also the join. 
\end{enumerate}

\noindent Additionally, algebra maps are precisely the continuous
functions. Thus we may use as category $\Dcpo = \Alg(\Idl)$.

The monad $\Idl$ on $\PoSets$ induces a comonad on $\Dcpo$, written
$\overline{\Idl}$, with counit $\varepsilon = \bigvee \colon \Idl(X)
\rightarrow X$ and comultiplication $\delta = \Idl(\downset) \colon
\Idl(X)\rightarrow \Idl^{2}(X)$, so that $\delta(U) =
\downset\set{\downset x}{x\in U}$. In order to characterise coalgebras
of this comonad $\overline{\Idl}$ we need the following. In a dcpo
$X$, the way below relation $\ll$ is defined as: for $x,y \in X$,
$$\begin{array}{rcl}
x \ll y
& \Longleftrightarrow &
\mbox{for each directed }U\subseteq X, \mbox{ if } 
      y\leq {\textstyle \bigvee U} \mbox{ then }
      \exin{z}{U}{x\leq z}.
\end{array}$$

\noindent A \emph{continuous poset} is then a dcpo in which for each
element $x\in X$ the set $\ddownset x = \setin{y}{X}{y \ll x}$ is
directed and has $x$ as join. These elements way-below $x$ may be seen
as a (local) basis.

\auxproof{
It is easy to see that $x\ll y \implies x \leq y$. Thus $\ll$ is
anti-symmetric. It is obviously transitive. Further, if $x' \leq x \ll
y \leq y'$ then $x' \ll y'$. We shall make use below of the following
interpolation result, which is a mild generalization of~\cite[VII,
  Lemma 2.4]{Johnstone82}.

\begin{lem}
\label{InterpolationLem}
For each continuous function $f\colon X\rightarrow Y$ between
continuous posets $X,Y$ one has:
$$\begin{array}{rcl}
y \ll f(x) 
& \Longrightarrow &
\exin{z}{X}{y \ll f(z) \mbox{ and } z\ll x}.
\end{array}$$
\end{lem}

\begin{myproof}
Given $x\in X$, consider the subset $W = \bigcup\set{\ddownset f(z)}{z
  \in \ddownset x} \subseteq Y$ which is directed and has $f(x)$ as
supremum. Thus, if $y \ll f(x) = \bigvee W$, then $y \leq w$, for some
$w \in W$. But then $w \in \ddownset f(z)$ for some $z\in \ddownset
x$. Hence $y \ll f(z)$ and $z\ll x$. \qed

\auxproof{
We check that $W$ is directed: if $y_{1},y_{2}\in W$, say $y_{i}\ll
f(z_{i})$ for $z_{i}\ll x$, then, because $X$ is continuous, there is
a $z\ll x$ with $z_{i} \leq z$. Then $f(z_{i}) \leq f(z)$ and thus
$y_{i} \ll f(z)$. Since $Y$ is also continuous there is an $y\in Y$
with $y_{i} \leq y$ and $y\ll f(z)$. Hence $y\in W$.

Then:
$$\begin{array}{rcll}
\bigvee W
& = &
\bigvee \bigcup\set{\ddownset f(z)}{z \in \ddownset x} \\
& = &
\bigvee \downset\set{\bigvee\ddownset f(z)}{z \in \ddownset x} 
   \qquad & \mbox{since $\bigvee$ is an algebra} \\
& = &
\bigvee \downset\set{f(z)}{z \in \ddownset x} \\
& = &
\bigvee \set{f(z)}{z \in \ddownset x} \\
& = &
f(\bigvee\set{z}{z\in\ddownset x})
   & \mbox{since $f$ is continuous} \\
& = &
f(x).
\end{array}$$
}

\end{myproof}
}

The following equivalence formed the basis for~\cite[Thm.~4.2]{Kock95}
(of which Theorem~\ref{KZCoAlgThm} is a special case).  The
equivalence of points~(1) and~(2) is known from the literature, see
e.g.~\cite[VII, Proposition~2.1]{Johnstone82},
\cite[Proposition~2.3]{Hoffmann79},
or~\cite[Theorem~I-1.10]{GierzHKLMS03}. The equivalence of points~(2)
and~(3) is given by Theorem~\ref{KZCoAlgThm}.

For a dcpo $X$, the following statements are equivalent.
\begin{enumerate}[(1)]
\item $X$ is a continuous poset;

\item The counit $\bigvee\colon\Idl(X)\rightarrow X$ of the comonad
  $\overline{\Idl}$ on \Dcpo has a left adjoint (in \Dcpo); it is
  $x\mapsto \ddownset x$.

\item $X$ carries a (necessarily unique) $\overline{\Idl}$-coalgebra
  structure $X \rightarrow \Idl(X)$, which is also
  $\ddownset(-)$. 
\end{enumerate}

%

\auxproof{
We shall prove the equivalence of the first two points. In one
direction, if $X$ is a continuous dcpo, then $\ddownset\colon X
\rightarrow \Idl(X)$ is a continuous function: for a directed subset
$U\subseteq X$ one has:
$$\begin{array}{rcccccl}
y\in \ddownset U
& \Longleftrightarrow &
y \ll \bigvee U 
& \smash{\stackrel{(*)}{\Longleftrightarrow}} &
\exin{x}{U}{y\ll x}
& \Longleftrightarrow &
y\in \bigcup\set{\ddownset x}{x\in U}.
\end{array}$$

\noindent The implication $\smash{\stackrel{(*)}{\Longleftarrow}}$ is
obvious; for $\smash{\stackrel{(*)}{\Longrightarrow}}$, assume $y\ll
\bigvee U$; then by interpolation (Lemma~\ref{InterpolationLem}, with
$f = \idmap$), there is a $z$ with $y \ll z \ll \bigvee U$. But then
$z \leq x$ for some $x \in U$ and thus $y\ll x$. Next we have to
prove the adjoint correspondence:
$$\begin{prooftree}
\ddownset x \;\subseteq\; U
\Justifies
x \;\leq\; \textstyle \bigvee U
\end{prooftree}$$

\noindent where $U\in\Idl(X)$. For the downward direction, if
$\ddownset x \subseteq U$, then $x = \bigvee\ddownset x \leq \bigvee
U$. Upwards: if $x\leq\bigvee$ and $y\in\ddownset x$, then $y\ll x$
and so $y\leq z$ for some $z\in U$. But then $y\in U$ since $U$ is a
downset.

Now assume the existence of a left adjoint $c\dashv \bigvee$ in \Dcpo.
Hence $c$ is a continuous function $c\colon X\rightarrow \Idl(X)$ with
$c(x)\subseteq U \Leftrightarrow x\leq \bigvee U$, for
$U\in\Idl(X)$. We then have $\bigvee c(x) = x$, since in one
direction, $c(x)\subseteq c(x)$ yields $x\leq \bigvee c(x)$. And
conversely, from $x\leq \bigvee\downset x$ we get $c(x)\subseteq
\downset x$, and so $\bigvee c(x) \leq \bigvee\downset x = x$.

It thus suffices to prove $y\in c(x) \iff y\ll x$.
\begin{iteMize}{$\bullet$}
\item[($\Rightarrow$)] Let $y\in c(x)$. In order to show $y\ll x$,
  assume $x\leq \bigvee U$ for $U\subseteq X$ directed. Then $\downset
  U \in \Idl(X)$ with $\bigvee U = \bigvee\downset U$. Since
  $x\leq\bigvee U = \bigvee\downset U$, the adjunction $c\dashv
  \bigvee$ yields $c(x) \subseteq \downset U$. Hence $y\in\downset U$,
  which gives a $z\in U$ with $y\leq z$.

\item[($\Leftarrow$)] If $y\ll x = \bigvee c(x)$, then $y\leq z$ for
some $z\in c(x)$. But then $y\in c(x)$, because the latter is
a downset. \qed
\end{iteMize}
}


\auxproof{
A map $f\colon X\rightarrow Y$ in $\Dcpo$ is an
$\overline{T}$-coalgebra map if and only if for each $x\in X$,

$$\ddownset f(x)
=
\Idl(f)(\ddownset x)
=
\downset\set{f(y)}{y\in\ddownset x}.$$

\noindent First assume $f$ is such a coalgebra map.  If $y\ll x$, then
$f(y)$, being in the set on the right, is also in the set on the
left. Thus $f(y) \ll f(x)$, and $f$ is monotone.

In the reverse direction, assume $f$ is monotone with respect to
$\ll$. We show that the above equation holds.
$$\!\!\!\begin{array}[b]{rcl}
y\in\ddownset f(x)
& \Longrightarrow &
y \ll f(x) \\[1mm]
& \Longrightarrow &
\ex{z}{y \ll f(z) \mbox{ and } z\ll x} \quad\mbox{by interpolation}
   \quad\quad \\[1mm]
& \Longrightarrow &
\ex{z}{y \leq f(z) \mbox{ and } z\in\ddownset x} \\[1mm]
& \Longrightarrow &
y\in\downset\set{f(z)}{z\in\ddownset x} \\[2mm]
y\in\downset\set{f(z)}{z\in\ddownset x}
& \Longrightarrow &
\ex{z}{y \leq f(z) \mbox{ and } z\ll x} \\[1mm]
& \Longrightarrow &
\ex{z}{y \leq f(z) \mbox{ and } f(z)\ll f(x)}  
     \quad\rlap{since $f$ is monotone}\\[1mm]
& \Longrightarrow &
y \ll f(x) \\[1mm]
& \Longrightarrow &
y\in\ddownset f(x). 
\end{array}\eqno{\qEd}$$
}

\noindent Theorem~\ref{KZCoAlgBasisThm} says that another iteration
$\smash{\overline{\overline{\Idl}}}$ yields nothing new.

\subsection{Frames over semi-lattices}\label{FrmMSLSubsec}

For a poset $X$, the set of its downsets:
$$\begin{array}{rcl}
\Dwn(X)
& = &
\set{U\subseteq X}{U \mbox{ is downclosed}}
\end{array}$$

\noindent is a frame (or complete Heyting algebra, or locale),
see~\cite{Johnstone82}. If the poset $X$ has finite meets
$\top,\conjun$, then the downset map $\downset\; \colon X\rightarrow
\Dwn(X)$ preserves meets: $\downset\top = X$ and $\downset(x\conjun y)
= \downset x \cap \downset y$. Hence it is a morphism in the category
\MSL of meet semi-lattices. It is not hard to see that $\Dwn$ is a
monad on \MSL that is of Kock-Z{\"o}berlein type. For a (meet)
semi-lattice $X = (X, \top, \wedge)$ the following are equivalent.
\begin{enumerate}[(1)]
\item $X$ is a frame: $X$ has arbitrary joins and its finite meets
  distribute over these joins: $x \conjun \big(\bigvee_{i}y_{i}\big) =
  \bigvee_{i}(x \conjun y_{i})$;

\item The unit $\downset\; \colon  X \rightarrow  \Dwn(X)$
  has a left adjoint in \MSL---which is the join;

\item $X$ carries a (necessarily unique) algebra structure $\Dwn(X)
  \rightarrow X$ in \MSL, which is also the join.
\end{enumerate}

\noindent Moreover, the algebra maps are precisely the frame maps,
preserving arbitrary joins and finite meets; thus $\Frm = \Alg(\Dwn)$.

\auxproof{
Despite Theorem~\ref{KZAlgThm}, we include an explicit (older, already
existing) proof.  For the implication $(1)\Rightarrow(2)$, assume $X$
is a frame. We show that the join $\bigvee$ forms a map $\Dwn(X)
\rightarrow X$ in \MSL. Clearly, $\bigvee\top = \bigvee X =
\top$. Preservation of meets follows because $X$ is a frame: for 2
downsets $U,V\in\Dwn(X)$ we have,
$$\begin{array}{rcl}
(\bigvee U) \conjun (\bigvee V)
& = &
\bigvee\set{(\bigvee U)\conjun y}{y\in V} \\
& = &
\bigvee\set{\bigvee\set{x\conjun y}{x\in U}}{y\in V} \\
& = &
\bigvee(U\cap V),
\end{array}$$

\noindent where the last equation holds because:

$(\leq)$ If $x\in U$ and $y\in V$, then $x\conjun y \in U\cap V$
because $U,V$ are downsets, so $x\conjun y \leq \bigvee(U\cap
V)$. Hence $\bigvee\set{x\conjun y}{x\in U} \leq \bigvee(U\cap V)$,
and also $\bigvee\set{\bigvee\set{x\conjun y}{x\in U}}{y\in V} \leq
\bigvee(U\cap V)$.

$(\geq)$ If $z\in U\cap V$, then $z\leq \bigvee\set{x\conjun z}{x\in
  U}$, since $z\conjun z = z$. Thus also $z \leq
\bigvee\set{\bigvee\set{x\conjun y}{x\in U}}{y\in V}$. Hence
$\bigvee(U\cap V) \leq \bigvee\set{\bigvee\set{x\conjun y}{x\in
    U}}{y\in V}$.

Next, the adjunction $\bigvee \dashv \downset$ involves the standard
correspondence:
$$\begin{prooftree}
\bigvee U \;\leq\; x
\Justifies
U \;\subseteq\; \downset x
\end{prooftree}$$

For the implication $(2)\Rightarrow(3)$ assume a left adjoint
$a\dashv\downset$. The adjoint correspondence immediately yields
that $a = \bigvee$, which in this case is a meet preserving map.
We need to show that it satisfies the algebra equations:
$$\begin{array}{rcl}
\big(\bigvee \after \downset\big)(x)
& = &
\bigvee\downset x \\
& = &
x \\
\big(\bigvee \after \Dwn(\bigvee)\big)(A)
& = &
\bigvee\downset\set{\bigvee U}{U\in A} \\
& \smash{\stackrel{(*)}{=}} &
\bigvee\bigcup A \\
& = &
\big(\bigvee \after \bigcup\big)(A),
\end{array}$$

\noindent where the marked equation holds for the following
reasons.

$(\leq)$ Suppose $x\leq \bigvee U$ for $U\in A$. Then $U\subseteq
\bigcup A$ and thus $x \leq \bigvee U \leq \bigvee \bigcup A$.

$(\geq)$ Suppose $x\in\bigcup A$, say $x\in U\in A$. Then
$x \leq \bigvee U$, and thus $x\in \downset\set{\bigvee U}{U\in A}$.
Hence $x\leq \bigvee\downset\set{\bigvee U}{U\in A}$.

For the implication $(3)\Rightarrow(1)$ assume an algebra
$a\colon\Dwn(X) \rightarrow X$. It satisfies $a(\downset x) = x$ and
$a(\downset\set{a(U)}{U\in A}) = a(\bigvee A)$ for $A\in\Dwn^{2}(X)$.
We first prove that $a(U) = \bigvee U$.
\begin{iteMize}{$\bullet$}
\item $a(U)$ is an upperbound of $U$: if $x\in U$, then $\downset x
\subseteq U$ and thus $x = a(\downset x) \leq a(U)$.

\item $a(U)$ is the least upperbound: if $x\leq y$ for each $x\in U$,
then $U\subseteq\downset y$, and thus $a(U) \leq a(\downset y) = y$.
\end{iteMize}

We still have to show $a(U) \conjun y = a(\set{x\conjun y}{x\in U})$.
This follows because $a$ preserves meets:
$$\begin{array}{rcl}
a(U) \conjun y
& = &
a(U) \conjun a(\downset y) \\
& = &
a(U \cap \downset y) \\
& = &
a(\set{x\conjun y}{x\in U}),
\end{array}$$

\noindent where the latter equation holds because:
$$\begin{array}{rcl}
U \cap \downset y
& = &
\set{x\conjun y}{x\in U}
\end{array}$$

\noindent since:

$(\subseteq)$ If $x\in U \cap \downset y$, then $x\in U$ and
$x \leq y$. Hence $x \conjun y = x\in U$.

$(\supseteq)$ If $x\in U$, then $x\conjun y\in U$ and $x\conjun y
\in \downset y$, so that $x\conjun y \in U \cap \downset y$. \qed
}

In a next step, for a frame $X$, the following statements are
equivalent.
\begin{enumerate}[(1)]
\item $X$ is a stably continuous frame, \textit{i.e.}~a frame that is
  continuous as a dcpo, in which $\top\ll \top$, and also $x\ll y$ and
  $x\ll z$ implies $x\ll y\conjun z$;

\item The counit $\bigvee\colon\Dwn(X)\rightarrow X$ of the
comonad $\overline{\Dwn}$ on \Frm has a left adjoint in \Frm;
it is $x\mapsto \ddownset x$.

\item $X$ carries a (necessarily unique) $\overline{\Dwn}$-coalgebra
  structure $X \rightarrow \Dwn(X)$, which is also
  $\ddownset(-)$. \qed
\end{enumerate}
One can show that coalgebra homomorphisms are the \textit{proper}
frame homomorphisms (from~\cite{BanaschewskiB88}) that preserve $\ll$.
We recall from~\cite[VII, 4.5]{Johnstone82} that for a sober
topological space $X$, its opens $\Omega(X)$ form a continuous lattice
iff X is a locally compact space.  Further, the stably continuous
frames are precisely the retracts of frames of the form $\Dwn(X)$, for
$X$ a meet semi-lattice---here via the coreflection $\ddownset \dashv
\bigvee$.

\section{Examples in effectful programming}\label{ExceptionSec}

Since~\cite{Moggi91a} it is standard to describe the semantics of
(sequential) programs in the Kleisli category of a (strong) monad
$T$. The monad captures the computational effect involved, such as
partial computation via the lift monad, non-deterministic computation
via powerset $\Pow$, probabilistic computation via distribution
$\Dstr$, exceptions via $E+(-)$, \textit{etc.} The combination of such
effects has also been studied, in terms of \emph{monad transformers},
see \textit{e.g.}~\cite{LiangHJ95,JaskelioffaM10}. Probably the most
well-known monad transformer is $T \mapsto T(E+-)$, which sends a
monad $T$ capturing some computational effect to the monad $T(E+-)$
which additionally incorporates exceptions (via a fixed exception
object $E$).

It has been observed before that this monad transformer $\E$, given by
$\E(T) = T(E+-)$, is a monad itself, on the category of monads. Hence
we can proceed as in Section~\ref{ComonadSec}, study its category
$\Alg(\E)$ of $\E$-algebras, with induced comonad $\overline{\E}
\colon \Alg(\E) \rightarrow \Alg(\E)$, and with category of coalgebras
$\CoAlg(\overline{\E})$. This has been done in~\cite{Levy06},
resulting in an intruiguing description of the `raise' and `handle'
operations associated with exceptions. Here we briefly recall these
main points, simply because the approach fits very well in the setting
of the current paper. We do not add any new material. The situation is
analogous to previous examples if one sees the algebra as the relevant
introduction rule and the coalgebra as the associated elimination rule
(for a new language construct).

Assume a distributive category $\cat{A}$, that is, a category with
finite products $(\times, 1)$ and coproducts $(+,0)$ such that
products distribute over coproducts, via (canonical) isomorphisms
$\smash{(X\times Y) + (X\times Z) \conglongrightarrow X\times (Y+Z)}$
and $\smash{0 \conglongrightarrow X\times 0}$.  We write
$\StMnd(\cat{A})$ for the category of strong monads on $\cat{A}$, with
monad maps commuting with strength as morphisms. Strength $\st\colon
T(X) \times Y \rightarrow T(X\times Y)$ is standardly assumed in the
theory of monadic computation (see~\cite{Moggi91a}), where it is used
to handle computations in contexts.

\begin{thm}
\label{LevyThm}
In the setting described above,
\begin{enumerate}[\em(1)]
\item the mapping $\E(T) = T(E+-)$ is a monad on the category
  $\StMnd(\cat{A})$ of strong monads on $\cat{A}$, giving rise to a
  situation:
$$\xymatrix@C-1pc{
& \Alg(\E)\ar@(ul,ur)^-{\overline{\E}=FU}\ar[dl]_{\dashv}\ar@/^3.5ex/[dr]
& \\
\StMnd(\cat{A})\ar@(dl,dr)[]_-{\E}\ar@/^3.5ex/[ur]^{F}
   & & \CoAlg\rlap{$(\overline{\E})$}\ar[ul]_{\dashv}
}$$

\item an $\E$-algebra corresponds to a ``throw'' map $E \rightarrow
  T(0)$;

\item a $\overline{\E}$-coalgebra corresponds to a `handle' (or
  `catch') family of maps $T(X) \rightarrow T(X+E)$, satisfying the
  equations for exception handling, see~\cite{Levy06}. \qed
\end{enumerate}
\end{thm}

The second point involves a bijective correspondence:
$$\hspace*{-5em}\begin{prooftree}
{\xymatrix{ \llap{$\E(T)=\;$}T(E+-)\ar@{=>}[r]^-{\sigma} & T 
   \rlap{\quad\small map of monads, as $\E$-algebra}}}
\Justifies
{\xymatrix{ E\ar[r]_-{r} & T(0)}}
\end{prooftree}$$

\noindent This works as follows.
\begin{iteMize}{$\bullet$}
\item Given $\sigma$, take:
$$\xymatrix{
\widehat{\sigma} = \Big(E\ar[r]^-{\eta} & 
   T(E)\ar[r]^-{T(\kappa_{1})}_-{\cong} & T(E+0)\ar[r]^-{\sigma_0} & 
   T(0)\Big).
}$$

\item And given $r\colon E\rightarrow T(0)$, define
$\widehat{r}\colon T(E+-) \Rightarrow T$ with components:
$$\xymatrix{
\widehat{r}_{X} = \Big(T(E+X)\ar[rr]^-{T([T(!)\after r,\eta])} & &
   T^{2}(X)\ar[r]^-{\mu} & T(X)\Big).
}$$
\end{iteMize}

\noindent The coalgebra $T(X) \rightarrow T(X+E)$ in the third point
indeed does a catch, since after this coalgebra one can combine a
cotuple of a normal computation $f\colon X \rightarrow T(Y)$ with an
exception handler $g\colon E\rightarrow T(Y)$ in cotuple $[f,g] \colon
X+E \rightarrow T(Y)$, yielding a catch map $T(X) \rightarrow T(Y)$.

We refer to~\cite{Levy06} (and also~\cite{SchroderM04}) for more
information.

\auxproof{
We write $\eta,\mu$ of the monad of $T$, and define
$\eta^{\E(T)}$ and $\mu^{\E(T)}$ for $\E(T) = T(E+-)$,
in the following way.
$$\xymatrix@R-1.5pc{
\eta^{\E(T)} = \Big(X\ar[r]^-{\eta} & T(X)\ar[r]^-{T(\kappa_{2})} &
   T(E+X)\Big) \\
\mu^{\E(T)} = \Big(T(E + T(E + X))
   \ar[rr]^-{T([T(\kappa_{1})\after\eta, \idmap])}
   & & T^{2}(E+X)\ar[r]^-{\mu} & T(E+X)\Big)
}$$

\noindent Then:
$$\begin{array}{rcl}
\big(\mu^{\E(T)}_{X} \after \eta^{\E(T)}_{T(E+X)}\big)
& = &
\mu \after T([T(\kappa_{1})\after\eta, \idmap]) \after T(\kappa_{2})
   \after \eta \\
& = &
\mu \after T(\idmap) \after \eta \\
& = &
\idmap \\
\big(\mu^{\E(T)}_{X} \after T(E+\eta^{\E(T)}_{X})\big)
& = &
\mu \after T([T(\kappa_{1})\after\eta, \idmap]) \after 
   T([\kappa_{1}, \kappa_{2} \after T(\kappa_{2}) \after \eta]) \\
& = &
\mu \after T([T(\kappa_{1})\after\eta, T(\kappa_{2}) \after \eta]) \\
& = &
\mu \after T([\eta \after \kappa_{1}, \eta \after \kappa_{2}]) \\
& = &
\mu \after T(\eta) \after T([\kappa_{1}, \kappa_{2}]) \\
& = &
\idmap \after T(\idmap) \\
& = &
\idmap \\
\big(\mu^{\E(T)}_{X} \after T(E+\mu^{\E(T)}_{X})\big) 
& = &
\mu \after T([T(\kappa_{1})\after\eta, \idmap]) \after 
   T([\kappa_{1}, \kappa_{2} \after \mu \after 
      T([T(\kappa_{1})\after\eta, \idmap])]) \\
& = &
\mu \after T([T(\kappa_{1})\after\eta, \mu \after 
      T([T(\kappa_{1})\after\eta, \idmap])]) \\
& = &
\mu \after T([\mu \after \eta \after T(\kappa_{1})\after\eta, 
   \mu \after T([T(\kappa_{1})\after\eta, \idmap])]) \\
& = &
\mu \after T(\mu) \after T([T(T(\kappa_{1})\after\eta) \after \eta, 
   T([T(\kappa_{1})\after\eta, \idmap])]) \\
& = &
\mu \after \mu \after T^{2}([T(\kappa_{1})\after\eta, \idmap]) \after
   T([T(\kappa_{1})\after\eta, \idmap]) \\
& = &
\mu \after T([T(\kappa_{1})\after\eta, \idmap]) \after
   \mu \after T([T(\kappa_{1})\after\eta, \idmap]) \\
& = &
\big(\mu^{\E(T)}_{X} \after \mu^{\E(T)}_{T(E+X)}\big)
\end{array}$$

\noindent As strength map $\st^{\E(T)}$ for $T(E+-)$ we take:
$$\xymatrix@C-1pc{
T(E+X)\times Y\ar[r]^-{\st} &
   T((E+X)\times Y)\ar[r]^-{T(\dis)}_-{\cong} & 
   T((E\times Y)+(X\times Y))\ar[rr]^-{T(\pi_{1}+\idmap)} & &
   T(E+(X\times Y)
}$$

\noindent This is a strength map indeed. We start with the
functor-properties of $\st^{\E(T)}$.
$$\begin{array}{rcl}
\lefteqn{\E(T)(f\times g) \after \st^{\E(T)}} \\
& = &
T(\idmap+(f\times g)) \after
   T(\pi_{1}+\idmap) \after T(\dis) \after \st \\
& = &
T(\pi_{1}+(f\times g)) \after T(\dis) \after \st \\
& = &
T(\pi_{1}+\idmap) \after 
   T((\idmap\times g)+(f\times g)) \after T(\dis) \after \st \\
& = &
T(\pi_{1}+\idmap) \after T(\dis) \after 
   T((\idmap+f)\times g) \after \st \\
& = &
T(\pi_{1}+\idmap) \after T(\dis) \after \st \after
   (T(\idmap+f)\times g) \\
& = &
\st^{\E(T)} \after (\E(T)(f)\times g) \\
\lefteqn{\E(T)(\pi_{1}) \after \st^{\E(T)}} \\
& = &
T(\idmap+\pi_{1}) \after T(\pi_{1}+\idmap) \after T(\dis) \after \st \\
& = &
T(\pi_{1}+\pi_{1}) \after T(\dis) \after \st \\
& = &
T(\pi_{1}) \after \st \\
& = &
\pi_{1} \\
\lefteqn{\E(T)(\alpha^{-1}) \after \st^{\E(T)} \after 
   (\st^{\E(T)}\times\idmap)} \\
& = &
T(\idmap+\alpha^{-1}) \after T(\pi_{1}+\idmap) \after T(\dis) \after \st 
   \after 
   \big((T(\pi_{1}+\idmap) \after T(\dis) \after \st)\times\idmap\big) \\
& = &
T(\pi_{1}+\alpha^{-1}) \after T(\dis) \after 
   T((\pi_{1}+\idmap)\times\idmap) \after T(\dis\times\idmap) \after
   \st \after (\st\times\idmap) \\
& = &
T(\pi_{1}+\alpha^{-1}) \after T((\pi_{1}\times\idmap)+(\idmap\times\idmap))
   \after T(\dis) \after T(\dis\times\idmap) \after
   \st \after (\st\times\idmap) \\
& = &
T((\pi_{1}\after\pi_{1})+\alpha^{-1}) 
   \after T(\dis) \after T(\dis\times\idmap) \after
   \st \after (\st\times\idmap) \\
& = &
T(\pi_{1}+\idmap) \after T(\alpha^{-1}+\alpha^{-1}) 
   \after T(\dis) \after T(\dis\times\idmap) \after
   \st \after (\st\times\idmap) \\
& = &
T(\pi_{1}+\idmap) \after T(\dis) \after T(\alpha^{-1}) \after 
  \st \after (\st\times\idmap) \\
& = &
T(\pi_{1}+\idmap) \after T(\dis) \after \st \after \alpha^{-1} \\
& = &
\st^{\E(T)} \after \alpha^{-1}.
\end{array}$$

\noindent We turn to strenght, and $\eta,\mu$:
$$\begin{array}{rcl}
\lefteqn{\st^{\E(T)} \after (\eta^{\E(T)}\times\idmap)} \\
& = &
T(\pi_{1}+\idmap) \after T(\dis) \after \st \after 
   \big((T(\kappa_{2})\after \eta)\times\idmap\big) \\
& = &
T(\pi_{1}+\idmap) \after T(\dis) \after T(\kappa_{2}\times\idmap) \after
   \st \after (\eta\times\idmap) \\
& = &
T(\pi_{1}+\idmap) \after T(\kappa_{2}) \after \eta \\
& = &
T(\kappa_{2}) \after \eta \\
& = &
\eta^{\E(T)} \\
\lefteqn{\st^{\E(T)} \after (\mu^{\E(T)}\times\idmap)} \\
& = &
T(\pi_{1}+\idmap) \after T(\dis) \after \st \after 
   ((\mu \after T([T(\kappa_{1})\after\eta, \idmap]))\times\idmap) \\
& = &
T(\pi_{1}+\idmap) \after T(\dis) \after \mu \after  T(\st) \after \st
   \after (T([T(\kappa_{1})\after\eta, \idmap])\times\idmap) \\
& = &
\mu \after T^{2}(\pi_{1}+\idmap) \after T^{2}(\dis) \after T(\st) 
   \after T([T(\kappa_{1})\after\eta, \idmap]\times\idmap) \after \st \\
& = &
\mu \after T^{2}(\pi_{1}+\idmap) \after T^{2}(\dis) \after T(\st) 
   \after T(\nabla\times\idmap) \after
   T((T(\kappa_{1})\after\eta) + \idmap)\times\idmap) \after \st \\
& = &
\mu \after T^{2}(\pi_{1}+\idmap) \after T^{2}(\dis) \after T(\st) 
   \after T(\nabla) \after T(\dis) \after 
   T((T(\kappa_{1})\after\eta) + \idmap)\times\idmap) \after \st \\
& = &
\mu \after T^{2}(\pi_{1}+\idmap) \after T^{2}(\dis) \after T(\st) 
   \after T(\nabla) \after 
   T(((T(\kappa_{1})\after\eta)\times\idmap) + (\idmap\times\idmap))
   \after T(\dis) \after \st \\
& = &
\mu \after T^{2}(\pi_{1}+\idmap) \after T^{2}(\dis) \after T(\st) 
   \after T([(T(\kappa_{1})\after\eta)\times\idmap, \idmap])
   \after T(\dis) \after \st \\
& = &
\mu \after T([T(\pi_{1}+\idmap) \after T(\dis) \after \st
   \after (T(\kappa_{1})\times\idmap) \after (\eta\times\idmap), 
   T(\pi_{1}+\idmap) \after T(\dis) \after \st])
   \after T(\dis) \after \st \\
& = &
\mu \after T([T(\pi_{1}+\idmap) \after 
   T(\dis \after (\kappa_{1}\times\idmap)) \after \st
   \after (\eta\times\idmap), 
   T(\pi_{1}+\idmap) \after T(\dis) \after \st])
   \after T(\dis) \after \st \\
& = &
\mu \after T([T(\pi_{1}+\idmap) \after T(\kappa_{1}) \after \eta,
   T(\pi_{1}+\idmap) \after T(\dis) \after \st])
   \after T(\dis) \after \st \\
& = &
\mu \after T([T(\kappa_{1} \after \pi_{1}) \after \eta,
   T(\pi_{1}+\idmap) \after T(\dis) \after \st])
   \after T(\dis) \after \st \\
& = &
\mu \after T([T(\kappa_{1})\after\eta \after \pi_{1},
   T(\pi_{1}+\idmap) \after T(\dis) \after \st]) \after 
   T(\dis) \after \st \\
& = &
\mu \after T([T(\kappa_{1})\after\eta, \idmap]) \after 
   T(\pi_{1}+(T(\pi_{1}+\idmap) \after T(\dis) \after \st)) \after 
   T(\dis) \after \st \\
& = &
\mu \after T([T(\kappa_{1})\after\eta, \idmap]) \after 
   T(\idmap+(T(\pi_{1}+\idmap) \after T(\dis) \after \st)) \after 
   T(\pi_{1}+\idmap) \after T(\dis) \after \st \\
& = &
\mu^{\E(T)} \after T(E+\st^{\E(T)}) \after 
\st^{\E(T)}.
\end{array}$$

Next we show that $\E$, given by $\E(T) = T(E+-)$
is a monad on $\StMnd(\cat{A})$. For a map of strong monads
$\sigma\colon T\Rightarrow S$ one has $\E(\sigma) \colon
\E(T) \rightarrow \E(S)$ with components:
$$\begin{array}{rcl}
\E(\sigma)_{X}
& = &
\sigma_{E+X} \colon T(E+X) \longrightarrow S(E+X).
\end{array}$$

\noindent This $\E(\sigma)$ is a map of monads:
$$\begin{array}{rcl}
\E(\sigma) \after \eta^{\E(T)}
& = &
\sigma \after T(\kappa_{2}) \after \eta \\
& = &
S(\kappa_{2}) \after \sigma \after \eta \\
& = &
S(\kappa_{2}) \after \eta \\
& = &
\eta^{\E(S)} \\
\E(\sigma) \after \mu^{\E(T)}
& = &
\sigma \after \mu \after T([T(\kappa_{1})\after\eta, \idmap]) \\
& = &
\mu \after \sigma \after T(\sigma) \after 
   T([T(\kappa_{1})\after\eta, \idmap]) \\
& = &
\mu \after \sigma \after 
   T([\sigma \after T(\kappa_{1})\after\eta, \sigma]) \\
& = &
\mu \after S([S(\kappa_{1})\after \sigma \after\eta, \sigma]) 
   \after \sigma \\
& = &
\mu \after S([S(\kappa_{1})\after \eta, \sigma]) 
   \after \sigma \\
& = &
\mu \after S([S(\kappa_{1})\after\eta, \idmap]) 
   \after S(\idmap+\sigma) \after \sigma \\
& = &
\mu \after S([S(\kappa_{1})\after\eta, \idmap]) \after \sigma
   \after T(\idmap+\sigma) \\
& = &
\mu^{\E(S)} \after \E(\sigma) \after 
   \E(T)(\E(\sigma)) \\
\st^{\E(T)} \after (\E(\sigma)\times\idmap) 
& = &
T(\pi_{1}+\idmap) \after T(\dis) \after \st \after (\sigma\times\idmap) \\
& = &
T(\pi_{1}+\idmap) \after T(\dis) \after \sigma \after \st \\
& = &
\sigma \after S(\pi_{1}+\idmap) \after S(\dis) \after \st \\
& = &
\E(\sigma) \after \st^{\E(S)}.
\end{array}$$

The unit $\eta^{\E}_{T} \colon
T \Rightarrow \E(T)$ has components:
$$\begin{array}{rcl}
\eta^{\E}_{T,X}
& = &
T(\kappa_{2}) \colon T(X) \longrightarrow T(E+X).
\end{array}$$

\noindent Naturality means that for a map of strong monads
$\sigma\colon T\Rightarrow S$ one has:
$$\begin{array}{rcl}
\E(\sigma) \after \eta^{\E}_{T} 
& = &
\sigma \after T(\kappa_{2}) \\
& = &
S(\kappa_{2}) \after \sigma \\
& = &
\eta^{\E}_{S} \after \sigma.
\end{array}$$

\noindent Next, we check that $\eta^{\E}$ is a map of strong 
monads:
$$\begin{array}{rcl}
\eta^{\E} \after \eta
& = &
T(\kappa_{2}) \after \eta \\
& = &
\eta^{\E(T)} \\
\mu^{\E(T)} \after \eta^{\E} \after T(\eta^{\E})
& = &
\mu \after T([T(\kappa_{1})\after\eta, \idmap]) \after T(\kappa_{2})
   \after T^{2}(\kappa_{2}) \\
& = &
\mu \after T^{2}(\kappa_{2}) \\
& = &
T(\kappa_{2}) \after \mu \\
& = &
\eta^{\E} \after \mu \\
\st^{\E(T)} \after (\eta^{\E}\times\idmap) 
& = &
T(\pi_{1}+\idmap) \after T(\dis) \after \st \after 
   (T(\kappa_{2})\times\idmap) \\
& = &
T(\pi_{1}+\idmap) \after T(\dis) \after T(\kappa_{2}\times\idmap) 
  \after \st \\
& = &
T(\pi_{1}+\idmap) \after T(\kappa_{2}) \after \st \\
& = &
T(\kappa_{2}) \after \st \\
& = &
\eta^{\E} \after \st
\end{array}$$

\noindent Similarly there is a multiplication $\mu^{\E}_{T}
\colon \E^{2}(T) \Rightarrow \E(T)$, with
components:
$$\begin{array}{rcl}
\mu^{\E}_{T,X} 
& = &
T([\kappa_{1},\idmap])\colon T(E + (E+X)) \longrightarrow T(E+X)
\end{array}$$

\noindent This $\mu^{\E}$ is natural:
$$\begin{array}{rcl}
\mu^{\E} \after \E^{2}(\sigma)
& = &
T([\kappa_{1},\idmap]) \after \sigma \\
& = &
\sigma \after S([\kappa_{1},\idmap]) \\
& = &
\E(\sigma) \after \mu^{\E}.
\end{array}$$

\noindent Moreover, $\mu^{\E}$ is a map of strong monads:
$$\begin{array}{rcl}
\mu^{\E} \after \eta^{\E^{2}(T)}
& = &
T([\kappa_{1},\idmap]) \after \E(T)(\kappa_{2}) \after
   \eta^{\E(T)} \\
& = &
T([\kappa_{1},\idmap]) \after T(\idmap+\kappa_{2}) \after
   T(\kappa_{2}) \after \eta \\
& = &
T([\kappa_{1},\kappa_{2}]) \after T(\kappa_{2}) \after \eta \\
& = &
T(\kappa_{2}) \after \eta \\
& = &
\eta^{\E(T)}.
\end{array}$$

\noindent For the multiplication law we need to clarify our
notation. We have:
$$\xymatrix@R-1.5pc@C+1pc{
T(E+T(E+-)) = \E(T)^{2}\ar@{=>}[r]^-{\mu^{\E(T)}} &
   \E(T) \\
T(E+(E+-)) = \E^{2}(T)\ar@{=>}[r]^-{\mu^{\E}_{T}} &
   \E(T) \\
}$$

\noindent The following diagram thus needs to commute.
$$\xymatrix@C+1pc{
\E^{2}(T)\E^{2}(T)\ar[d]_{\mu^{\E^{2}(T)}}
      \ar[r]^-{\mu^{\E}_{\E^{2}(T)}} & 
   \E(T)\E^{2}(T)
      \ar[r]^-{\E(T)\mu^{\E}_{T}} & 
   \E(T)\E(T)\ar[d]^{\mu^{\E(T)}} \\
\E^{2}(T)\ar[rr]_-{\mu^{\E}_{T}} & & \E(T)
}$$

\noindent The vertical multiplication $\mu^{\E^{2}(T)}$ on
the left is the following composite.
$$\xymatrix@R-1.5pc{
\E^{2}(T)\E^{2}(T)(X)\ar@{=}[d]
   \ar[rr]^-{\E(T)([\E(T)(\kappa_{1}) \after
       \eta^{\E(T)}_{E},\idmap])} & &
   \E(T)^{2}(E+X)\ar@{=}[d]\ar[r]^-{\mu^{\E(T)}_{E+X}} &
   \E^{2}(T)(X)\ar@{=}[d] \\
T(E+(E+T(E+(E+X))))\ar@/_4ex/[rr]_-{T(\idmap+([T(\idmap+\kappa_{1}) 
   \after T(\kappa_{1}) \after \eta_{E}, \idmap]))} & &
   T(E+T(E+(E+X)))\ar@/_4ex/[dr]_-{\mu \after 
       T([T(\kappa_{1})\after\eta,\idmap])}  & 
   T(E+(E+X))\ar@{=}[d] \\
& & & \E(T)(E+X)
}$$

$$\begin{array}{rcl}
\lefteqn{\mu^{\E(T)} \after \E(T)\mu^{\E}_{T} 
   \after \mu^{\E}_{\E^{2}(T)}} \\
& = &
\mu \after T([T(\kappa_{1})\after\eta, \idmap]) \after
   T(\idmap+T([\kappa_{1},\idmap])) \after T([\kappa_{1},\idmap]) \\
& = &
\mu \after T([T(\kappa_{1})\after\eta, T([\kappa_{1},\idmap])]) 
   \after T([\kappa_{1},\idmap]) \\
& = &
\mu \after T([T(\kappa_{1})\after\eta, 
   [T(\kappa_{1}) \after \eta, T([\kappa_{1},\idmap])]]) \\
& = &
\mu \after T^{2}([\kappa_{1},\idmap]) \after 
   T([T(\kappa_{1})\after\eta, [T(\kappa_{1}) \after \eta, \idmap]]) \\
& = &
\mu \after T^{2}([\kappa_{1},\idmap]) \after 
   T([T(\kappa_{1})\after\eta,\idmap]) \after
   T(\idmap+([T(\kappa_{1}) \after \eta, \idmap])) \\
& = &
T([\kappa_{1},\idmap]) \after \mu \after 
   T([T(\kappa_{1})\after\eta,\idmap]) \after
   T(\idmap+([T(\idmap+\kappa_{1}) \after T(\kappa_{1}) \after 
   \eta, \idmap])) \\
& = &
\mu^{\E} \after \mu^{\E^{2}(T)}
\end{array}$$

Next we prove that there is a correspondence between
$\E$-algebras with carrier $T\in\StMnd(\cat{A})$ and certain
maps $E\rightarrow T(0)$ in $\cat{A}$:
$$\begin{prooftree}
{\xymatrix{ \llap{$T(E+-)=\;$}\E(T)\ar@{=>}[r]^-{\sigma} & T
   \rlap{\quad algebra}}}
\Justifies
{\xymatrix{ E \ar[r]_-{r} & T(0) }}
\end{prooftree}$$

\noindent This works as follows (see~\cite{Levy06}).
\begin{iteMize}{$\bullet$}
\item Given $\sigma$, take:
$$\xymatrix{
\overline{\sigma} = \Big(E\ar[r]^{\eta} & 
   T(E)\ar[r]^-{T(\kappa_{1})}_-{\cong} & T(E+0)\ar[r]^-{\sigma_0} & 
   T(0)\Big).
}$$

\item Conversely, given $r\colon E\rightarrow T(0)$, define
$\overline{r}\colon T(E+-) \Rightarrow T$ with components:
$$\xymatrix{
\overline{r}_{X} = \Big(T(E+X)\ar[rr]^-{T([T(!)\after r,\eta])} & &
   T^{2}(X)\ar[r]^-{\mu} & T(X)\Big).
}$$

\noindent We start with naturality: for $f\colon X\rightarrow Y$,
$$\begin{array}{rcl}
\overline{r}_{Y} \after \E(T)(f)
& = &
\mu \after T([T(!)\after r,\eta]) \after T(\idmap+f) \\
& = &
\mu \after T([T(!)\after r,\eta \after f]) \\
& = &
\mu \after T([T(f) \after T(!)\after r, T(f) \after \eta]) \\
& = &
\mu \after T^{2}(f) \after T([T(!)\after r, \eta]) \\
& = &
T(f) \after \mu \after T([T(!)\after r, \eta]) \\
& = &
T(f) \after \overline{r}_{X}
\end{array}$$

\noindent Further, $\overline{r}$ is a map of monads:
$$\begin{array}{rcl}
\overline{r} \after \eta^{\E(T)}
& = &
\mu \after T([T(!)\after r,\eta]) \after T(\kappa_{2}) \after \eta \\
& = &
\mu \after T(\eta) \after \eta \\
& = &
\eta \\
\overline{r} \after \mu^{\E(T)}
& = &
\mu \after T([T(!)\after r,\eta]) \after \mu \after
   T([T(\kappa_{1}) \after \eta, \idmap]) \\
& = &
\mu \after \mu \after T^{2}([T(!)\after r,\eta]) \after 
   T([T(\kappa_{1}) \after \eta, \idmap]) \\
& = &
\mu \after T(\mu) \after 
   T([T(T(!)\after r) \after \eta, T([T(!)\after r,\eta])]) \\
& = &
\mu \after T([\mu \after T^{2}(!)\after T(r) \after \eta, 
   \mu \after T([T(!)\after r,\eta])]) \\
& = &
\mu \after T([T(!)\after \mu \after \eta \after r, 
   \mu \after T([T(!)\after r,\eta])]) \\
& = &
\mu \after T([\mu \after T(\eta \after\, !)\after r, 
   \mu \after T([T(!)\after r,\eta])]) \\
& = &
\mu \after T(\mu) \after 
   T([T(!)\after r, T([T(!)\after r,\eta])]) \\
& = &
\mu \after \mu \after T^{2}([T(!)\after r,\eta]) \after 
   T([T(!)\after r, \idmap]) \\
& = &
\mu \after T([T(!)\after r,\eta]) \after \mu \after 
   T([T(!)\after r, \idmap]) \\
& = &
\mu \after T([T(!)\after r,\eta]) \after \mu \after 
   T(\mu) \after T([T(\eta \after !)\after r,\eta]) \\
& = &
\mu \after T([T(!)\after r,\eta]) \after \mu \after 
   \mu \after T([T(!)\after r,\eta]) \\
& = &
\mu \after \mu \after T^{2}([T(!)\after r,\eta]) \after 
   \mu \after T([T(!)\after r,\eta]) \\
& = &
\mu \after T(\mu \after T([T(!)\after r,\eta])) \after 
   \mu \after T([T(!)\after r,\eta]) \\
& = &
\mu \after T(\overline{r}) \after \overline{r} \\
\st \after (\overline{r}\times\idmap)
& = &
\st \after ((\mu \after T([T(!)\after r,\eta]))\times\idmap) \\
& = &
\mu \after T(\st) \after \st \after (T([T(!)\after r,\eta])\times\idmap) \\
& = &
\mu \after T(\st) \after T([T(!)\after r,\eta]\times\idmap) \after \st \\
& = &
\mu \after T(\st) \after T(\nabla\times\idmap) \after 
    T(((T(!)\after r) + \eta)\times\idmap) \after \st \\
& = &
\mu \after T(\st) \after T(\nabla) \after T(\dis) \after 
    T(((T(!)\after r) + \eta)\times\idmap) \after \st \\
& = &
\mu \after T(\st) \after T(\nabla) \after 
    T(((T(!)\after r)\times\idmap) + (\eta\times\idmap))
   \after T(\dis) \after \st \\
& = &
\mu \after T([\st \after (T(!)\times\idmap) \after (r\times\idmap),
   \st \after (\eta\times\idmap)]) \after T(\dis) \after \st \\
& \smash{\stackrel{(*)}{=}} &
\mu \after T([T(!) \after \pi_{1} \after (r\times\idmap),
   \st \after (\eta\times\idmap)]) \after T(\dis) \after \st \\
& = &
\mu \after T([T(!)\after r \after \pi_{1},\eta]) \after
   T(\dis) \after \st \\
& = &
\mu \after T([T(!)\after r,\eta]) \after T(\pi_{1}+\idmap) \after
   T(\dis) \after \st \\
& = &
\overline{r} \after \st^{\E(T)}.
\end{array}$$

\noindent The marked equation $(*)$ holds because $\st \after
(T(!)\times\idmap) = T(!) \after \pi_{1}$. This follows from the
following diagram which commutes, since $0\times Y \cong 0$ by
distributivity.
$$\xymatrix{
& T(0)\times Y\ar[d]_{\idmap\times\,!}\ar@/_4ex/[ddl]_{\pi_1}\ar[dr]^{\st}
      \ar[rrr]^-{T(!)\times\idmap} & & & T(X)\times Y\ar[d]^{\st} \\
& T(0)\times 1\ar[dl]_{\pi_1}\ar[dr]^{\st} & 
   T(0\times Y)\ar[d]^{\cong}\ar[rr]^-{T(!\times\idmap)} & & 
   T(X\times Y) \\
T(0)\ar@/_12ex/[rrrru]^-{T(!)} & & 
   T(0\times 1)\ar[ll]_-{T(\pi_{1})}^-{\cong}
}$$

Finally, we also have to check that $\overline{r}$ satisfies
the equations for an $\E$-algebra:
$$\begin{array}{rcl}
\overline{r} \after \eta^{\E}
& = &
\mu \after T([T(!) \after r, \eta]) \after T(\kappa_{2}) \\
& = &
\mu \after T(\eta) \\
& = &
\idmap \\
\overline{r} \after \mu^{\E}
& = &
\mu \after T([T(!) \after r, \eta]) \after T([\kappa_{1},\idmap]) \\
& = &
\mu \after T([T(!) \after r, [T(!) \after r, \eta]]) \\
& = &
\mu \after T([\mu \after T(\eta\after\,!) \after r, 
   \mu \after \eta \after [T(!) \after r, \eta]]) \\
& = &
\mu \after T(\mu) \after  
   T([T(!) \after r, T([T(!) \after r, \eta]) \after \eta]) \\
& = &
\mu \after \mu \after T^{2}([T(!) \after r, \eta]) \after 
   T([T(!) \after r, \eta]) \\
& = &
\mu \after T([T(!) \after r, \eta]) \after \mu \after 
   T([T(!) \after r, \eta]) \\
& = &
\overline{r} \after \E(\overline{r}).
\end{array}$$

\end{iteMize}

\noindent We do have a bijective correspondence, since:
$$\begin{array}{rcl}
\overline{\overline{\sigma}}_{X}
& = &
\mu \after T([T(!)\after \overline{\sigma}, \eta]) \\
& = &
\mu \after T([T(!)\after \sigma_{0} \after T(\kappa_{1}) \after \eta, 
   \eta]) \\
& = &
\mu \after T([\sigma_{X} \after T(\idmap+\,!) \after T(\kappa_{1})
    \after \eta, \eta]) \\
& = &
\mu \after T([\sigma_{X} \after T(\kappa_{1}) \after \eta, \eta]) \\
& = &
\mu \after T([\sigma_{X} \after T(\kappa_{1}) \after \eta, 
   \sigma_{X} \after \eta^{\E(T)}]) \\
& = &
\mu \after T(\sigma_{X}) \after T([T(\kappa_{1}) \after \eta, 
   T(\kappa_{2}) \after \eta]) \\
& = &
\mu \after T(\sigma_{X}) \after T([\eta \after \kappa_{1}, 
   \eta \after \kappa_{2}]) \\
& = &
\mu \after T(\sigma_{X}) \after T(\eta) \\
& = &
\mu \after T(\sigma_{X}) \after \sigma_{X} \after \eta^{\E}
   \after T(\eta) \\
& = &
\sigma_{X} \after \mu^{\E(T)} \after T(\kappa_{2}) \after
   T(\eta) \\
& = &
\sigma_{X} \after \mu \after T([T(\kappa_{1}) \after \eta, \idmap]) 
   \after T(\kappa_{2}) \after T(\eta) \\
& = &
\sigma_{X} \after \mu \after T(\eta) \\
& = &
\sigma_{X} \\
\overline{\overline{r}}
& = &
\overline{r}_{0} \after T(\kappa_{1}) \after \eta \\
& = &
\mu_{0} \after T([r, \eta_{0}]) \after T(\kappa_{1}) \after \eta \\
& = &
\mu_{0} \after T(r) \after \eta \\
& = &
\mu_{0} \after \eta \after r  \\
& = &
r.
\end{array}$$
}

\section{Comonoids from bases}\label{ComonoidSec}

A recent insight, see~\cite{CoeckePV12}, is that orthonormal bases in
finite-dimensional Hilbert spaces can be described via so-called
Frobenius algebras. Orthonormal bases are very important in quantum
theory because they provide a `perspective' for a measurement on a
system. The algebraic re-description of bases in terms of Frobenius
algebras is influential because it gives rise to a diagrammatic
calculus for quantum protocols, see \textit{e.g.}~\cite{Coecke10a}.
In the present section we show how the coalgebra-as-basis perspective
gives rise to comonoidal structure for copy and delete --- and thus to
the essential part of a Frobenius algebra structure.

In general, such a Frobenius algebra consists of an object carrying
both a monoid and a comonoid structure that interact appropriately. In
the self-dual category of Hilbert spaces, it suffices to have either a
monoid or a comonoid, since the dual is induced by the dagger /
adjoint transpose $(-)^{\dag}$. In this section we show that the kind
of coalgebras (on algebras) considered in this paper also give rise to
comonoids, assuming that the category of algebras has monoidal
(tensor) structure.

In a (symmetric) monoidal category $\cat{A}$ a comonoid is the dual of
a monoid, given by maps $\smash{I \stackrel{u}{\leftarrow} X
  \stackrel{d}{\rightarrow} X\otimes X}$ satisfying the duals of the
monoid equations. Such comonoids are used for copying and deletion, in
linear and quantum logic. If $\otimes$ is cartesian product $\times$,
each object carries a unique comonoid structure
$\smash{1\stackrel{!}{\leftarrow} X \stackrel{\Delta}{\rightarrow}
  X\times X}$. The no-cloning theorem in quantum mechanics (due to
Dieks, Wootters and Zurek) says that copying arbitrary states is
impossible. But copying wrt.\ a basis is allowed,
see~\cite{NielsenC00,CoeckePV12,Abramsky10b}.

If a monad $T$ on a symmetric monoidal category $\cat{A}$ is a
commutative (aka.\ symmetric monoidal) monad, and the category
$\Alg(T)$ has enough coequalisers, then it is also symmetric monoidal,
and the free functor $F\colon\cat{A}\rightarrow\Alg(T)$ preserves this
monoidal structure (via the isomorphism $\xi$ used below). This
classical result goes back to~\cite{Kock71a,Kock71b}. We shall use it
for the special case where the monoidal structure on the base category
$\cat{A}$ is cartesian.

\begin{prop}
\label{ComonoidProp}
In the setting described above, assume the category of algebras
$\Alg(T)$ is symmetric monoidal, for a commutative monad $T$ on a
cartesian category $\cat{A}$. Each $\overline{T}$-coalgebra / basis
$b\colon X\rightarrow T(X)$, say on algebra $a\colon T(X)\rightarrow
X$, gives rise to a commutative comonoid in $\Alg(T)$ by:
\begin{equation}
\label{ComonoidDefEqn}
\begin{array}{rcl}
d_{b}
& = &
{\xymatrix@C-.5pc{
\Big(X\ar[r]^-{b} & TX\ar[r]^-{T(\Delta)} &
   T(X\times X)\ar[r]^-{\xi^{-1}}_-{\cong} & 
   T(X)\otimes T(X)\ar[r]^-{a\otimes a} & X\otimes X\Big)
}} \\[-.3em]
u_{b}
& = &
{\xymatrix@C-.5pc{
\Big(X\ar[r]^-{b} & T(X)\ar[r]^-{T(!)} & T(1)=I\Big),
}}
\end{array}
\end{equation}

\noindent where we use the underlying comonoid structure
$\smash{1\stackrel{!}{\leftarrow} X \stackrel{\Delta}{\rightarrow}
  X\times X}$ on $X$ in the underlying category $\cat{A}$.
\end{prop}

\begin{myproof}
It is not hard to see that these $d_b$ and $u_b$ are maps of algebras:
$$\mu_{1} \after T(u_{b})
=
\mu_{1} \after T^{2}(!) \after T(b)
=
T(!) \after \mu_{X} \after T(b)
=
T(!) \after b \after a
=
u_{b} \after a.$$

\noindent The verification of the comonoid properties involves lengthy
calculations, which are basically straightforward. We just show that
$u$ is neutral element for $d$, using the equations from
Definition~\ref{BasisDef}.
$$\begin{array}[b]{rcl}
(u_{b}\otimes\idmap) \after d_{d}
& = &
(T(!)\otimes\idmap) \after (b\otimes\idmap) \after (a \otimes a) 
   \after \xi^{-1} \after T(\Delta) \after b \\
& = &
(T(!)\otimes\idmap) \after (\mu\otimes\idmap) \after (T(b)\otimes a) 
   \after \xi^{-1} \after T(\Delta) \after b \\
& = &
(T(!)\otimes a) \after (\mu\otimes T(a)) \after (T(b)\otimes T(b)) 
   \after \xi^{-1} \after T(\Delta) \after b \\
& = &
(T(!)\otimes a) \after (\mu\otimes \mu) \after \xi^{-1} \after 
   T(b\times b) \after T(\Delta) \after b \\
& = &
(T(!)\otimes a) \after (\mu\otimes \mu) \after \xi^{-1} 
   \after T(\Delta) \after T(b) \after b \\
& = &
(T(!)\otimes a) \after (\mu\otimes \mu) \after \xi^{-1} 
   \after T(\Delta) \after T(\eta) \after b \\
& = &
(T(!)\otimes a) \after (\mu\otimes \mu) \after \xi^{-1} \after 
   T(\eta\times \eta) \after T(\Delta) \after b \\
& = &
(T(!)\otimes a) \after (\mu\otimes \mu) \after (T(\eta)\otimes T(\eta)) \after 
   \xi^{-1} \after T(\Delta) \after b \\
& = &
(T(!)\otimes a) \after \xi^{-1} \after T(\Delta) \after b \\
& = &
(\idmap\otimes a) \after (T(!)\otimes \idmap) \after \xi^{-1} 
   \after T(\Delta) \after b \\
& = &
(\idmap\otimes a) \after \xi^{-1} \after T(!\times \idmap) 
   \after T(\Delta) \after b \\
& = &
(\idmap\otimes a) \after \xi^{-1} \after T(\lambda^{-1}) \after b 
   \qquad \mbox{where }\lambda\colon 1\times X \conglongrightarrow X \\
& = &
(\idmap\otimes a) \after \lambda^{-1} \after b 
   \qquad \mbox{since $\xi$ is monoidal, where $\lambda \colon 
   I \otimes X  \conglongrightarrow X$} \\
& = &
\lambda^{-1} \after a \after b \\
& = &
\lambda^{-1} \colon X\conglongrightarrow I\otimes X.
\end{array}\eqno{\qEd}$$

\end{myproof}


\auxproof{ 
We first check that $u_{b}, d_{b}$ are maps of algebras:
$$\begin{array}{rcl}
\mu \after T(u_{b})
& = &
\mu \after T^{2}(!) \after T(b) \\
& = &
T(!) \after \mu \after T(b) \\
& = &
T(!) \after b \after a \\
& = &
u_{b} \after a \\
d_{b} \after a 
& = &
(a\otimes a) \after \xi^{-1} \after T(\Delta) \after b \after a \\
& = &
(a\otimes a) \after \xi^{-1} \after T(\Delta) \after \mu \after T(b) \\
& = &
(a\otimes a) \after \xi^{-1} \after \mu \after T^{2}(\Delta) \after T(b) \\
& = &
(a\otimes a) \after (\mu\otimes\mu) \after \xi^{-1} \after T(\xi^{-1}) 
   \after T(T(\Delta) \after b) \\
& = &
(a\otimes a) \after (T(a)\otimes T(a)) \after \xi^{-1} \after T(\xi^{-1}
   \after T(\Delta) \after b) \\
& = &
(a\otimes a) \after \xi^{-1} \after T(a\otimes a) \after T(\xi^{-1}
   \after T(\Delta) \after b) \\
& = &
(a\boxtimes a) \after T(d_{b}).
\end{array}$$


The verification of the comonoid properties involves lengthy
calculations, which are basically straightforward. We just show that
$u$ is neutral element for $d$, using the equations from
Definition~\ref{BasisDef}.
$$\begin{array}{rcl}
\lefteqn{(u_{b}\otimes\idmap) \after d_{d}} \\
& = &
(T(!)\otimes\idmap) \after (b\otimes\idmap) \after (a \otimes a) 
   \after \xi^{-1} \after T(\Delta) \after b \\
& = &
(T(!)\otimes\idmap) \after (\mu\otimes\idmap) \after (T(b)\otimes a) 
   \after \xi^{-1} \after T(\Delta) \after b \\
& = &
(T(!)\otimes a) \after (\mu\otimes T(a)) \after (T(b)\otimes T(b)) 
   \after \xi^{-1} \after T(\Delta) \after b \\
& = &
(T(!)\otimes a) \after (\mu\otimes \mu) \after \xi^{-1} \after 
   T(b\otimes b) \after T(\Delta) \after b \\
& = &
(T(!)\otimes a) \after (\mu\otimes \mu) \after \xi^{-1} 
   \after T(\Delta) \after T(b) \after b \\
& = &
(T(!)\otimes a) \after (\mu\otimes \mu) \after \xi^{-1} 
   \after T(\Delta) \after T(\eta) \after b \\
& = &
(T(!)\otimes a) \after (\mu\otimes \mu) \after \xi^{-1} \after 
   T(\eta\times \eta) \after T(\Delta) \after b \\
& = &
(T(!)\otimes a) \after (\mu\otimes \mu) \after (T(\eta)\times T(\eta)) \after 
   \xi^{-1} \after T(\Delta) \after b \\
& = &
(T(!)\otimes a) \after \xi^{-1} \after T(\Delta) \after b \\
& = &
(\idmap\otimes a) \after (T(!)\otimes \idmap) \after \xi^{-1} 
   \after T(\Delta) \after b \\
& = &
(\idmap\otimes a) \after \xi^{-1} \after T(!\times \idmap) 
   \after T(\Delta) \after b \\
& = &
(\idmap\otimes a) \after \xi^{-1} \after T(\lambda^{-1}) \after b 
   \qquad \mbox{where }\lambda\colon 1\times X \congrightarrow X \\
& = &
(\idmap\otimes a) \after \lambda^{-1} \after b 
   \qquad \mbox{since $\xi$ is monoidal, where $\lambda \colon 
   I \otimes X  \congrightarrow X$} \\
& = &
\lambda^{-1} \after a \after b \\
& = &
\lambda^{-1}.
\end{array}$$

Associativity:
$$\begin{array}{rcl}
\lefteqn{\alpha \after (\idmap\otimes d) \after d} \\
& = &
\alpha \after (\idmap \otimes (a \otimes a)) \after (\idmap \otimes \xi^{-1}) 
   \after (\idmap \otimes T(\Delta)) \after (\idmap \otimes b) \after \\
& & \qquad
   (a \otimes a) \after \xi^{-1} \after T(\Delta) \after b \\
& = &
\alpha \after (a \otimes (a \otimes a)) \after (\idmap \otimes \xi^{-1}) 
   \after (\idmap \otimes T(\Delta)) \after (\idmap \otimes \mu) \after \\
& & \qquad
   (\idmap \otimes T(b)) \after \xi^{-1} \after T(\Delta) \after b \\
& = &
\alpha \after (a \otimes (a \otimes a)) \after (\idmap \otimes \xi^{-1}) 
   \after (\idmap \otimes T(\Delta)) \after (T(a) \otimes \mu) \after \\
& & \qquad
   (T(b) \otimes T(b)) \after \xi^{-1} \after T(\Delta) \after b \\
& = &
\alpha \after (a \otimes (a \otimes a)) \after (\idmap \otimes \xi^{-1}) 
   \after (\idmap \otimes T(\Delta)) \after (\mu \otimes \mu) \after \\
& & \qquad
   \xi^{-1} \after T(b \times b) \after T(\Delta) \after b \\
& = &
((a \otimes a) \otimes a) \after \alpha \after (\idmap \otimes \xi^{-1}) 
   \after (\idmap \otimes T(\Delta)) \after (\mu \otimes \mu) \after \\
& & \qquad
   \xi^{-1} \after T(\Delta) \after T(b) \after b \\
& = &
((a \otimes a) \otimes a) \after \alpha \after (\idmap \otimes \xi^{-1}) 
   \after (\idmap \otimes T(\Delta)) \after (\mu \otimes \mu) \after \\
& & \qquad
   \xi^{-1} \after T(\Delta) \after T(\eta) \after b \\
& = &
((a \otimes a) \otimes a) \after \alpha \after (\idmap \otimes \xi^{-1}) 
   \after (\idmap \otimes T(\Delta)) \after (\mu \otimes \mu) \after \\
& & \qquad
   \xi^{-1} \after T(\eta\times\eta) \after T(\Delta) \after b \\
& = &
((a \otimes a) \otimes a) \after \alpha \after (\idmap \otimes \xi^{-1}) 
   \after (\idmap \otimes T(\Delta)) \after (\mu \otimes \mu) \after \\
& & \qquad
   (T(\eta)\otimes T(\eta)) \after \xi^{-1} \after T(\Delta) \after b \\
& = &
((a \otimes a) \otimes a) \after \alpha \after (\idmap \otimes \xi^{-1}) 
   \after (\idmap \otimes T(\Delta)) \after \\
& & \qquad
   \xi^{-1} \after T(\Delta) \after b \\
& = &
((a \otimes a) \otimes a) \after \alpha \after (\idmap \otimes \xi^{-1}) 
   \after \xi^{-1} \after T(\idmap \times \Delta) \after T(\Delta) \after b \\
& = &
((a \otimes a) \otimes a) \after (\xi^{-1} \otimes \idmap) \after \xi^{-1} 
   \after T(\alpha) \after T(\idmap \times \Delta) \after T(\Delta) \after b \\
& = &
((a \otimes a) \otimes a) \after (\xi^{-1} \otimes \idmap) \after \xi^{-1} 
   \after T(\Delta \times \idmap) \after T(\Delta) \after b.
\end{array}$$

\noindent In a similar way one obtains the same outcome by starting from:
$$\begin{array}{rcl}
(d\otimes\idmap) \after d
& = &
((a \otimes a) \otimes\idmap) \after (\xi^{-1}\otimes\idmap) 
   \after (T(\Delta)\otimes\idmap) \after (b\otimes\idmap) \after \\
& & \qquad
   (a \otimes a) \after \xi^{-1} \after T(\Delta) \after b \\
& = &
((a \otimes a) \otimes a) \after (\xi^{-1}\otimes\idmap) 
   \after (T(\Delta)\otimes\idmap) \after (\mu\otimes\idmap) \after \\
& & \qquad
   (T(b)\otimes\idmap) \after \xi^{-1} \after T(\Delta) \after b \\
& = &
((a \otimes a) \otimes a) \after (\xi^{-1}\otimes\idmap) 
   \after (T(\Delta)\otimes\idmap) \after (\mu\otimes T(a )) \after \\
& & \qquad
   (T(b)\otimes T(b)) \after \xi^{-1} \after T(\Delta) \after b \\
& = &
((a \otimes a) \otimes a) \after (\xi^{-1}\otimes\idmap) 
   \after (T(\Delta)\otimes\idmap) \after (\mu\otimes\mu) \after \\
& & \qquad
   \xi^{-1} \after T(b\times b) \after T(\Delta) \after b \\
& = &
((a \otimes a) \otimes a) \after (\xi^{-1}\otimes\idmap) 
   \after (T(\Delta)\otimes\idmap) \after (\mu\otimes\mu) \after \\
& & \qquad
   \xi^{-1} \after T(\Delta) \after T(b) \after b \\
& = &
((a \otimes a) \otimes a) \after (\xi^{-1}\otimes\idmap) 
   \after (T(\Delta)\otimes\idmap) \after (\mu\otimes\mu) \after \\
& & \qquad
   \xi^{-1} \after T(\Delta) \after T(\eta) \after b \\
& = &
((a \otimes a) \otimes a) \after (\xi^{-1}\otimes\idmap) 
   \after (T(\Delta)\otimes\idmap) \after (\mu\otimes\mu) \after \\
& & \qquad
   \xi^{-1} \after T(\eta\times \eta) \after T(\Delta) \after b \\
& = &
((a \otimes a) \otimes a) \after (\xi^{-1}\otimes\idmap) 
   \after (T(\Delta)\otimes\idmap) \after (\mu\otimes\mu) \after \\
& & \qquad
   (T(\eta)\otimes T(\eta)) \after \xi^{-1} \after T(\Delta) \after b \\
& = &
((a \otimes a) \otimes a) \after (\xi^{-1}\otimes\idmap) 
   \after (T(\Delta)\otimes\idmap) \after \\
& & \qquad
   \xi^{-1} \after T(\Delta) \after b \\
& = &
((a \otimes a) \otimes a) \after (\xi^{-1} \otimes \idmap) \after \xi^{-1} 
   \after T(\Delta \times \idmap) \after T(\Delta) \after b.
\end{array}$$

\noindent This comultiplication $d$ is commutative:
$$\begin{array}{rcl}
\gamma \after d
& = &
\gamma \after (a \otimes a) \after \xi^{-1} \after T(\Delta) \after b \\
& = &
(a \otimes a) \after \gamma \after \xi^{-1} \after T(\Delta) \after b \\
& = &
(a \otimes a) \after \xi^{-1} \after T(\gamma) \after T(\Delta) \after b \\
& = &
(a \otimes a) \after \xi^{-1} \after T(\Delta) \after b \\
& = &
d.
\end{array}$$

For functoriality, assume two algebras $\smash{T(X_{i})
  \stackrel{a_i}{\rightarrow} X_{i}}$, for $i=1,2$, carrying
$\smash{\overline{T}}$-coalgebras $X_{i} \stackrel{b_i}{\rightarrow}
T(X_{i})$. A map of coalgebras $f\colon b_{1} \rightarrow b_{2}$ is
given by a map $f\colon X_{1} \rightarrow X_{2}$ satisfying $f \after
a_{1} = a_{2} \after T(f)$ and $b_{2} \after f = T(f) \after b_{1}$.
This $f$ is then a map $(d_{1},u_{1}) \rightarrow (d_{2}, u_{2})$
between the induced comonoid. For instance,
$$u_{2} \after f
=
F(!) \after b_{2} \after f 
=
F(!) \after T(f) \after b_{1}
=
F(!) \after b_{1}
=
u_{1}.\eqno{\qEd}$$

$$\begin{array}{rcl}
d_{2} \after f
& = &
(a_{2} \otimes a_{2}) \after \xi^{-1} \after T(\Delta) \after b_{2}
   \after f \\
& = &
(a_{2} \otimes a_{2}) \after \xi^{-1} \after T(\Delta) \after T(f)
   \after b_{1} \\
& = &
(a_{2} \otimes a_{2}) \after \xi^{-1} \after T(f\times f) \after T(\Delta) 
   \after b_{1} \\
& = &
(a_{2} \otimes a_{2}) \after (T(f)\otimes T(f)) \after \xi^{-1} \after T(\Delta) 
   \after b_{1} \\
& = &
(f \otimes f) \after (a_{1}\otimes a_{1}) \after \xi^{-1} \after T(\Delta) 
   \after b_{1} \\
& = &
(f \otimes f) \after d_{1}.
\end{array}$$

It is not hard to see that when the monoidal structure on $\cat{A}$ is
cartesian, applying the constructions~(\ref{ComonoidDefEqn}) to the
coalgebra $\delta = T(\eta) \colon T(Y)\rightarrow T^{2}(X)$ on a free
algebra $T(Y)$, as in Lemma~\ref{FreeAlgCoalgLem}, yields essentially
this comonoid:
\begin{equation}
\label{FreeAlgCoalgComonoidDiag}
\vcenter{\xymatrix{
T(1) & & T(Y)\ar[ll]_-{u = T(!)}\ar[rr]^-{d = \xi^{-1} \after T(\Delta)}
   & & T(Y)\otimes T(Y)
}}
\end{equation}

The unit is obtained as $u = T(!) \after \delta = T(!) \after T(\eta)
= T(! \after \eta) = T(!)$, and the comultiplication as:
$$\begin{array}{rcl}
d
& = &
(\mu \otimes\mu) \after \xi^{-1} \after T(\Delta) \after \delta \\
& = &
(\mu \otimes\mu) \after \xi^{-1} \after T(\Delta) \after T(\eta) \\
& = &
(\mu \otimes\mu) \after \xi^{-1} \after T(\eta\times\eta) \after T(\Delta) \\
& = &
(\mu \otimes\mu) \after (T(\eta)\otimes T(\eta)) \after \xi^{-1} \after 
   T(\Delta) \\
& = &
\xi^{-1} \after T(\Delta).
\end{array}$$
}

\begin{exa}
\label{ComonoidEx}
To make the comonoid construction~(\ref{ComonoidDefEqn}) more
concrete, let $V$ be a vector space, say over the complex numbers
$\mathbb{C}$, with a Hamel basis, described as a coalgebra $b\colon
V\rightarrow \Mlt_{\mathbb{C}}(V)$ like in
Theorem~\ref{OperatorCoalgThm}, with basic elements $(e_{j})$,
satisfying $b(e_{j}) = 1e_{j}$. The counit $u_{b} =
\Mlt_{\mathbb{C}}(!)  \after b \colon V\rightarrow \mathbb{C}$
from~\eqref{ComonoidDefEqn} first represents a vector wrt.\ this
basis, and then adds the (finitely many) coefficients:
$$\textstyle v
\longmapsto
\sum_{j}v_{j}e_{j}
\longmapsto
\sum_{j}v_{j}.$$ 

\noindent Similarly, the comultiplication $d_{b} \colon V \rightarrow
V\otimes V$ as in~\eqref{ComonoidDefEqn} is the composite:
$$\textstyle v
\longmapsto
\sum_{j}v_{j}e_{j}
\longmapsto
\sum_{j}v_{j}(e_{j}\sotimes e_{j}),$$

\noindent like in~\cite{CoeckePV12}. 

(For Hilbert spaces one uses orthonormal bases instead of Hamel bases;
the counit $u$ of the comonoid then exists only in the
finite-dimensional case. The comultiplication $d$ seems more relevant,
see also below, and may thus also be studied on its own, like
in~\cite{AbramskyH12}, without finiteness restriction.)


Another example is the ideal monad $\Idl\colon\PoSets \rightarrow
\PoSets$ from Subsection~\ref{DcpoSubsec}. It preserves finite
products, and as a result, the induced monoidal structure on the
category of algebras $\Alg(\Idl) = \Cat{Dcpo}$ is cartesian. Hence the
comonoid structure~(\ref{ComonoidDefEqn}) is given by actual diagonals
and (unique) maps to the final object. For instance, when $\otimes =
\times$ on algebras:
$$\begin{array}{rcll}
d
& = &
(a\times a) \after \xi^{-1} \after T(\Delta) \after b \\
& = &
(a\times a) \after \Delta \after b &
   \quad \mbox{since } \xi^{-1} = \tuple{T(\pi_{1}), T(\pi_{2})} \\
& = &
\Delta \after a \after b \\
& = &
\Delta.
\end{array}$$
\end{exa}

\noindent In general, given a comonoid $\smash{I\stackrel{u}{\leftarrow} X
  \stackrel{d}{\rightarrow} X\otimes X}$, an endomap $f\colon
X\rightarrow X$ may be called \emph{diagonalised}---wrt.~this
comonoid, or actually, comultiplication $d$---if there is a ``map of
eigenvalues'' $v\colon X\rightarrow I$ such that $f$ equals the
composite:
\begin{equation}
\label{ComonoidDiagEqn}
\vcenter{\xymatrix{
X\ar[r]^-{d} & X\otimes X\ar[r]^-{v\otimes\idmap} & 
   I\otimes X\ar[r]^-{\lambda}_-{\cong} & X.
}}
\end{equation}

\noindent In case the diagonal $d$ is part of a comonoid, with a
counit $u\colon X \rightarrow I$, then this eigenvalue map $v$ equals
$u \after f$.

\auxproof{
Assume $f$ is diagonalised as above. Then:
$$\begin{array}{rcl}
u \after f
& = &
u \after \lambda_{X} \after (v\otimes\idmap) \after d \\
& = &
\lambda_{I} \after (\idmap\otimes v) \after (v\otimes\idmap) \after d \\
& = &
\rho_{I} \after (v\otimes\idmap) \after (\idmap\otimes v) \after d \\
& = &
\rho_{I} \after (v\otimes\idmap) \after \rho_{X}^{-1} \\
& = &
v.
\end{array}$$
}

In the special case where the comonoid comes from a
coalgebra (basis) $b\colon X\rightarrow T(X)$, like
in~(\ref{ComonoidDefEqn}), an endomap of algebras $f\colon
X\rightarrow X$, say on $a\colon T(X)\rightarrow X$, is diagonalised
if there is a map of algebras $v\colon X\rightarrow I = T(1)$ such
that $f$ is:
$$\vcenter{\xymatrix@C-.5pc{
X\ar[r]^-{b} &
T(X)\ar[rr]^-{T(\tuple{v,\idmap})} & &
T(T(1)\times X)\ar[r]^-{T(\st)} &
T^{2}(1\times X) \ar[r]^-{\cong} &
T^{2}(X)\ar[r]^-{\mu} &
T(X)\ar[r]^-{a} &
X,
}}$$

\noindent where $\st$ is a strength map of the form $T(X)\times Y
\rightarrow T(X\times Y)$, which exists because the monad $T$ is
assumed to be commutative.

\begin{exa}
\label{BifMRelEx}
Recall the multiset monad $\Mlt_S$ from
Subsection~\ref{VectorSpaceSubsec}, where $S$ is a semiring.
In~\cite{Jacobs11a} it is used to defined a dagger category
$\BifMRel_S$ of ``bifinite multirelations''.  Objects are sets $X$,
and maps $X \rightarrow Y$ in $\BifMRel_S$ are multirelations $r\colon
X\times Y \rightarrow S$ which factor both as
$X\rightarrow\Mlt_{S}(Y)$ and as $Y\rightarrow\Mlt_{S}(X)$. This means
that for each $x$ there are finitely many $y$ with $r(x,y)\neq 0$, and
similarly, for each $y$ there are finitely many $x$ with $r(x,y)\neq
0$. Composition of $r\colon X\rightarrow Y$ and $s\colon Y\rightarrow
Z$ is done via matrix multiplication: $(s\after r)(x,z) = \sum_{y}
s(y,z)\cdot r(x,y)$. The identity $\idmap\colon X \rightarrow X$ is
the given by $\idmap(x,x) = 1$ and $\idmap(x,x')=0$ if $x\neq x'$. 

Here we don't need the dagger $(-)^{\dag}$ on $\BifMRel_{S}$, but for
completeness we briefly mention how it arises, assuming that $S$
carries an involution $\overline{(-)}\colon S\rightarrow S$, like
conjugation on the complex numbers. For a map $r\colon X\rightarrow Y$
there is an associated map $r^{\dag}\colon Y\rightarrow X$ in the
reverse direction, obtained by swapping arguments and involution:
$r^{\dag}(y,x) = \overline{r(x,y)}$, like in a conjugate
transpose. This makes $\BifMRel_{S}$ a dagger category.

The category $\BifMRel_{S}$ is symmetric monoidal, with $\times$ as
tensor and $1 = \{*\}$ as tensor unit. Coproducts $(+,0)$ give
biproducts. Interestingly, each object $X$ carries a (canonical)
diagonal $d\colon X \rightarrow X\otimes X$ given by:
$$\xymatrix{
X\times (X\times X)\ar[r]^-{d} & S, 
\qquad
(x,(y,y'))\ar@{|->}[r] & {\left\{\begin{array}{ll}
   1 & \mbox{ if }x=y=y' \\ 0 & \mbox{ otherwise.} \end{array}\right.}
}$$

\noindent There is in general no associated counit $u\colon X
\rightarrow 1$.

Now let's see what it means that a map $r\colon X\rightarrow X$ in
$\BifMRel_{S}$ is diagonalised wrt.\ this $d$. It would require an
eigenvalue map $v\colon X \rightarrow 1$, that is, a function $v\colon
X \rightarrow S$ in $\Sets$, so that $r\colon X \times X \rightarrow S$
satisfies:
$$\begin{array}{rcl}
r(x,x')
& = &
\left\{\begin{array}{ll}
v(x) & \mbox{ if }x=x' \\
0 & \mbox{ otherwise.}
\end{array}\right.
\end{array}$$

\noindent Hence such a diagonalised map is a diagonal matrix.
\end{exa}

What precisely is a diagonalised form depends on the diagonalisation
(comonoid) map $d$ involved. This is clear in the following example,
involving Pauli matrices.

\begin{exa}
\label{PauliEx}
We consider the set $\mathbb{C}^{2}$ as vector space over
$\mathbb{C}$, and thus as algebra of the (commutative) multiset monad
$\Mlt_{\mathbb{C}}\colon \Sets \rightarrow \Sets$ via the map
$\smash{\Mlt_{\mathbb{C}}(\mathbb{C}^{2})
  \stackrel{a}{\longrightarrow} \mathbb{C}^{2}}$ that sends a formal
sum $s_{1}(z_{1}, w_{1}) + \cdots + s_{n}(z_{n}, w_{n})$ of pairs in
$\mathbb{C}^{2}$ to the pair of sums $(s_{1}\cdot z_{1} + \cdots +
s_{n}\cdot z_{n}, s_{1}\cdot w_{1} + \cdots + s_{n}\cdot w_{n})
\in\mathbb{C}^{2}$.

The familiar Pauli spin functions $\sigma_{\textsf{x}},
\sigma_{\textsf{y}}, \sigma_{\textsf{z}} \colon \mathbb{C}^{2}
\rightarrow \mathbb{C}^{2}$ are given by:
$$\sigma_{\textsf{x}}(z,w) = (w,z)
\qquad\quad
\sigma_{\textsf{y}}(z,w) = (-iw, iz)
\qquad\quad
\sigma_{\textsf{z}}(z,w) = (z,-w).$$

\noindent We concentrate on $\sigma_{\textsf{x}}$; it satisfies
$\sigma_{\textsf{x}}(1,1) = (1,1)$ and $\sigma_{\textsf{x}}(1,-1) =
-(1,-1)$. These eigenvectors $(1,1)$ and $(1,-1)$ are organised in a
basis $b_{\textsf{x}}\colon \mathbb{C}^{2} \rightarrow
\Mlt_{\mathbb{C}}(\mathbb{C}^{2})$, as in Definition~\ref{BasisDef},
via the following formal sum.
$$\begin{array}{rcl}
b_{\textsf{x}}(z,w)
& = &
\frac{z+w}{2}(1,1) + \frac{z-w}{2}(1,-1).
\end{array}$$

\noindent It expresses an arbitrary element of
$\mathbb{C}^{2}$ in terms of this basis of eigenvectors. It is not
hard to see that $b_{\textsf{x}}$ is a
$\overline{\Mlt_{\mathbb{C}}}$-coalgebra; for instance:
$$\textstyle\big(a \after b_{\textsf{x}})(z,w)
=
a\big(\frac{z+w}{2}(1,1) + \frac{z-w}{2}(1,-1)\big) 
=
(\frac{z+w}{2} + \frac{z-w}{2}, \frac{z+w}{2} - \frac{z-w}{2}) 
=
(z,w).$$

\auxproof{
$$\begin{array}{rcl}
\big(\Mlt_{\mathbb{C}}(b_{\textsf{x}}) \after b_{\textsf{x}}\big)(z,w)
& = &
\frac{z+w}{2}b_{\textsf{x}}(1,1) + \frac{z-w}{2}b_{\textsf{x}}(1,-1) \\
& = &
\frac{z+w}{2}\big((1(1,1) + 0(1,-1)\big) + 
   \frac{z-w}{2}\big(0(1,1) + 1(1,-1)\big) \\
& = &
\frac{z+w}{2}\big((1(1,1)\big) + \frac{z-w}{2}\big(1(1,-1)\big) \\
& = &
\frac{z+w}{2}\big(\eta(1,1)\big) + \frac{z-w}{2}\big(\eta(1,-1)\big) \\
& = &
\big(\Mlt_{\mathbb{C}}(\eta) \after b_{\textsf{x}}\big)(z,w) \\
& = &
\big(\delta \after b_{\textsf{x}}\big)(z,w).
\end{array}$$

We also check that $b_{\textsf{x}}$ is a map of algebras:
$$\begin{array}{rcl}
\lefteqn{\textstyle\big(\mu \after \Mlt_{\mathbb{C}}(b_{\textsf{x}})\big)
   (\sum_{j}s_{j}(z_{j},w_{j}))} \\
& = &
\mu(\sum_{j}s_{j}b_{\textsf{x}}(z_{j},w_{j})) \\
& = &
\mu(\sum_{j}s_{j}(\frac{z_{j}+w_{j}}{2}(1,1) + \frac{z_{j}-w_{j}}{2}(1,-1))) \\
& = &
\sum_{j}(s_{j}\cdot\frac{z_{j}+w_{j}}{2})(1,1) + 
   (s_{j}\cdot\frac{z_{j}-w_{j}}{2})(1,-1) \\
& = &
\sum_{j}\frac{s_{j}\cdot z_{j}+s_{j}\cdot w_{j}}{2}(1,1) + 
   \sum_{j}\frac{s_{j}\cdot z_{j}-s_{j}\cdot w_{j}}{2}(1,-1) \\
& = &
{\displaystyle
\frac{(\sum_{j}s_{j}\cdot z_{j})+(\sum_{j}s_{j}\cdot w_{j})}{2}(1,1) + 
   \frac{(\sum_{j}s_{j}\cdot z_{j})-(\sum_{j}s_{j}\cdot w_{j})}{2}(1,-1)} \\
& = &
b_{\textsf{x}}(\sum_{j}s_{j}\cdot z_{j}, \sum_{j}s_{j}\cdot w_{j}) \\
& = &
\big(b_{\textsf{x}} \after a\big)(\sum_{j}s_{j}(z_{j},w_{j})).
\end{array}$$

The equaliser~(\ref{BasisEqualiserDiag}) for the basic elements
yields the expected outcome:
$$\begin{array}{rcl}
X_{b}
& = &
\set{(z,w)}{b_{\textsf{x}}(z,w) = \eta(z,w)} \\
& = &
\set{(z,w)}{\frac{z+w}{2}(1,1) + \frac{z-w}{2}(1,-1) = 1(z,w)} \\
& = &
\{(1,1), (1,-1)\}.
\end{array}$$
}

\noindent The comonoid structure $\smash{\mathbb{C}
  \xleftarrow{u_{\textsf{x}}} \mathbb{C}^{2}
  \xrightarrow{d_{\textsf{x}}} \mathbb{C}^{2}\otimes \mathbb{C}^{2}}$
induced by $b_{\textsf{x}}$ as in~(\ref{ComonoidDefEqn}) is given by
$u_{\textsf{x}}(z,w) = z$ and $d_{\textsf{x}}(z,w) =
\frac{z+w}{2}\big((1,1)\sotimes(1,1)\big) +
\frac{z-w}{2}\big((1,-1)\sotimes(1,-1)\big)$. The eigenvalue map
$v_{\textsf{x}}\colon \mathbb{C}^{2} \rightarrow \mathbb{C}$ is given
by $v_{\textsf{x}}(z,w) = w$. The eigenvalues $1,-1$ appear by
application to the basic elements: $v_{\textsf{x}}(1,1) = 1$ and
$v_{\textsf{x}}(1,-1) = -1$. Further, the Pauli function
$\sigma_{\textsf{x}}$ is diagonalised as in~(\ref{ComonoidDiagEqn})
via these $d_{\textsf{x}}, v_{\textsf{x}}$, since:
$$\begin{array}{rcl}
\lefteqn{\big(\lambda \after (v_{\textsf{x}}\otimes\idmap) \after 
   d_{\textsf{x}}\big)(z,w)} \\
& = &
\big(\lambda \after (v_{\textsf{x}}\otimes\idmap)\big)
   \Big(\frac{z+w}{2}\big((1,1)\sotimes(1,1)\big) + 
    \frac{z-w}{2}\big((1,-1)\sotimes(1,-1)\big)\Big) \\
& = &
\lambda\Big(\frac{z+w}{2}\big(1\sotimes(1,1)\big) + 
    \frac{z-w}{2}\big(-1\sotimes(1,-1)\big)\Big) \\
& = &
\frac{z+w}{2}(1,1) - \frac{z-w}{2}(1,-1) \\
& = &
(w, z) \\
& = &
\sigma_{\textsf{x}}(z,w).
\end{array}$$

\auxproof{
$$\begin{array}{rcl}
u_{\textsf{x}}(z,w)
& = &
\big(\Mlt_{\mathbb{C}}(!) \after b_{\textsf{x}}\big)(z,w) \\
& = &
\Mlt_{\mathbb{C}}(!)\Big(\frac{z+w}{2}(1,1) + \frac{z-w}{2}(1,-1)\Big) \\
& = &
\frac{z+w}{2} + \frac{z-w}{2} \\
& = &
z \\
d_{\textsf{x}}(z,w)
& = &
\big((a\otimes a) \after \xi^{-1} \after \Mlt_{\mathbb{C}}(\Delta)
   \after b_{\textsf{x}}\big)(z,w) \\
& = &
\big((a\otimes a) \after \xi^{-1} \after \Mlt_{\mathbb{C}}(\Delta)\big)
   \Big(\frac{z+w}{2}(1,1) + \frac{z-w}{2}(1,-1)\Big) \\
& = &
\big((a\otimes a) \after \xi^{-1}\big)
   \Big(\frac{z+w}{2}\tuple{(1,1),(1,1)} + 
      \frac{z-w}{2}\tuple{(1,-1),(1,-1)}\Big) \\
& = &
\big(a\otimes a\big)
   \Big(\frac{z+w}{2}\big((1,1)\sotimes(1,1)\big) + 
      \frac{z-w}{2}\big((1,-1)\sotimes(1,-1)\big)\Big) \\
& = &
\frac{z+w}{2}\big((1,1)\sotimes(1,1)\big) + 
    \frac{z-w}{2}\big((1,-1)\sotimes(1,-1)\big).
\end{array}$$

Alternatively, elaborating the formula~(\ref{AlgDiagEqn}) yields:
$$\begin{array}{rcl}
\lefteqn{\big(a \after \mu \after \Mlt_{\mathbb{C}}^{2}(\pi_{1}) \after 
   \Mlt_{\mathbb{C}}(\st) \after \Mlt_{\mathbb{C}}(\tuple{\idmap,\lambda})
   \after b\big)(z,w)} \\
& = &
\big(a \after \mu \after \Mlt_{\mathbb{C}}^{2}(\pi_{1}) \after 
   \Mlt_{\mathbb{C}}(\st) \after \Mlt_{\mathbb{C}}(\tuple{\idmap,\lambda})\big)
   \big(\frac{z+w}{2}(1,1) + \frac{z-w}{2}(1,-1)\big) \\
& = &
\big(a \after \mu \after \Mlt_{\mathbb{C}}^{2}(\pi_{1}) \after 
   \Mlt_{\mathbb{C}}(\st)\big)
   \big(\frac{z+w}{2}\tuple{\idmap,\lambda}(1,1) + 
   \frac{z-w}{2}\tuple{\idmap,\lambda}(1,-1)\big) \\
& = &
\big(a \after \mu \after \Mlt_{\mathbb{C}}^{2}(\pi_{1})\big)
   \big(\frac{z+w}{2}\st(\tuple{(1,1), 1*}) + 
   \frac{z-w}{2}\st(\tuple{(1,-1), -1*})\big) \\
& = &
\big(a \after \mu \after \Mlt_{\mathbb{C}}^{2}(\pi_{1})\big)
   \big(\frac{z+w}{2}(1\tuple{(1,1), *}) + 
   \frac{z-w}{2}(-1\tuple{(1,-1), *})\big) \\
& = &
\big(a \after \mu\big) \big(\frac{z+w}{2}(1(1,1)) + 
   \frac{z-w}{2}(-1(1,-1))\big) \\
& = &
a\big(\frac{z+w}{2}(1,1) + \frac{w-z}{2}(1,-1)\big) \\
& = &
(\frac{z+w}{2} + \frac{w-z}{2}, \frac{z+w}{2} - \frac{w-z}{2}) \\
& = &
(w, z) \\
& = &
\sigma_{\textsf{x}}(z,w).
\end{array}$$
}

\noindent In a similar way one defines for the other Pauli functions
$\sigma_{\textsf{y}}$ and $\sigma_{\textsf{z}}$:
$$\begin{array}{rclcrcl}
b_{\textsf{y}}(z,w)
& = &
\frac{iz+w}{2}(-i,1) + \frac{iz-w}{2}(i,1)
& \qquad &
v_{\textsf{y}}(z,w)
& = &
iz \\
b_{\textsf{z}}(z,w)
& = &
z(1,0) - w(0,1)
& \qquad &
v_{\textsf{z}}(z,w)
& = &
z-w.
\end{array}$$


\end{exa}

\noindent The situation that we have is similar to what one finds in categorical
models of linear logic, where the exponential $!A$, giving arbitrarily
many copies of $A$, is interpreted via a comonad $!$ carrying a
comonoid structure $!A \leftarrow\, !A\otimes !A \rightarrow I$ for
weakening and contraction. In the current situation we have a basis as
a coalgebra $A\rightarrow\, !A$, so that we get a comonoid structure
on $A$, instead of on $!A$.

The next result can be interpreted informally as: base vectors are
copyable.

\begin{prop}
\label{BaseCopyProp}
Assume an algebra $\smash{T(X)\stackrel{a}{\rightarrow} X}$ with an
`element' $\smash{I\stackrel{x}{\rightarrow} X}$ in $\Alg(T)$ that is
in the basis of a coalgebra $b\colon X\rightarrow T(X)$, as in the
equaliser diagram~(\ref{BasisEqualiserDiag}):
$$\xymatrix@C+0.5pc@R-1pc{
X_{b}\ar@{ >->}[r]^-{e} & X \ar@<+.5ex>[r]^-{b}\ar@<-.5ex>[r]_-{\eta} & T(X) \\
I\ar@{-->}[u]\ar[ur]_{x}
}$$

\noindent This $x$ is then copyable, in the sense that the
following diagram commutes
$$\xymatrix@R-1pc{
I\ar[r]^-{x}\ar[d]_{\cong} & X\ar[d]^{d} \\
I\otimes I\ar[r]_-{x\otimes x} & X\otimes X
}$$

\noindent where $d$ is the comultiplication associated with $b$
as in~(\ref{ComonoidDefEqn}).
\end{prop}

\begin{myproof}
Since $I = T(1)$ is a free algebra, the map of algebras $x\colon
I\rightarrow X$ can be written as $x = a \after T(x')$, for the
element $x' = x \after \eta \colon 1 \rightarrow X$ in the underlying
category. Then:
$$\begin{array}{rcll}
b \after x 
\hspace*{\arraycolsep} = \hspace*{\arraycolsep}
b \after a \after T(x') 
& = &
\mu \after T(b) \after T(x \after \eta) 
   & \mbox{see Definition~\ref{BasisDef}} \\
& = &
\mu \after T(\eta) \after T(x \after \eta) \quad
   & \mbox{since $x$ is in the basis $X_b$} \\
& = &
T(x \after \eta) \\
& = &
T(x').
\end{array}$$

\noindent Now we use that $\xi$ is a monoidal isomorphism in:
$$\xymatrix@R-1pc{
\llap{$I=\;$}T(1)\ar[d]_{\rho}^{\cong}
   \ar@/^2ex/[rd]^-{T(\rho) = T(\Delta)} \\
T(1)\otimes T(1)\ar[r]_-{\xi}^-{\cong} & T(1\times 1)
}$$

\noindent in order to prove that $x$ is copyable:
$$\begin{array}[b]{rcll}
(x\otimes x) \after \rho
& = &
(a\otimes a) \after (T(x')\otimes T(x')) \after \xi^{-1} \after T(\Delta) \\
& = &
(a\otimes a) \after \xi^{-1}  \after T(x'\times x')\after T(\Delta) \\
& = &
(a\otimes a) \after \xi^{-1} \after T(\Delta) \after T(x') \\
& = &
(a\otimes a) \after \xi^{-1} \after T(\Delta) \after b \after x
   & \mbox{as just shown} \\
& = &
d \after x & \mbox{with $d$ as in~(\ref{ComonoidDefEqn}).}
\end{array}\eqno{\qEd}$$
\end{myproof}

\noindent The converse of this result is not true: for the ideal monad in
Example~\ref{ComonoidEx} every element is copyable, but not every
element is compact, \textit{i.e.}~in a basis, see
Subsection~\ref{DcpoSubsec}.

Comonoids make tensors cartesian, see~\cite{Fox76,CoeckeP08}, so
that they bring us into the classical world. This cartesian structure
already exists for coalgebras, as the next result shows.

\begin{prop}
\label{CoalgComonoidProductProp}
The comonoid structure~(\ref{ComonoidDefEqn}) restricts to a comonoid
in the category $\smash{\CoAlg(\overline{T})}$ of bases. Therefore, this
category has finite products.  Moreover, the restriction of the free
functor $F\colon \cat{A} \rightarrow \Alg(T)$ to $F\colon \cat{A}
\rightarrow \CoAlg(\overline{T})$, as in Lemma~\ref{FreeAlgCoalgLem},
preserves finite products.
\end{prop}

\begin{myproof}
First one checks that the maps $d,u$ in~(\ref{ComonoidDefEqn}) are
homomorphisms of coalgebras. Then one uses that the comonad
$\smash{\overline{T}} \colon \Alg(T) \rightarrow \Alg(T)$ is monoidal,
see~\cite[Prop.~5.7]{Jacobs94a}, so that the products of coalgebras
can be defined as:
$$\begin{array}{rclcrcl}
1
& = &
\left(\vcenter{\xymatrix@R-1.3pc{
T^{2}(1) \\
T(1)\ar[u]_{\delta}
}}\right)
& \qquad\quad &
\left(\vcenter{\xymatrix@R-1.3pc{
\overline{T}(X_{1}) \\
X_{1}\ar[u]_{b_1}
}}\right)
\times 
\left(\vcenter{\xymatrix@R-1.3pc{
\overline{T}(X_{2}) \\
X_{2}\ar[u]_{b_2}
}}\right)
& = &
\left(\vcenter{\xymatrix@R-1.3pc{
\overline{T}(X_{1}\otimes X_{2}) \\
\overline{T}(X_{1})\otimes \overline{T}(X_{2})\ar[u] \\
X_{1}\otimes X_{2}\ar[u]_{b_{1}\otimes b_{2}}
}}\right)
\end{array}$$

\noindent where on the right-hand-side we use the map:
$$\xymatrix{
FU(X_{1})\otimes FU(X_{2}) \ar[r]^-{\xi}_-{\cong} &
   F\big(U(X_{1})\otimes U(X_{2})\big)\ar[r]^-{F(\sotimes)} & 
   FU(X_{1}\otimes X_{2}),
}$$

\noindent in which $\sotimes$ is the universal bi-homomorphism. \qed
\end{myproof}

\auxproof{
Given a coalgebra $b\colon X\rightarrow \overline{T}(X)$ on $a\colon T(X)
\rightarrow X$, the unit map $u_{b} = T(!) \after b \colon X \rightarrow T(1)$
is a map of algebras \& coalgebras:
$$\begin{array}{rcl}
\mu \after T(u_{b})
& = &
\mu \after T(T(!) \after b) \\
& = &
T(!) \after \mu \after T(b) \\
& = &
T(!) \after b \after a \\
& = &
u_{b} \after a \\
\delta \after u_{b} 
& = &
\delta \after T(!) \after b \\
& = &
T^{2}(!) \after \delta \after b \\
& = &
T^{2} \after T(b) \after b \\
& = &
T(T(!) \after b) \after b \\
& = &
T(u_{b}) \after b.
\end{array}$$

\noindent Moreover, it is the unique one, since if $f\colon
X\rightarrow T(1)$ is also a map of $\overline{T}$-coalgebras, then it
is a map of comonoids, by Proposition~\ref{CoalgComonoidProp}, and
thus:
$$\begin{array}{rcl}
f
& = &
\idmap[T(1)] \after f \\
& = &
u_{T(1)} \after f \qquad \mbox{by~(\ref{FreeAlgCoalgComonoidDiag})} \\
& = &
u_{b} \\
& = &
T(!) \after b.
\end{array}$$

We turn to products, and define projections:
$$\xymatrix{
\pi_{1} = \Big(X_{1}\otimes X_{2}\ar[r]^-{\idmap\otimes u_{2}} &
   X_{1}\otimes T(1)\ar[r]^-{\rho}_-{\cong} & X_{1}\Big)
}$$

\noindent and similarly for $\pi_2$. This gives maps of algebras,
by construction:
$$\xymatrix@R-1pc@C+1pc{
T(X_{1}\otimes X_{2})\ar[d]_{a_{1}\otimes a_{2}}\ar[r]^-{T(\idmap\otimes u_{2})} &
   T(X_{1}\otimes T(1))\ar[d]^{a_{1}\otimes \mu_{1}}\ar[r]^-{T(\rho)} &
   T(X_{1})\ar[d]^{a_{1}} \\
X_{1}\otimes X_{2}\ar[r]_-{\idmap\otimes u_{2}} &
   X_{1}\otimes T(1)\ar[r]_-{\rho} & X_{1}
}$$

\noindent where we have overloaded the $\otimes$-notation. It is
also a map of coalgebras:
$$\begin{array}{rcl}
T(\pi_{1}) \after (b_{1}\times b_{2})
& = &
T(\rho) \after T(\idmap\otimes u_{2}) \after \xi \after (b_{1}\otimes b_{2}) \\
& = &
T(\rho) \after \xi \after (\idmap\otimes T(u_{2})) \after (b_{1}\otimes b_{2}) \\
& = &
T(\rho) \after \xi \after (b_{1}\otimes \delta_{1}) \after 
   (\idmap\otimes u_{2}) \\
& = &
\rho \after (b_{1}\otimes \delta_{1}) \after (\idmap\otimes u_{2}) \\
& = &
b_{1} \after \rho \after (\idmap\otimes u_{2}) \\
& = &
b_{1} \after \pi_{1}
\end{array}$$

Next, assume homomorphisms $f_{i} \colon b \rightarrow b_{i}$. Take as
tuple $\tuple{f_{1}, f_{2}} = (f_{1}\otimes f_{2}) \after d_{b} \colon
X \rightarrow X_{1}\otimes X_{2}$. We need to check that
comultiplications $d$ are maps of coalgebras $b\rightarrow
b\otimes b$. This requires some care, and a better understanding of
the monoidal transformations involved. We recall that the map $\xi
\colon FA \otimes FB \congrightarrow F(A \times B)$, for $A,B\in\cat{A}$,
is obtained in:
$$\xymatrix{
T(X)\times T(Y)\ar[rr]^-{\sotimes = t\after \eta}\ar[drr]_{\dst = \xi^T} & & 
   T(X)\otimes T(Y)\ar@{-->}[d]^{\xi}_{\cong} \\
& & T(X\times Y)
}$$

\noindent with inverse for $\xi$ given by:
$$\xymatrix{
\sigma \stackrel{\textrm{def}}{=} \Big(T(X\times Y)\ar[r]^{T(\eta\times\eta)} &
   T(T(X)\times T(Y))\ar[r]^-{t} & T(X)\otimes T(Y)\Big).
}$$

\noindent The map $\varphi\colon \overline{T}(X_{1})\otimes \overline{T}(X_{2})
\rightarrow \overline{T}(X_{1}\otimes X_{2})$ used in the description
of the product coalgebra is then the composite:
$$\xymatrix{
\varphi = \Big(FU(X_{1}) \otimes FU(X_{2})\ar[r]^-{\xi}_-{\cong} &
   F(U(X_{1})\times U(X_{2}))\ar[r]^-{T(\sotimes)} & FU(X_{1}\otimes X_{2})\Big)
}$$

\noindent where we consider the $X_i$ as algebras.

In order to see that $d$ is a map in $\CoAlg(\overline{T})$ we
need to prove:
$$(b\times b) \after d = T(d) \after b.$$

\noindent We proceed in two steps.
$$\begin{array}{rcl}
(b\times b) \after d
& = &
T(\sotimes) \after \xi \after (b\otimes b) \after
   (a\otimes a) \after \xi^{-1} \after T(d) \after b \\
& = &
T(\sotimes) \after \xi \after (\mu\otimes \mu) \after
   (T(b)\otimes T(b)) \after \xi^{-1} \after T(d) \after b \\
& = &
T(\sotimes) \after \xi \after (\mu\otimes \mu) \after
   \xi^{-1} \after T(b\times b) \after T(d) \after b \\
& = &
T(\sotimes) \after \xi \after (\mu\otimes \mu) \after
   \xi^{-1} \after T(d) \after T(b) \after b \\
& = &
T(\sotimes) \after \xi \after (\mu\otimes \mu) \after
   \xi^{-1} \after T(d) \after T(\eta) \after b \\
& = &
T(\sotimes) \after \xi \after (\mu\otimes \mu) \after
   \xi^{-1} \after T(\eta\times \eta) \after T(d) \after b \\
& = &
T(\sotimes) \after \xi \after (\mu\otimes \mu) \after
   (T(\eta)\otimes T(\eta)) \after \xi^{-1} \after T(d) \after b \\
& = &
T(\sotimes) \after \xi \after \xi^{-1} \after T(d) \after b \\
& = &
T(\sotimes) \after T(d) \after b.
\end{array}$$

\noindent Similarly,
$$\begin{array}{rcl}
T(d) \after b
& = &
T(a\otimes a) \after T(\xi^{-1}) \after T^{2}(d) \after T(b) \after b \\
& = &
T(a\otimes a) \after T(\xi^{-1}) \after T^{2}(d) \after T(\eta) \after b \\
& = &
T(a\otimes a) \after T(\sigma) \after T(\eta) \after T(d) \after b \\
& & \qquad \mbox{with } \sigma = \xi^{-1} = t \after T(\eta\times\eta)
   \mbox{ as described above} \\
& = &
T(a\otimes a) \after T(t) \after T^{2}(\eta\times\eta) \after T(\eta) \after 
   T(d) \after b \\
& = &
T(a\otimes a) \after T(t) \after T(\eta) \after T(\eta\times\eta) \after 
   T(d) \after b \\
& = &
T(a\otimes a) \after T(\sotimes) \after T(\eta\times\eta) \after 
   T(d) \after b \\
& = &
T(\sotimes) \after T(a\times a) \after T(\eta\times\eta) \after 
   T(d) \after b \\
& & \qquad\mbox{because $a$ is a homomorphism of algebras} \\
& = &
T(\sotimes) \after T(d) \after b.
\end{array}$$

We can now prove the familiar tuple equations:
$$\begin{array}{rcl}
\pi_{1} \after \tuple{f_{1}, f_{2}}
& = &
\rho \after (\idmap\otimes u_{2}) \after (f_{1}\otimes f_{2}) \after d \\
& = &
\rho \after (f_{1}\otimes\idmap) \after (\idmap\otimes u) \after d \\
& = &
f_{1} \after \rho \after \rho^{-1} \\
& = &
f_{1} \\
\tuple{f_{1} \after g, f_{2} \after g}
& = &
((f_{1} \after g) \otimes (f_{2} \after g)) \after c \\
& = &
(f_{1}\otimes f_{2}) \after (g\otimes g) \after c \\
& = &
(f_{1}\otimes f_{2}) \after b \after g \\
& & \qquad \mbox{since $g$ is a map of coalgebras, and thus of comonoids} \\
& = &
\tuple{f_{1}, f_{2}} \after g.
\end{array}$$

The final equation requires more work:
$$\begin{array}{rcl}
\tuple{\pi_{1}, \pi_{2}}
& = &
(\pi_{1}\otimes\pi_{2}) \after d_{b_{1}\times b_{2}} \\
& = &
(\rho\otimes\lambda) \after 
   ((\idmap\otimes u_{2}) \otimes (u_{1} \otimes \idmap)) \after \\
& & \qquad
   ((a_{1}\boxtimes a_{2}) \otimes (a_{1} \boxtimes a_{2})) \after
   \xi^{-1} \after T(\Delta) \after (b_{1}\times b_{2}) \\
& = &
(\rho\otimes\lambda) \after 
   ((a_{1}\boxtimes F(1)) \otimes (F(1) \boxtimes a_{2})) \after \\
& & \qquad
   (T(\idmap\otimes u_{2}) \otimes T(u_{1} \otimes \idmap)) \after 
   \xi^{-1} \after T(\Delta) \after \varphi \after (b_{1}\otimes b_{2}) \\
& = &
(a_{1}\otimes a_{2}) \after (T(\rho)\otimes T(\lambda)) \after \xi^{-1} \after \\
& & \qquad
   T((\idmap\otimes u_{2}) \times (u_{1} \otimes \idmap)) \after 
   T(\Delta) \after \varphi \after (b_{1}\otimes b_{2}) \\
& = &
(a_{1}\otimes a_{2}) \after \xi^{-1} \after T(\rho\times \lambda) \after
   T((a_{1}\otimes T(!)) \times (T(!) \otimes a_{2})) \after \\
& & \qquad
   T((b_{1}\otimes b_{2}) \times (b_{1}\otimes b_{2})) \after 
   T(\Delta) \after \varphi \after (b_{1}\otimes b_{2}) \\
& = &
(a_{1}\otimes a_{2}) \after \xi^{-1} \after T(\rho\times \lambda) \after
   T((a_{1}\otimes T(!)) \times (T(!) \otimes a_{2})) \after \\
& & \qquad
   T(\Delta) \after T(b_{1}\otimes b_{2}) \after 
   \varphi \after (b_{1}\otimes b_{2}) \\
& = &
(a_{1}\otimes a_{2}) \after \xi^{-1} \after T(\rho\times \lambda) \after
   T((a_{1}\otimes T(!)) \times (T(!) \otimes a_{2})) \after \\
& & \qquad
   T(\Delta) \after \varphi \after (T(b_{1}) \otimes T(b_{2})) \after 
   (b_{1}\otimes b_{2}) \\
& = &
(a_{1}\otimes a_{2}) \after \xi^{-1} \after T(\rho\times \lambda) \after
   T((a_{1}\otimes T(!)) \times (T(!) \otimes a_{2})) \after \\
& & \qquad
   T(\Delta) \after \varphi \after (T(\eta) \otimes T(\eta)) \after 
   (b_{1}\otimes b_{2}) \\
& = &
(a_{1}\otimes a_{2}) \after \xi^{-1} \after T(\rho\times \lambda) \after
   T((a_{1}\otimes T(!)) \times (T(!) \otimes a_{2})) \after \\
& & \qquad
   T(\Delta) \after T(\eta\otimes\eta) \after 
   \varphi \after (b_{1}\otimes b_{2}) \\
& = &
(a_{1}\otimes a_{2}) \after \xi^{-1} \after T(\rho\times \lambda) \after
   T((a_{1}\otimes T(!)) \times (T(!) \otimes a_{2})) \after \\
& & \qquad
   T((\eta\otimes\eta) \times (\eta\otimes\eta)) \after 
   T(\Delta) \after T(\sotimes) \after \xi \after (b_{1}\otimes b_{2}) \\
& = &
(a_{1}\otimes a_{2}) \after \xi^{-1} \after T(\rho\times \lambda) \after
   T((\idmap\otimes \eta) \times (\eta \otimes \idmap)) \after \\
& & \qquad
   T((\idmap\otimes\,!) \times (!\otimes\idmap)) \after 
   T(\sotimes\times\sotimes) \after T(\Delta) \after \xi \after 
   (b_{1}\otimes b_{2}) \\
& = &
(a_{1}\otimes a_{2}) \after \xi^{-1} \after T(\rho\times \lambda) \after
   T(\sotimes\times\sotimes) \after 
   T((\idmap\times \eta) \times (\eta \times \idmap)) \after \\
& & \qquad
   T((\idmap\times\,!) \times (!\times\idmap)) \after 
   T(\Delta) \after \xi \after (b_{1}\otimes b_{2}) \\
& = &
(a_{1}\otimes a_{2}) \after \xi^{-1} \after 
   T(\rho^{\times}\times \lambda^{\times}) \after \\
& & \qquad
   T((\idmap\times\,!) \times (!\times\idmap)) \after 
   T(\Delta) \after \xi \after (b_{1}\otimes b_{2}) \\
& & \qquad\qquad \mbox{where }
   \rho^{\times} \colon A\times 1 \congrightarrow A, 
   \mbox{see the `auxproof' of Theorem~\ref{MonadAlgStructThm}} \\
& = &
(a_{1}\otimes a_{2}) \after \xi^{-1} \after \xi \after (b_{1}\otimes b_{2}) \\
& = &
\idmap.
\end{array}$$

The restriction of the free functor $F\colon \cat{A} \rightarrow \Alg(T)$
to $F\colon \cat{A} \rightarrow \CoAlg(\overline{T})$, as in
Lemma~\ref{FreeAlgCoalgLem}, preserves finite products, since it
may be described as:
$$\xymatrix{
F(A) = \Big(\mu_{A}\ar[rr]^-{\delta_{\mu_A} = T(\eta_{A})} & & \mu_{TA}\Big)
}$$

\noindent so that $F(1)$ is the final object $1 = \delta_{\mu_1}$ as
described in the proof of~\ref{CoalgComonoidCor}. For binary
products, consider the following diagram in $\cat{A}$, forming
an isomorphism between coalgebras.
$$\hspace*{4em}\xymatrix@R-1pc@C-1pc{
\llap{$F(A\times B) = \Big($}T(A\times B)\ar[rrrr]^-{T(\eta_{A\times B})}
      \ar[drrr]^{T(\eta_{A}\times \eta_{B})}
   & & & & T^{2}(A\times B)\,\Big) \\
\llap{$F(A)\times F(B) = \Big($}T(A)\otimes T(B)\ar[u]^{\xi}_{\cong}
   \ar[rr]_-{T(\eta_{A})\otimes T(\eta_{B})} & &
   T^{2}(A)\otimes T^{2}(B)\ar[r]_-{\xi} &
   T(TA\times TB)\ar[r]_-{T(\sotimes)}\ar[ur]^{T(\dst)} &
   T(TA\otimes TB)\ar[u]_{T(\xi)}^{\cong}\,\Big)
}$$

\noindent The triangle on the right commutes by definition of
$\xi$, see the `auxproof' of Theorem~\ref{MonadAlgStructThm}.
Moreover, this isomorphism $\xi$ commutes with the projections:
$$\xymatrix{
T(A\times B)\ar[r]_-{T(\idmap\times\,!)}\ar @/^4ex/[rr]^-{T(\pi)} &
   T(A\times 1)\ar[r]_-{T(\rho)} & T(A)\ar@{=}[d] \\
T(A)\otimes T(B)\ar[u]^{\xi}_{\cong}\ar[r]^-{\idmap\otimes T(!)}
      \ar @/_4ex/[rr]_-{\pi} &
   T(A)\otimes T(1)\ar[u]^{\xi}_{\cong}\ar[r]^-{\rho} & T(A)
}$$
}

\section{Conclusions}

This paper elaborates the novel view that coalgebras-on-algebras are
bases, in a very general sense. This applies to coalgebras of the
comonad that is canonically induced on a category of algebras of a
monad. Various set-theoretic and order-theoretic examples support this
view. It remains to be investigated to what extent this view also
applies in program semantics, beyond the example of the exception
monad transformer. Also, the connection between bases and copying,
that is so important in quantum mechanics, exists in the current
abstract setting.

\subsection*{Acknoledgements} Thanks are due to Paul Blain Levy, Jorik
Mandemaker, and Paul Taylor, for helpful information and discussions
about the earlier version~\cite{Jacobs11d} of this paper.

\bibliographystyle{plain} 

\begin{thebibliography}{10}

\bibitem{Abramsky10b}
S.~Abramsky.
\newblock No-cloning in categorical quantum mechanics.
\newblock In S.~Gay and I.~Mackie, editors, {\em Semantical Techniques in
  Quantum Computation}, pages 1--28. Cambridge Univ. Press, 2010.

\bibitem{AbramskyH12}
S.~Abramsky and C.~Heunen.
\newblock {H$^\star$-algebras} and nonunital {Frobenius} algebras: first steps
  in infinite-dimensional categorical quantum mechanics.
\newblock {\em Clifford Lectures, AMS Proceedings of Symposia in Applied
  Mathematics}, 71:1--24, 2012.

\bibitem{BanaschewskiB88}
B.~Banaschewski and G.~Br{\"u}mmer.
\newblock Stably continuous frames.
\newblock {\em Math. Proc. Cambridge Phil. Soc.}, 114:7--19, 1988.

\bibitem{Barr69}
M.~Barr.
\newblock Coalgebras in a category of algebras.
\newblock In {\em Category Theory, Homology Theory and their Applications {I}},
  number~86 in Lect. Notes Math., pages 1--12. Springer, Berlin, 1969.

\bibitem{Borceux94}
F.~Borceux.
\newblock {\em Handbook of Categorical Algebra}, volume 50, 51 and 52 of {\em
  Encyclopedia of Mathematics}.
\newblock Cambridge Univ. Press, 1994.

\bibitem{Coecke10a}
B.~Coecke.
\newblock Quantum picturalism.
\newblock {\em Contemp. Physics}, 51(1):59–--83, 2010.

\bibitem{CoeckeP08}
B.~Coecke and D.~Pavlovi{\'c}.
\newblock Quantum measurements without sums.
\newblock In G.~Chen, L.~Kauffman, and S.~Lamonaco, editors, {\em Mathematics
  of Quantum Computing and Technology}, pages 559--596. Taylor and Francis,
  2008.

\bibitem{CoeckePV12}
B.~Coecke, D.~Pavlovi{\'c}, and J.~Vicary.
\newblock A new description of orthogonal bases.
\newblock {\em Math. Struct. in Comp. Sci.}, pages 1--13, 2012.

\bibitem{CoumansJ13}
D.~Coumans and B.~Jacobs.
\newblock Scalars, monads and categories.
\newblock In C.~Heunen, M.~Sadrzadeh, and E.~Grefenstette, editors, {\em
  Quantum Physics and Linguistics. A Compositional, Diagrammatic Discourse},
  pages 184--216. Oxford Univ. Press, 2013.

\bibitem{Escardo98}
M.~Escard{\'o}.
\newblock Properly injective spaces and function spaces.
\newblock {\em Topology and its Applications}, 98(1-2):75--120, 1999.

\bibitem{Fox76}
T.~Fox.
\newblock Coalgebras and cartesian categories.
\newblock {\em Communic. in Algebra}, 4(7):665--667, 1976.

\bibitem{GierzHKLMS03}
G.~Gierz, K.H. Hofmann, , K.~Keimel, J.D. Lawson, M.~Mislove, and D.~Scott.
\newblock {\em Continuous Lattices and Domains}, volume~93 of {\em Encyclopedia
  of Mathematics}.
\newblock Cambridge Univ. Press, 2003.

\bibitem{Hoffmann79}
R.-E. Hoffmann.
\newblock Continuous posets and adjoint sequences.
\newblock {\em Semigroup Forum}, 18:173--188, 1979.

\bibitem{Jacobs94b}
B.~Jacobs.
\newblock Coalgebras and approximation.
\newblock In A.~Nerode and Yu.~V. Matiyasevich, editors, {\em Logical
  Foundations of Computer Science}, number 813 in Lect. Notes Comp. Sci., pages
  173--183. Springer, Berlin, 1994.

\bibitem{Jacobs94a}
B.~Jacobs.
\newblock Semantics of weakening and contraction.
\newblock {\em Ann. Pure \& Appl. Logic}, 69(1):73--106, 1994.

\bibitem{Jacobs10e}
B.~Jacobs.
\newblock Convexity, duality, and effects.
\newblock In C.~Calude and V.~Sassone, editors, {\em IFIP Theoretical Computer
  Science 2010}, number 82(1) in IFIP Adv. in Inf. and Comm. Techn., pages
  1--19. Springer, Boston, 2010.

\bibitem{Jacobs11d}
B.~Jacobs.
\newblock Bases as coalgebras.
\newblock In A.~Corradini, B.~Klin, and C.~C{\"\i}rstea, editors, {\em
  Conference on Algebra and Coalgebra in Computer Science (CALCO 2011)}, number
  6859 in Lect. Notes Comp. Sci., pages 237–--252. Springer, Berlin, 2011.

\bibitem{Jacobs11a}
B.~Jacobs.
\newblock Coalgebraic walks, in quantum and {Turing} computation.
\newblock In M.~Hofmann, editor, {\em Foundations of Software Science and
  Computation Structures}, number 6604 in Lect. Notes Comp. Sci., pages 12--26.
  Springer, Berlin, 2011.

\bibitem{JaskelioffaM10}
M.~Jaskelioffa and E.~Moggi.
\newblock Monad transformers as monoid transformers.
\newblock {\em Theor. Comp. Sci.}, 51-52:4441–--4466, 2010.

\bibitem{Johnstone82}
P.~Johnstone.
\newblock {\em Stone Spaces}.
\newblock Number~3 in Cambridge Studies in Advanced Mathematics. Cambridge
  Univ. Press, 1982.

\bibitem{Kock71b}
A.~Kock.
\newblock Bilinearity and cartesian closed monads.
\newblock {\em Math. Scand.}, 29:161--174, 1971.

\bibitem{Kock71a}
A.~Kock.
\newblock Closed categories generated by commutative monads.
\newblock {\em Journ. Austr. Math. Soc.}, XII:405--424, 1971.

\bibitem{Kock95}
A.~Kock.
\newblock Monads for which structures are adjoint to units.
\newblock {\em Journ. of Pure \& Appl. Algebra}, 104:41--59, 1995.

\bibitem{Levy06}
P.~Levy.
\newblock Monads and adjunctions for global exceptions.
\newblock In {\em Math. Found. of Programming Semantics}, number 158 in Elect.
  Notes in Theor. Comp. Sci., pages 261--287. Elsevier, Amsterdam, 2006.

\bibitem{LiangHJ95}
S.~Liang, P.~Hudak, and M.~Jones.
\newblock Monad transformers and modular interpreters.
\newblock In {\em Principles of Programming Languages}, pages 333–--343. ACM
  Press, 1995.

\bibitem{Mesablishvili06}
B.~Mesablishvili.
\newblock Monads of effective descent type and comonadicity.
\newblock {\em Theory and Applications of Categories}, 16(1):1--45, 2006.

\bibitem{Moggi88}
E.~Moggi.
\newblock Partial morphisms in categories of effective objects.
\newblock {\em Inf. \& Comp.}, 76(2/3):250--277, 1988.

\bibitem{Moggi91a}
E.~Moggi.
\newblock Notions of computation and monads.
\newblock {\em Inf. \& Comp.}, 93(1):55--92, 1991.

\bibitem{NielsenC00}
M.~Nielsen and I.~Chuang.
\newblock {\em Quantum Computation and Quantum Information}.
\newblock Cambridge Univ. Press, 2000.

\bibitem{RoseburghW91}
R.~Rosebrugh and R.~Wood.
\newblock Constructive complete distributivity {II}.
\newblock {\em Math. Proc. Cambridge Phil. Soc.}, 10:245--249, 1991.

\bibitem{SchroderM04}
L.~Schr{\"o}der and T.~Mossakowski.
\newblock Generic exception handling and the {Java} monad.
\newblock In C.~Rattray, S.~Maharaj, and C.~Shankland, editors, {\em Algebraic
  Methods and Software Technology}, number 3116 in Lect. Notes Comp. Sci.,
  pages 443--459. Springer, Berlin, 2004.

\end{thebibliography}

\vspace{-30 pt}
\end{document}